\documentclass[nohyper]{JHEP3}
\pdfoutput=1
\usepackage{amsmath}
\usepackage{amsfonts}
\usepackage{slashed}
\usepackage{graphicx}
\usepackage{verbatim}
\usepackage{subfigure}
\usepackage{cite}

\newcommand \e{\epsilon}

\title{Webs in multiparton scattering using the replica trick}

\author{Einan Gardi\\
School of Physics and Astronomy, The University of Edinburgh, Kings Buildings, Mayfield Road,
 Edinburgh EH9 3JZ, Scotland, UK\\
 E-mail: \email{Einan.Gardi@ed.ac.uk}}
\author{Eric Laenen\\
ITFA, University of Amsterdam, Science Park 904, 1018 XE Amsterdam; \\
ITF, Utrecht University, Leuvenlaan 4, 3584 CE Utrecht;\\
Nikhef Theory Group, Science Park 105, 1098 XG Amsterdam, The Netherlands\\
  E-mail: \email{Eric.Laenen@nikhef.nl}}
\author{Gerben Stavenga\\
Fermi National Accelerator Laboratory, Batavia, IL 60510, USA\\
 E-mail: \email{stavenga@fnal.gov}}
\author{Chris D. White\\
Department of Physics and Astronomy, University of Glasgow, Glasgow G12 8QQ, Scotland, UK;\\
Institute for Particle Physics Phenomenology, Department of Physics, Durham University, Durham DH1 3LE, United Kingdom\\
  E-mail: \email{c.white@physics.gla.ac.uk}}

\abstract{Soft gluon exponentiation in non-abelian gauge theories can be described in terms of webs. 
So far this description has been restricted to amplitudes with two hard partons, where webs were defined as the colour-connected subset of diagrams.  
Here we generalise the concept of webs to the multi-leg case, where the hard interaction involves non-trivial colour flow.
Using the replica trick from statistical physics we solve the combinatorial problem of non-abelian exponentiation to all orders. In particular, we derive an algorithm for computing the colour factor associated with any given diagram in the exponent. 
The emerging result is exponentiation of a sum of webs, where each web is a linear combination of a subset of diagrams that are mutually related by permuting the eikonal gluon attachments to each hard parton. These linear combinations are responsible for partial cancellation of subdivergences, conforming with the renormalization of a multi-leg eikonal vertex. 
We also discuss the generalisation of exponentiation properties to beyond the eikonal approximation. 

}


\preprint{
Edinburgh 2010/20, ITFA-10-18\\
ITP-UU-10/22, IPPP/10/54\\
FERMILAB-PUB-10-234-T\\
NIKHEF/2010-018
}

\begin{document}

\section{Introduction\label{sec:introduction}}

\subsection{Wilson lines and exponentiation\label{sec:Wilson_lines}}

Wilson lines and their renormalisation properties are crucial ingredients in a variety of formal and phenomenological applications of quantum field theory. It is well known that the vacuum expectation value of a product of Wilson line operators, which we may denote generically by ${\cal S}$, renormalises multiplicatively to all orders in perturbation theory~\cite{Polyakov:1980ca,Arefeva:1980zd,Dotsenko:1979wb,Brandt:1981kf}. As a consequence, such quantities ${\cal S}$ obey linear evolution equations of the form
\begin{equation}
\label{evolution}
\mu \frac{d{\cal S}}{d\mu}=-{\cal S}\, \Gamma_{\cal S} , 
\end{equation}  
where $\Gamma_{\cal S}$ is the corresponding anomalous dimension. The solution of this equation has the form
\begin{equation}
{\cal S}=  \,\,\,{\cal P} \,\exp\bigg\{-\frac12 \int_0^{\mu^2} \frac{d\lambda^2}{\lambda^2} \Gamma_{\cal S}(\lambda^2)\bigg\}\,,
\label{Ssol}
\end{equation}
so correlators of Wilson lines generically have an exponential form. 

\begin{figure}[htb]
\begin{center}
\scalebox{0.9}{\includegraphics{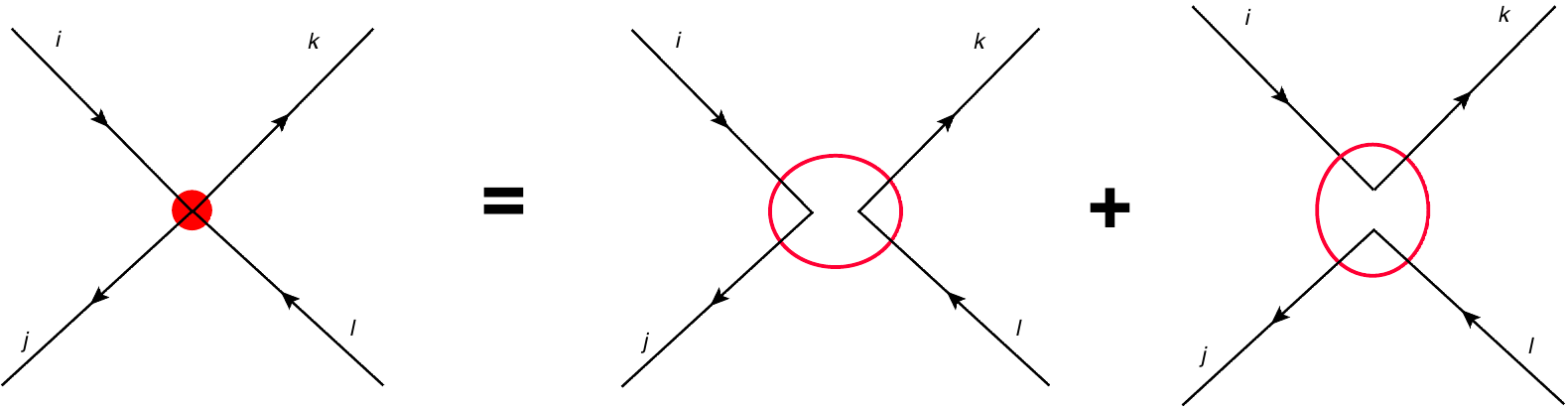}}
\caption{The process $q\bar{q}\to q\bar{q}$ as a simple example of multiple colour flows: 
the two diagrams on the {\it r.h.s.} represent the two possible contractions of colour indices in the hard interaction, namely, $c_1^{ijkl}=\delta_{ij}\delta_{kl}$ and $c_2^{ijkl}=\delta_{ik}\delta_{jl}$.
 Note that gluon exchange between the quark lines mixes between the two hard-scattering colour-flow components, leading to a matrix evolution equation in the two-dimensional colour-flow space defined by the two tensors $c_1$ and $c_2$. }
\label{fig:colour_flows}
\end{center}
\end{figure}

In an Abelian gauge theory, the anomalous dimension is simply a scalar function of the renormalized coupling
and the kinematics and charges of the Wilson lines.
In a non-Abelian gauge theory, things are more complicated due to the non-trivial colour structure. Specifically, for a correlator involving four or more eikonal lines joining at a hard interaction vertex, or equivalently a self-intersecting Wilson loop, there are several different possible contractions of the colour indices, corresponding to different colour flows at the hard vertex, as illustrated in a simple case in figure \ref{fig:colour_flows}. Because gluons carry colour charge, these colour flows mix under evolution, implying that (\ref{evolution}) is a matrix equation. The exponential solution for ${\cal S}$ in (\ref{Ssol}) is then defined through its Taylor expansion and the matrices in each term are ordered in correspondence with the scale $\lambda$ (this is indicated by the symbol~${\cal P}$). 

\begin{figure}[htb]
\begin{center}
\scalebox{0.6}{\includegraphics{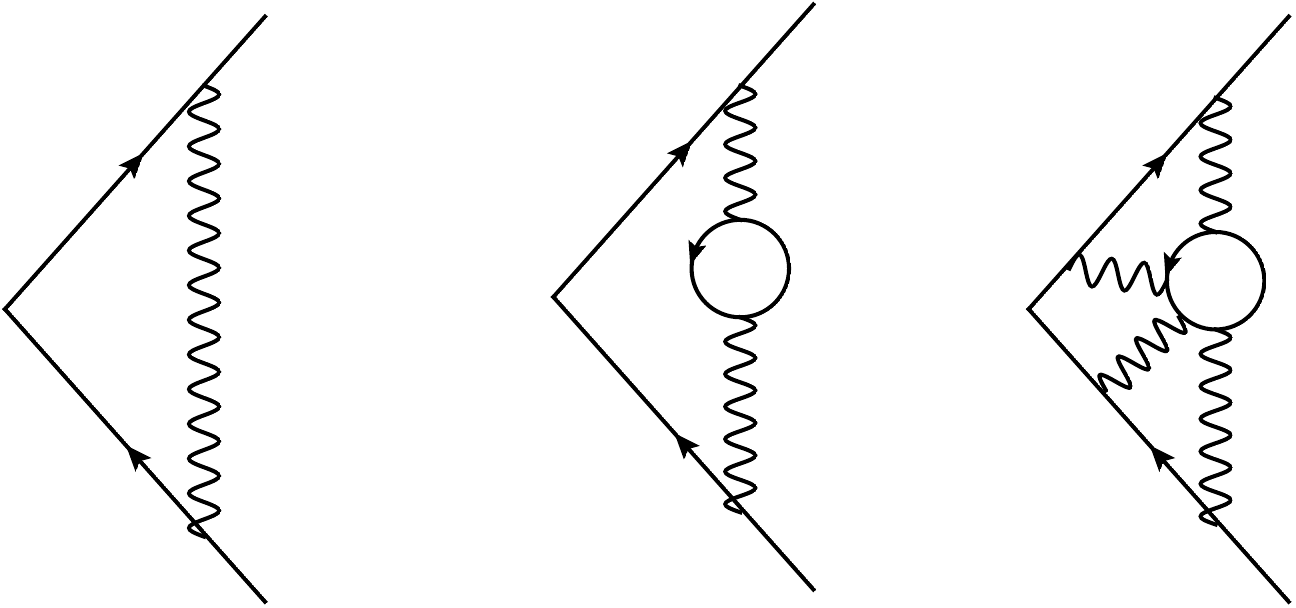}}
\caption{Examples of connected diagrams in an abelian gauge theory; these diagrams exponentiate. The straight lines represent semi-infinite Wilson lines meeting with a cusp at the origin, where a charge-conserving hard interaction takes place. }
\label{connex}
\end{center}
\end{figure}

\begin{figure}[htb]
\begin{center}
\scalebox{0.6}{\includegraphics{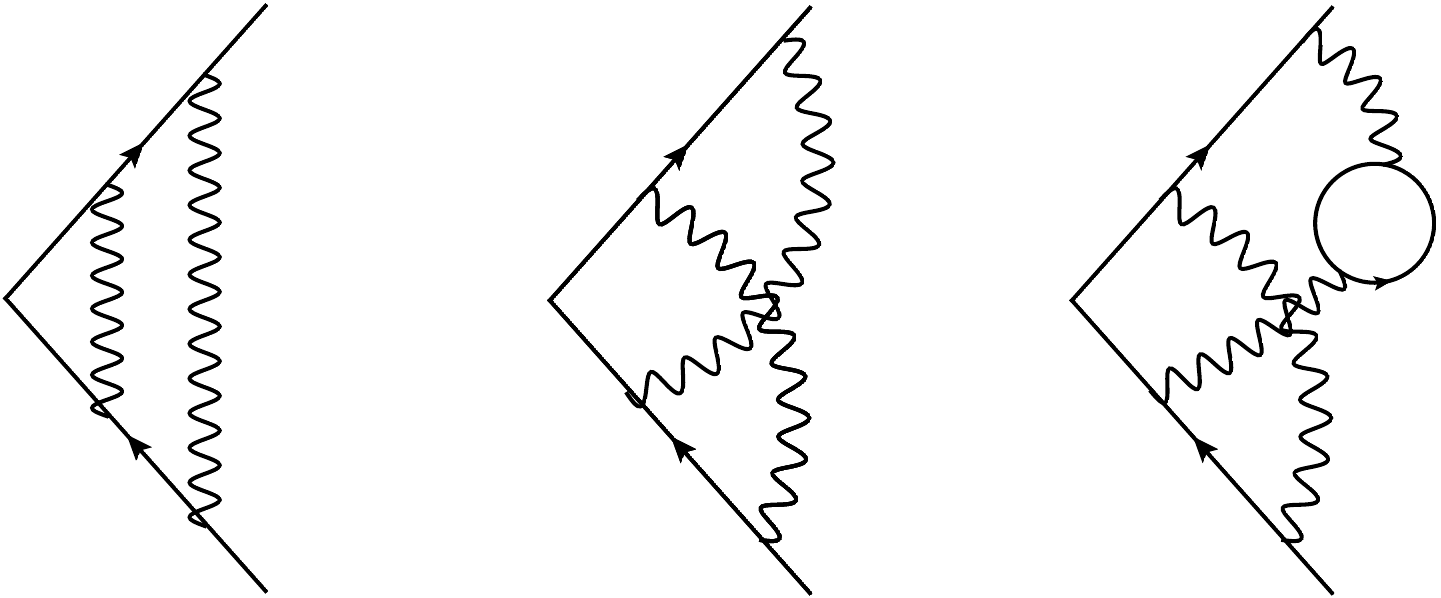}}
\caption{Examples of non-connected diagrams in an abelian gauge theory: these diagrams do \emph{not} contribute to the exponent. The straight lines represent semi-infinite Wilson lines as in figure \ref{connex}.   }
\label{Abelian_non_connected}
\end{center}
\end{figure}

A complementary picture by which exponentiation can be understood is that of `webs'. Considering the configuration of two eikonal lines meeting at a hard vertex (as exemplified by the diagrams of figure~\ref{connex}), webs in an abelian theory are simply the set of connected\footnote{Note that the eikonal lines themselves are excluded: subdiagrams with gluons that are attached to the same eikonal line(s) are not considered connected.\label{fotenote:connected}} diagrams as depicted in the figure. Multiple photon exchange diagrams between the Wilson lines, such as the ladder or crossed diagrams of figure~\ref{Abelian_non_connected}, are generated by exponentiation~\cite{Yennie:1961ad}: they are fully reproduced by higher-order terms in the expansion of the exponential of a single exchange. Therefore, the exponent itself does not include any contributions from multiple-exchange diagrams, and its structure is very simple.

\begin{figure}[htb]
\begin{center}
\scalebox{0.6}{\includegraphics{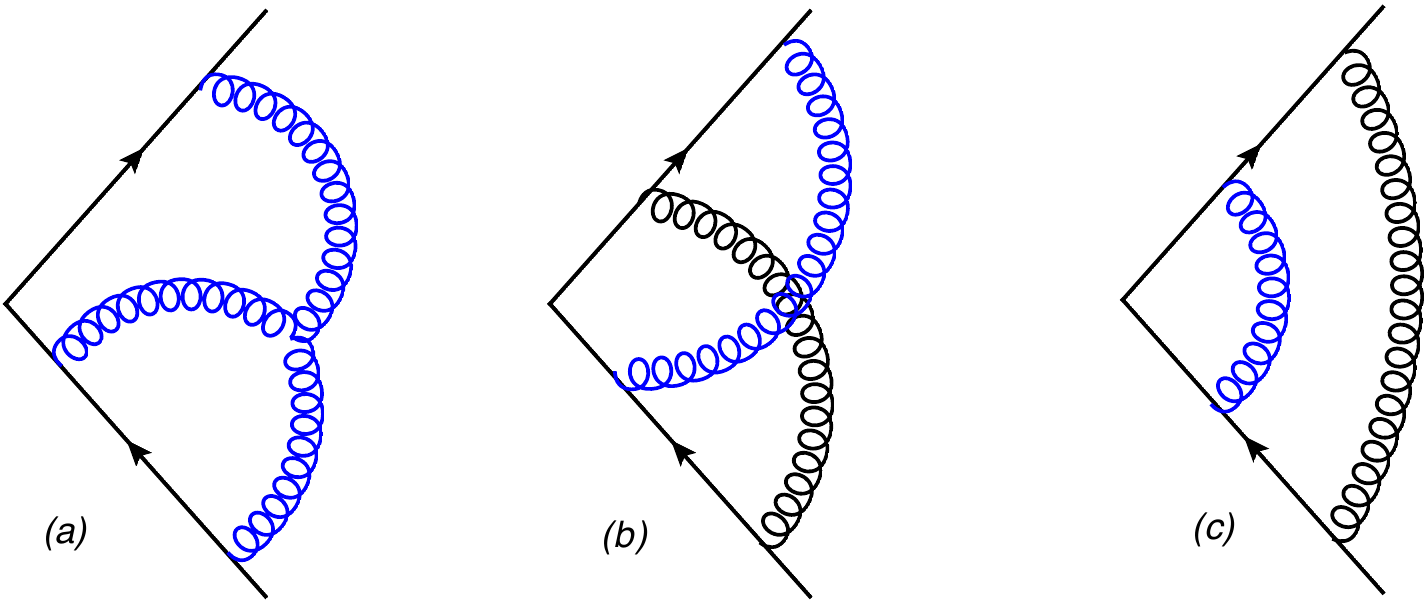}}
\caption{Examples of two-loop diagrams in a non-abelian gauge theory:
(a) connected; (b) colour connected; and (c) non colour connected.
 The straight lines represent semi-infinite Wilson lines (here taken in the fundamental representation of the gauge group) meeting with a cusp at the origin, where a colour-conserving hard interaction takes place. 
The ladder diagram (c) does not contribute to the exponent, while diagrams (a) and (b) do.
However, diagram (b) contributes with an exponentiated colour factor $\widetilde{C}{(b)}=-C_AC_F/2$, instead of its original colour factor $C{(b)}=C_F(C_F-C_A/2)$, whereas for diagram (a) the exponentiated colour factor equals the original one: $\widetilde{C}{(a)}=C{(a)}$. }
\label{Non_Abelian_non_connected}
\end{center}
\end{figure}

In a non-abelian theory the situation is complicated: multiple interactions with a given Wilson line involve colour generators, and therefore do not commute. Nevertheless, it is known for the specific case of two eikonal lines (with a colour-singlet hard interaction) that exponentiation in terms of webs can indeed be generalised to the non-Abelian case~\cite{Sterman:1981jc,Gatheral:1983cz,Frenkel:1984pz}. That is, the relevant correlator of Wilson lines has the form
\begin{equation}
\label{web_exponentiation}
{\cal S}=  \,\, \,\exp \bigg\{\sum_{D} {\cal F}(D)\,\widetilde{C}(D)\,\bigg\}\,,
\end{equation}
wher ${\cal F}(D)$ and $\widetilde{C}(D)$ denote, respectively, the kinematic dependence and the ``Exponentiated Colour Factor'' (ECF) of a given Feynman diagram $D$.
This expression has the same schematic form as the abelian case of figure~\ref{connex}, but the contributing class of diagrams is larger, and has a more involved structure: this class comprises all ``colour-connected" diagrams, namely the ones that cannot be partitioned by cutting only the eikonal lines.  Two-loop examples for connected, colour-connected and non-colour-connected diagrams in a non-abelian theory are shown in figure~\ref{Non_Abelian_non_connected}.

While the kinematic dependence of a given diagram, ${\cal F}(D)$ in (\ref{web_exponentiation}), is computed as usual using the eikonal Feynman rules, the corresponding colour factor $\widetilde{C}(D)$ differs from the original colour factor associated with that diagram, $C(D)$. Gatheral~\cite{Gatheral:1983cz} derived an iterative formula by which these ECFs $\widetilde{C}(D)$ can be computed, and characterized them as ``maximally non-abelian". Soon thereafter Frenkel and Taylor~\cite{Frenkel:1984pz} clarified the precise meaning of this term, defining a ``web'' as a  ``colour-connected" diagram.

The conclusion of Ref.~\cite{Sterman:1981jc,Gatheral:1983cz,Frenkel:1984pz} is then that in the two-eikonal line case, the diagrams $D$ that contribute 
to the exponent, that is the ``webs'', are the {\it two-eikonal-line irreducible} diagrams. We shall see that the definition of webs will have to be revised in the multi-leg case.

Despite the fact that webs provide a direct graphical description of the exponent, they have so far been defined only in the particular case of a colour-singlet hard interaction with two eikonal lines\footnote{The three eikonal line case was analysed in Sec. 3.3 of Ref.~\cite{Berger:2003zh}. Owing to the fact that for three lines (in the fundamental or adjoint representations) there is a single colour flow, this case ends up being a straightforward generalization of the two-eikonal line case, where only irreducible diagrams appear in the exponent.}. 
The definitions, and indeed the formalism for computing the ECF of  
Refs.~\cite{Gatheral:1983cz,Frenkel:1984pz} apply to a Wilson loop with any number of cusps or colour-singlet hard interactions, but they do not apply to self-intersecting Wilson loops which give rise to multiple colour flows.
 
The purpose of the present study is to extend the concept of webs to multi-parton eikonal interactions, or equivalently to self-intersecting Wilson loops with multiple colour flows. We will consider correlators of an arbitrary number of semi-infinite Wilson lines joining up at a single vertex (corresponding to a single hard interaction with an arbitrary colour exchange), and show that webs can be systematically defined, leading to exponentiation in the form of (\ref{web_exponentiation}). Furthermore, we will present a general algorithm by which the ECF of any multi-loop (and multi-leg) diagram can be directly computed as a linear combination of ordinary colour factors of a definite set of diagrams, and present explicit results in several non-trivial cases.

\subsection{Motivation}

While our method is completely general, our motivation stems from concrete applications to scattering amplitudes. A major long-term goal is to have a complete understanding of long-distance singularities in multi-leg gauge-theory scattering amplitudes with arbitrary kinematics and colour representations. It is known that long-distance singularities exponentiate, where the exponent admits an all-order structure which is far simpler than the amplitude as a whole. The singularities therefore provide a window by which the all-order structure of perturbation theory can be explored.
The general singularity structure is also universal amongst different gauge theories. 
In particular, a large class of singularities in any amplitude are controlled by the cusp anomalous dimension 
$\gamma_K (\alpha_s)$~\cite{Polyakov:1980ca,Korchemsky:1985xj,Ivanov:1985np,Korchemsky:1987wg,Korchemsky:1988hd,Korchemsky:1988si},
associated with the renormalization of Wilson lines with a cusp. This object is by now well under control in QCD~\cite{Moch:2004pa}, and even more so in the case of ${\cal N}=4$ super-Yang-Mills, where owing to integrability, a connection was established between the weak and strong coupling expansions~\cite{Alday:2007hr,Beisert:2005fw,Beisert:2006ez,Basso:2007wd}.
 
Beyond the obvious field-theoretic interest in long-distance singularities, determining their structure is essential for precision collider physics.
Indeed, the most difficult aspect of implementing cross-section computations with general kinematics beyond the tree level in QCD is the fact that infrared singularities only cancel after summing up real and virtual diagrams which are integrated over phase space with different numbers of partons.
Knowing the singularity structure is therefore a prior condition for any loop-level computation.  
Moreover, the singularity structure of gauge-theory amplitudes provides the key to resummation of large logarithms~\cite{Cacciari:2002xb,Gardi:2005yi,Korchemsky:1992xv,Korchemsky:1993uz,Korchemsky:1994jb,Beenakker:2010fw,Beneke:2010gm,Papaefstathiou:2010bw,Chien:2010kc,Idilbi:2009cc,Sjodahl:2009wx,Debove:2009ia,Almeida:2009jt,Czakon:2009zw,Moch:2009mu,Moch:2009my,Mantry:2009qz,Becher:2006mr,Becher:2007ty,Becher:2008cf,Ahrens:2008nc,Bozzi:2008bb,Kang:2008zzd,Kidonakis:2009ev,Catani:1996yz,Bonciani:1998vc,Bonciani:2003nt,Andersen:2005mj,Gardi:2007ma,Gardi:2002bg,Gardi:2001ny,Banfi:2010xy,Catani:1991kz,Catani:1990rp,Catani:1989ne,Sterman:1986aj,Laenen:2000ij,Bozzi:2007pn}, which in many cases is essential for providing precise predictions for cross sections.
So far resummation has mainly been applied to completely inclusive cross sections, where typically only two coloured partons participate in the hard interaction. LHC physics requires resummed calculations for less inclusive observables and more complex processes, where more partons participate in the hard interaction. Better understanding of long-distance singularities, and in particular of the correlation between momentum and colour flow, is essential for progress on this front.

\begin{figure}[htb]
\begin{center}
\scalebox{0.6}{\includegraphics{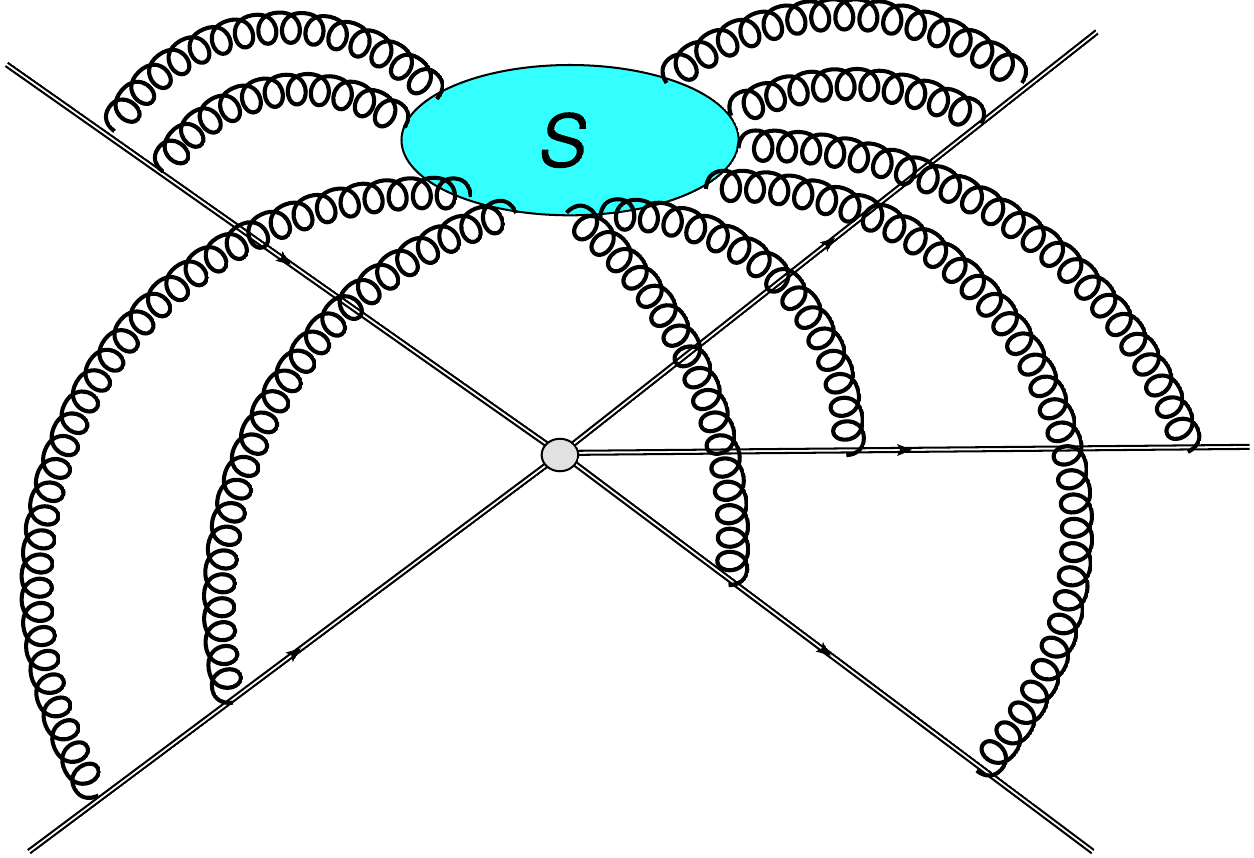}}
\caption{The soft function in the scattering amplitude involving five hard coloured partons. The hard interaction structure cannot be probed by soft (long wavelength) gluons, and therefore it reduces to a local vertex. Similarly, the partonic jets are not resolved and couple to the soft fields through eikonal vertices.}
\label{fig:soft_def}
\end{center}
\end{figure}
  
Over the past couple of years there has been much progress in understanding long-distance singularities of gauge-theory scattering amplitudes for both the
massless~\cite{Aybat:2006mzy,Dixon:2008gr,Gardi:2009qi,Dixon:2009gx,Becher:2009cu,Becher:2009qa,Dixon:2009ur,Gardi:2009zv,Dixon:2010hy,Gehrmann:2010ue} and massive~\cite{Mitov:2009sv,Kidonakis:2009ev,Becher:2009kw,Beneke:2009rj,Czakon:2009zw,Ferroglia:2009ep,Ferroglia:2009ii,Chiu:2009mg,Mitov:2010xw,Ferroglia:2010mi} cases, by considering the factorization of fixed-angle amplitudes into hard, jet and soft components. 
The soft component, which captures all non-collinear long-distance singularities, is defined as a correlator of semi-infinite eikonal lines joined together at a single interaction vertex, as shown schematically in figure \ref{fig:soft_def}.  
The soft function~${\cal S}$ is, in general, a matrix in colour space~\cite{Kidonakis:1998nf,Catani:1998bh,Sterman:2002qn,Kyrieleis:2005dt,Sjodahl:2008fz,Sjodahl:2009wx}, and it obeys an evolution equation of the form of eq.~(\ref{evolution}), leading to exponentiation in terms of the ``soft anomalous dimension'' matrix $\Gamma_{\cal S}$ (this object is also sometimes referred to as the ``cross anomalous dimension''~\cite{Brandt:1981kf,Korchemskaya:1994qp}).  
This anomalous dimension encodes all correlations between momentum and colour flow, which are only visible considering the full non-abelian theory at finite $N_c$.  
Concrete motivation to study this function beyond the two-loop level is provided by the fact that in the massless case its structure is strongly constrained~\cite{Gardi:2009qi,Becher:2009qa} by both factorization, and a rescaling symmetry with respect to the Wilson-line velocities. 
These constraints point to a remarkable possibility, namely that all soft singularities in any multi-leg amplitude take the form of a sum over colour dipoles formed by any pair of hard coloured partons~\cite{Gardi:2009qi,Becher:2009cu,Becher:2009qa}.
This very minimal structure, not involving any multi-parton correlations, is consistent with all explicit amplitude calculations done so far.
Only very particular types of corrections, depending on conformally--invariant cross ratios, may occur at the three-loop order~\cite{Gardi:2009qi,Becher:2009qa,Dixon:2009ur}. An additional type of multi-parton correlation may potentially show up starting from four loops, through higher Casimir terms.
Further progress in studying the soft anomalous dimension for multi-leg scattering at three-loops and beyond requires direct understanding of non-abelian exponentiation, for which webs are the basic ingredient. 

A further motivation is provided by the prospect of resumming logarithms near kinematic thresholds beyond the leading power in the momentum of emitted soft gluons, or equivalently, beyond the eikonal approximation~\cite{Grunberg:2009yi,Grunberg:2009vs,Moch:2009mu,Moch:2009my,Moch:2009hr,Soar:2009yh,Vogt:2010cv,Dokshitzer:2005bf,Laenen:2008ux,Laenen:2008gt}.  A systematic classification of the structure of next-to-eikonal corrections has been given in~\cite{Laenen:2008gt} for the case of two hard partons coupled by a colour singlet interaction (the setup originally considered by Gatheral). It was shown that a subclass of next-to-eikonal terms formally exponentiates, and the relevant diagrams constitute a well-defined generalisation of eikonal webs. These results suggest that it is also possible to classify those next-to-eikonal contributions which exponentiate in the multiparton case, provided one can also classify eikonal webs. We will see that this is indeed the case. Furthermore, webs are beneficial even in eikonal resummation applications, where an anomalous dimension description already exists. Essentially, webs contain {\it more information} than anomalous dimensions (see~\cite{Laenen:2000ij}), namely finite parts as well as singular terms. Classification of these terms is crucial to their resummation, which can be phenomenologically important. 

Finally, another theoretical motivation to have a direct handle on exponentiation via webs 
is provided by the observation that exponentiation of gauge-theory amplitudes, at least in some cases, appears to extend beyond the singular terms, and apply to the amplitude as a whole. 
An interesting case where complete exponentiation was conjectured to hold~\cite{Bern:2005iz} is the class of maximally helicity violating (MHV) amplitudes in ${\cal N}=4$ super-Yang-Mills (SYM).  While the very simple exponential form has later been shown to fail starting with the two-loop six-point function~\cite{Bern:2008ap}, ${\cal N} = 4$ SYM perturbative amplitudes certainly admit an iterative structure also in their finite parts. 
A further step in revealing the structure of these amplitudes was the discovery
of a surprising duality between scattering amplitudes in momentum space 
and expectation values of Wilson loops taken in an auxiliary coordinate 
space~\cite{Drummond:2007aua,Brandhuber:2007yx,Drummond:2007cf,Drummond:2007au,Drummond:2007bm,Drummond:2008aq}. Further, much progress was made on the same amplitude -- Wilson-loop relation at strong coupling, see in particular Refs.~\cite{Alday:2007hr,Alday:2008yw,Alday:2009dv}.
Importantly, all of this profound theoretical progress has been limited to the planar limit (see, however, \cite{Naculich:2009cv}), and it would be very interesting to extend it beyond this limit, where the non-trivial aspects of colour flow reveal themselves. 

Our goal in this paper, as stated above, is to extend the notion of web exponentiation to the case of multiparton scattering, or equivalently to self-intersecting Wilson lines.
To this end, we adopt the approach of~\cite{Laenen:2008gt}, which uses path integral methods to relate the amplitude for soft gluon emission in a given process to a field theory for the soft gauge field. 
In this way soft gluon exponentiation can be easily formulated to all orders. For example, in the abelian case it is equivalent to the known exponentiation of connected diagrams in quantum field theory. 
Furthermore, the path integral method allows a straightforward generalisation of exponentiation properties to beyond the eikonal approximation. 
In the non-Abelian case, the field theory obtained in \cite{Laenen:2008gt} 
for the soft gauge field was complicated by the non-commuting nature of the source vertices for soft gauge boson emission. It was still possible to classify which diagrams contribute in the exponent of the scattering amplitude, however, by using the {\it replica trick}, a combinatoric method often used in statistical physics applications (see e.g.~\cite{Replica}). The diagrams thus obtained were shown to be entirely equivalent to the webs of~\cite{Gatheral:1983cz,Frenkel:1984pz}. These results suggest that a similar combination of methods will prove fruitful in analysing the multiparton case. We will see that this is indeed true, once one generalises the replica trick argument from the two parton case considered in~\cite{Laenen:2008gt}.

\subsection{Results and outline of the paper}

We may summarise our main results as follows. Firstly, we provide an algorithmic method for determining which graphs enter the exponent of eikonal multiparton scattering amplitudes, and for computing their exponentiated colour factors (ECFs), to arbitrary order in perturbation theory. Secondly, we derive a combinatoric formula for the ECF of any graph. Thirdly, we show that the general structure of the exponent of the eikonal amplitude consists of sums of terms of the form
\begin{equation}
W_{(n_1,n_2,\ldots,n_L)}\,\equiv\, \sum_{D}{\cal F}(D)\,\, \widetilde{C}(D)
=\sum_{D,D'}{\cal F}(D)\,\,R_{DD'}\,\,C(D')\,,
\label{setmixintro}
\end{equation}
where the indices $D$ and $D'$ label Feynman diagrams, $C(D)$ and $\widetilde{C}(D)$ denote, respectively, the conventional and the exponentiated colour factors of diagram $D$, and ${\cal F}(D)$ the kinematic part. 
As is perhaps already clear from the structure of eq.~(\ref{setmixintro}), one finds closed sets of diagrams which mix with each other in the exponent. 
Each such set of diagrams can be labelled by the number of gluon attachments $n_i$ to each of the eikonal lines $i=1\ldots L$. Distinct diagrams $D$ in the set differ only by the order of attachments of the gluons to each eikonal line. 
Associated with each such set $(n_1,n_2,\ldots,n_L)$ is a mixing matrix $R_{DD'}$ describing how the exponentiated colour factors differ from the conventional ones. 
Combinations of terms such as that shown in eq.~(\ref{setmixintro}) are the generalisation of webs from the two-eikonal lines to the multi-line case. 
The study of web structure is then equivalent to the study of the mixing matrices $R_{DD'}$. In particular, we observe that these matrices have the property
\begin{equation}
\sum_{D'}R_{DD'}=0.
\label{Rdd0}
\end{equation}
As discussed in section~\ref{sec:fourloop}, this corresponds to the fact that the symmetric colour part of any diagram ({\it i.e.} that part of the colour factor that does not depend on the ordering of the soft gluon attachments to a given eikonal line) does not contribute to the exponent of the eikonal amplitude. This is the appropriate generalisation of the notion of antisymmetrisation of gluons previously noted at two loops.

We also relate further remarkable properties of the matrices $R$ to the cancellation of subdivergences in the exponent. We observe that $R$ is an idempotent matrix, $R^2=R$, and indeed it acts as a projection operator. Its eigenvalues are consisting only of 0 and~1. These pick out linear combinations of kinematic factors, where the zero eigenvalue combinations correspond to those which are generated by the exponentiation of lower order webs, and thus do not contribute to the exponent. The eigenvalue one combinations build up in the exponent. We find that the corresponding linear combination of kinematic factors are special in that they encode intricate cancellation of subdivergences, allowing the webs to conform with the renormalization of the multi-eikonal vertex~\cite{MSS}.

Finally, we argue that a subclass of soft gluon corrections exponentiates at next-to-eikonal order in multiparton scattering amplitudes. This generalises the previous results obtained for two parton scattering~\cite{Laenen:2008gt}, and combines the path integral approach of that paper with the generalised replica trick argument developed in the present paper. 

The structure of the paper is as follows. In section~\ref{sec:exponentiation}, we summarise in more detail previously known results on non-Abelian exponentiation, in particular the general singularity structure and the relationship between the soft anomalous dimension and web frameworks. In section~\ref{sec:replica} we set up the replica trick for multiparton scattering and establish a general algorithm for computing the ECF of a given diagram. 
We then use this in section~\ref{sec:formula} to provide explicit combinatoric results for ECFs in terms of conventional colour factors, as well as an inverse relation. 
In section~\ref{sec:special_cases} we provide several demonstrations of how to apply the new formalism; after showing how known results are reproduced, we analyse higher-order diagrams in perturbation theory and investigate the properties of the ECFs. Specifically, in section \ref{sec:ECF-3loops} and in Appendix \ref{sec:more-3loop-ECF} we compute all the non-trivial ECFs at three loops, and in section \ref{sec:fourloop} we analyse an interesting four-loop class of diagrams and then discuss some general features of the mixing matrices $R_{DD'}$. In section 6 we examine the cancellation of subdivergences in the exponent. In section~\ref{sec:next_to_eikonal} we discuss the exponentiation of next-to-eikonal corrections in multiparton scattering. In section~\ref{discuss} we present our conclusions.

\section{Non-abelian exponentiation via evolution equations\label{sec:exponentiation}}

Exponentiation is a fundamental property of correlators of Wilson lines, not of scattering amplitudes.
Nevertheless, it becomes relevant to the structure of amplitudes owing to the fact that amplitudes factorize, in such a way that their infrared singularities are captured by \emph{ultraviolet} singularities of corresponding correlators of eikonal lines~\cite{Korchemsky:1985xj,Ivanov:1985np,Korchemsky:1987wg,Korchemsky:1988hd,Korchemsky:1988si}, 
a concept which also underlies the formulation of soft-collinear effective theory \cite{Bauer:2000yr,Bauer:2001yt,Beneke:2002ph}.  
In this section we review the factorization of fixed-angle multiparton scattering amplitudes and then define and discuss the evolution of the corresponding ``soft function'' (or ``eikonal amplitude") which will be the central object we analyse in the remainder of this paper. 

The qualitative difference between the general $L$-parton scattering case we consider here and the two-parton case where the singularity structure is fully understood, is the presence of multiple colour flows, as illustrated in figure \ref{fig:colour_flows} above.  Let us therefore begin by introducing the notation for colour flow in the general case. 
The amplitude ${\cal M}$ describes the scattering of $L$ hard (massless or massive) gauge particles with momenta~$p_i$, plus any number of colour-singlet particles, so it is characterized by $L$ colour indices $\left\{a_i\right\}$, $i = 1,\ldots L$,	belonging to	arbitrary representations	of the gauge group; we denote the corresponding generators by ${\mathrm T}_i$.	
${\cal M}$~can be decomposed into components by choosing a basis of independent colour tensors with the same index structure. We denote these tensors by $(c_K)_{\left\{a_i\right\}}$, where $K$ runs over all irreducible representations of the gauge group that can be constructed with the given indices $\left\{a_i\right\}$. The decomposed amplitude in this colour-flow basis is
\begin{equation}
\label{colour_flow_decomposition_amp}
  {\cal M}_{\{a_i\}} \left(p_i, \alpha_s, \epsilon \right)  =  
 \sum_{K} {\cal M}_{K}  \left(p_i, \alpha_s, \epsilon \right)
  \, \left(c_K\right)_{\{a_i\}} \, ,
\end{equation}
where $\alpha_s=\alpha_s(\mu^2)$ is the renormalized coupling in $d=4-2\epsilon$ dimensions.
A simple example for such a decomposition is given in figure~\ref{fig:colour_flows}.

From the soft physics perspective, hard partons are seen merely as classical sources of radiation, eikonal lines whose kinematics is fixed: they do not recoil upon emitting soft gluons. This is the essence of factorization.
 
Having factorized the amplitude, the soft physics is described by the ultraviolet singularities of operators composed of products of eikonal lines. These operators renormalize multiplicatively, involving the so-called 
``soft anomalous dimension'' matrix (\ref{evolution}). Upon solving the corresponding evolution equations one recovers the infrared singularities of the amplitude. 
Much of what we know today about the all-order structure of infrared singularities is based on this formalism. 

In the following two subsections we briefly summarise the picture of factorization of scattering amplitudes in the massive and massless cases, and explain how it constrains the all-order structure of long-distance singularities, which are encoded in the soft anomalous dimension. 
In this discussion we use dimensional regularization throughout.
Subsequently we shall address the relation with the complementary picture of exponentiation through webs, which we develop in the rest of this paper.

\subsection{The singularity structure in the massive case}

Next we turn to discuss the singularity structure of ${\cal M}_{K}$ using the basic tool of factorization into hard, jet and soft components~\cite{Mueller:1979ih,Sen:1981sd,Sen:1982bt,Collins:1980ih,Sen:1982bt,Kidonakis:1998nf,Sterman:2002qn,Aybat:2006mz,Dixon:2008gr,Gardi:2009qi,Becher:2009cu,Becher:2009qa,Dixon:2009gx}.
For simplicity consider first the case where all the hard partons are massive. In this case long-distance singularities only arise from the exchange of soft gluons (gluons with vanishing energy), and power counting shows that there is at most one such singularity per loop, {\it i.e.} at 
${\cal O}(\alpha_s^n)$ the renormalized amplitude has poles ${\cal O}(1/\epsilon^k)$, with $k=1\dots n$. At each order, only the ${\cal O}(1/\epsilon^1)$ pole is ``new'', namely computing it requires an ${\cal O}(\alpha_s^n)$ calculation, whereas all the higher poles can be determined from lower orders using evolution equations.
\begin{figure}[htb]
\begin{center}
\scalebox{1}{\includegraphics{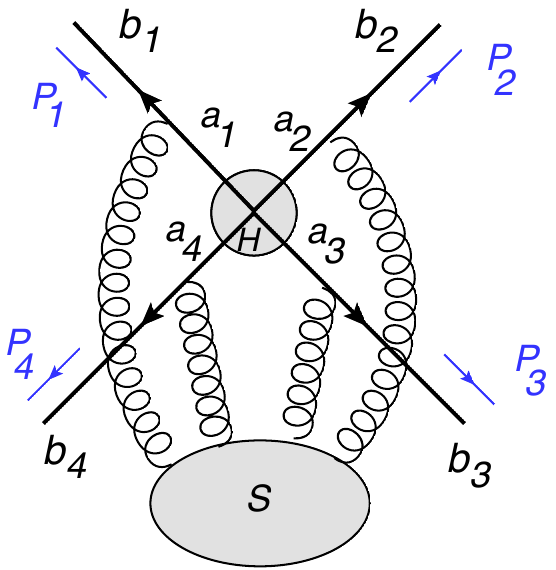}}
\caption{Factorization into soft and hard functions in the example of four parton scattering with momenta $p_i$. 
The four lines represent Wilson lines along the directions $p_i$ in the colour representation corresponding to the respective hard parton.
The labels $b_i$, for $i=1\ldots 4$, on the external lines are the colour indices of the eikonal amplitude, while the corresponding labels $a_i$ are the colour indices of the hard subprocess $H$ at the origin. }
\label{fig:soft_hard_factorization}
\end{center}
\end{figure}

If we assume that all kinematic invariants $p_i\cdot p_j$, including the masses $p_i^2=m_i^2$, are taken simultaneously large compared to $\Lambda_{\rm QCD}^2$, soft gluons completely decouple from the hard interaction: soft singularities are entirely independent of the overall hard scale.  
We may then factorise the amplitudes into a soft function ${\cal S}$, a matrix in the colour-flow space defined above, acting on the hard subprocess $H$, which similarly to ${\cal M}$ is a vector in this space.   
The essence of this factorization is the fact that all the singularities of the amplitude are captured by ${\cal S}$, while $H$ is finite. This separation is useful because ${\cal S}$ has a much simpler structure as compared to the full amplitude; for example, it does not depend on the spin of the scattered particles.

As illustrated in figure~\ref{fig:soft_hard_factorization}, the factorized amplitude takes the form:
\begin{align}
\label{M_SH_fact}
\begin{split}
{\cal M}_{J} \left(p_i, \alpha_s,
  \epsilon \right) & =  \sum_K
  {\cal S}_{J K} \left(\frac{(\beta_i \cdot \beta_j)^2}{\beta_i^2\,\beta_j^2}, \alpha_s, \epsilon 
  \right) \,  H_{K} \left( \frac{2 p_i \cdot p_j}{\mu^2},\frac{m_i^2}{\mu^2}, \alpha_s,\epsilon \right)\,,
\end{split}
\end{align}
where the soft function is defined as the vacuum expectation value of the following product of Wilson lines,
\begin{equation}
\left( c_J \right)_{\{a_k\}} {\cal S}_{J K} \left(\frac{(\beta_i \cdot \beta_j)^2}{\beta_i^2\,\beta_j^2}, \alpha_s(\mu^2), \epsilon \right) \equiv \sum_{\{b_k\}} \, \, \left\langle 0 \left|  \,
\prod_{i = 1}^L \Big[ \Phi_{\beta_i} (\infty, 0)_{a_kb_k} 
\Big]  \, \right| 0 \right\rangle  \, \left( c_K \right)_{\{b_k\}} \, .
\label{softcorr}
\end{equation}
The colour index structure of the Wilson lines is illustrated in figure~\ref{fig:soft_hard_factorization}. In (\ref{softcorr}) we projected the Wilson line correlator onto the colour-flow basis defined in terms of the tensors $\left\{ c_K \right\}$.
Each Wilson line is defined as usual, as a path-ordered exponential: 
\begin{equation}
\label{Wilson_line_def}
\Phi^{(l)}_{a_lb_l}\,\equiv\,
\left({\cal P}\exp\left[{\mathrm i}g_s \int_{{\cal C}_l} dz_l^\mu{A}_\mu\right]\right)_{a_lb_l}\,\,
=\,\,\left({\cal P}\exp\left[{\mathrm i}g_s \int_0^{\infty}dt \beta_l\cdot {A}(t\beta_l)\right]\right)_{a_lb_l}\,,
\end{equation}
going along the classical trajectory ${\cal C}_l$ of the hard parton $l$, that is starting at the origin and  going in a straight line, with 4-velocity $\beta_l$ proportional to its momentum $p_l$. 
The first argument of ${\cal S}$ indicates the dependence of this function on the kinematics: given $L$ legs, it depends on $L(L-1)/2$ Lorentz invariants. The ratio ${(\beta_i \cdot \beta_j)^2}/({\beta_i^2\,\beta_j^2})$ is the hyperbolic cosine of the cusp angle formed by the two Wilson lines $\beta_i$ and $\beta_j$ in Minkowski space-time. The dependence on this ratio -- rather than on $\beta_i\cdot \beta_j$ directly --- is dictated by the invariance of the eikonal Feynman rules with respect to rescaling the velocities.

The soft function ${\cal S}$ is defined as an operator composed of Wilson lines, and therefore it is multiplicatively renormalizable~\cite{Dotsenko:1979wb,Brandt:1981kf}. However, being a matrix in colour space its renormalization involves mixing~\cite{Brandt:1981kf}, 
so the evolution equation takes a matrix form:
\begin{equation}
\mu  \frac{d}{d \mu} {\cal S}_{J K} \left(\frac{(\beta_i \cdot \beta_j)^2}{\beta_i^2\,\beta_j^2}, \alpha_s(\mu^2), \epsilon \right) = - \, 
\, {\cal S}_{J I} \left(\frac{(\beta_i \cdot \beta_j)^2}{\beta_i^2\,\beta_j^2}, \alpha_s(\mu^2), \epsilon \right) \, \Gamma^{{\cal S}}_{I K} \left(\frac{(\beta_i \cdot \beta_j)^2}{\beta_i^2\,\beta_j^2}, \alpha_s(\mu^2)
\right) .
\label{renS}
\end{equation}
Here $\Gamma^{{\cal S}}_{I K}$ is the ``soft anomalous dimension" (it is called the ``cross anomalous dimension" in Ref.~\cite{Brandt:1981kf,Korchemskaya:1994qp}). It is a finite function of the kinematics and the $d=4-2\epsilon$ dimensional coupling constant.
As shown in \cite{Magnea:1990zb}, evolution equations of this type can be solved directly in 
dimensional regularization: the singularities in ${\cal S}$ are generated  upon integrating the evolution equation (\ref{renS}) from $\mu^2=0$. Thus, as anticipated, at each order there is one new coefficient to determine, the ${\cal O}(\alpha_s^n)$ term in $\Gamma^{{\cal S}}$, which would generate the  ${\cal O}(1/\epsilon^1)$ pole in ${\cal S}$ at this order in perturbation theory.

Very recently the soft anomalous dimension in the massive case, $\Gamma^{{\cal S}}$, has been explicitly computed to two-loop order~\cite{Mitov:2009sv,Kidonakis:2009ev,Becher:2009kw,Beneke:2009rj,Ferroglia:2009ep,Ferroglia:2009ii,Mitov:2010xw,Ferroglia:2010mi}, where a direct correlation between the colour and kinematic degrees of freedom for three partons first appeared. 

\subsection{The singularity structure in the massless case}

Let us turn now to discuss the singularity structure in the massless case. Here an additional source of long-distance singularities is present, namely gluons that are collinear with one of the hard massless partons. As a consequence, there are two extra singularities per loop:  that is, at a given order~${\cal O}(\alpha_s^n)$,  the amplitude contains~${\cal O}(1/\epsilon^k)$ singularities, where $k=1\ldots(2n)$. Here the real power of exponentiation becomes apparent: 
in the exponent for each extra power of $\alpha_s$ there is just one extra singularity.
At one loop the exponent has an~${\cal O}(1/\epsilon^2)$ singularity from overlapping soft and collinear divergences, as well as subleading~${\cal O}(1/\epsilon^1)$ singularities from either soft or collinear divergences; at order ${\cal O}(\alpha_s^n)$ it contains ${\cal O}(1/\epsilon^k)$ where $k=1\ldots(n+1)$. The absence of ${\cal O}(1/\epsilon^k)$ singularities for $n+1<k\leq 2n$ in the exponent corresponds to the fact that multiple exchange diagrams (e.g. ones where each gluon produces a soft as well as a collinear divergence) are not part of the exponent, but are instead reproduced by expanding it.
Similarly to the massive case, the only ``new'' singularity at any given order, which cannot be determined using the previous orders (knowing the cusp anomalous dimension and the $\beta$ function coefficients) is the~${\cal O}(1/\epsilon^1)$ pole. 
As in the massive case, this structure is dictated by evolution equations.
 
To present the factorized amplitude in the massless case, let us assume for simplicity that all the hard partons involved are massless, $p_l^2=0$ for any leg $l=1\ldots L$, and take the remaining kinematic invariants $p_i\cdot p_j$ to be simultaneously large compared to $\Lambda_{\rm QCD}^2$ (the fixed-angle scattering limit). Long-distance singularities can then be shown to emerge from separate collinear and soft regions; these decouple from the hard interaction as well as from each other (the latter step is rather subtle, see for example~\cite{Sterman:1995fz}), leading to the following factorised structure~\cite{Dixon:2008gr,Sterman:2002qn,Gardi:2009qi,Dixon:2009ur,Gardi:2009zv,Dixon:2010hy}:
\begin{align}
\label{fact_massless}
\begin{split}
{\cal M}_{J} \left(p_i, \alpha_s (\mu^2),
  \epsilon \right)  =   \sum_K 
  {\cal S}_{J K} &\left(\beta_i \cdot \beta_j, \alpha_s (\mu^2), \epsilon 
  \right) \,  H_{K} \left( \frac{2 p_i \cdot p_j}{\mu^2},
  \frac{(2 p_i \cdot n_i)^2}{n_i^2 \mu^2}, \alpha_s (\mu^2) ,\epsilon \right)
  \\ & \times
  \prod_{i = 1}^L \frac{{\displaystyle J_i 
  \left(\frac{(2 p_i \cdot n_i)^2}{n_i^2 \mu^2},
  \alpha_s (\mu^2), \epsilon \right)}}{{\displaystyle {\cal J}_i 
  \left(\frac{2 (\beta_i \cdot n_i)^2}{n_i^2}, \alpha_s (\mu^2), \epsilon \right)} \,} \,\,,
\end{split}
\end{align}
where the soft function is defined as in the massive case, (\ref{softcorr}), replacing each hard parton by a corresponding Wilson line:
\begin{equation}
\left( c_J \right)_{\{a_k\}} {\cal S}_{J K} \left(\beta_i \cdot \beta_j, \alpha_s(\mu^2), \epsilon \right) \equiv \sum_{\{b_k\}} \, \, \left\langle 0 \left|  \,
\prod_{i = 1}^L \Big[ \Phi_{\beta_i} (\infty, 0)_{a_kb_k} 
\Big]  \, \right| 0 \right\rangle  \, \left( c_K \right)_{\{b_k\}} \, .
\label{softcorr_massless}
\end{equation}
The evolution equation corresponding to ${\cal S}$ is:
\begin{equation}
\mu  \frac{d}{d \mu} {\cal S}_{J K} \left(\beta_i \cdot \beta_j, \alpha_s(\mu^2), \epsilon \right) = - \, 
\, {\cal S}_{J I} \left(\beta_i \cdot \beta_j, \alpha_s(\mu^2), \epsilon \right) \, \Gamma^{{\cal S}}_{I K} \left(\beta_i \cdot \beta_j, \alpha_s(\mu^2), \epsilon
\right) .
\label{renS_massless}
\end{equation}
There are two related aspects by which this equation differs from the one corresponding to the massive case (\ref{renS}):
the first is that the anomalous dimension $\Gamma^{{\cal S}}$ itself is now singular for $\epsilon\to 0$ owing to the collinear singularities, and the second is that this function, and thus ${\cal S}$ itself, depends on $\beta_i\cdot \beta_j$ --- not on the scale invariant ratio we encountered in the massive case --- 
which implies violation of the rescaling symmetry with respect to the  velocities $\beta_i$. 
The relation between these two aspects was explained in~\cite{Gardi:2009qi}: the singularities appearing in $\Gamma^{{\cal S}}$ and the violation of the rescaling symmetry are both consequences of the cusp anomaly for massless Wilson lines, and they are both controlled by
the cusp anomalous dimension $\gamma_K(\alpha_s)$ to all orders. 

In order to define the partonic jet function $J_i$ in (\ref{fact_massless}) we introduce an auxiliary vector $n_i$ (such that $n_i^2\neq0$)~\cite{Collins:1989bt} and form the following gauge-invariant transition amplitude~\cite{Dixon:2008gr,Gardi:2009qi}:
\begin{equation}
  \overline{u}(p)  \, J \left( \frac{(2p \cdot n)^2}{n^2 \mu^2}, \alpha_s(\mu^2), 
  \e \right) \, = \, \langle p \, | \, \overline{\psi} (0) \, \Phi_n (0, - \infty) \,  
  | 0 \rangle\, .
\label{Jdef}
\end{equation}
The overlapping soft and collinear divergences to jet $i$ are contained in both ${\cal S}$ and $J_i$. Therefore, to avoid the double counting of this overlap region we have divided each jet $J_i$ in (\ref{fact_massless}) by its eikonal part, ${\cal J}_i$, which is already contained in ${\cal S}$. The eikonal jet is defined by:
\begin{equation}
  {\cal J} \left( \frac{2 (\beta \cdot n)^2}{n^2}, \alpha_s(\mu^2), \e 
  \right)  \, = \, \langle 0 | \, \Phi_{\beta}(\infty, 0) \, 
  \Phi_{n} (0, - \infty) \, | 0 \rangle~.
\label{calJdef}
\end{equation} 
It follows then that the ``reduced soft function''~\cite{Dixon:2008gr,Gardi:2009qi}
\begin{equation}
  \overline{{\cal S}}_{J K} \left(\rho_{i j},\alpha_s(\mu^2), \epsilon \right) = 
  \frac{{\cal S}_{J K} \left(\beta_i 
  \cdot \beta_j, \alpha_s(\mu^2), \epsilon \right)}{\displaystyle \prod_{i = 1}^L 
  {\cal J}_i \left(\frac{2(\beta_i \cdot 
  n_i)^2}{n_i^2}, \alpha_s(\mu^2), \epsilon \right)}
\label{reduS}
\end{equation}
is free of collinear poles, and contains only infrared singularities originating 
from soft gluon radiation at large angles with respect to all external legs.
This function admits a similar evolution equation to (\ref{renS_massless}), 
\begin{equation}
\mu  \frac{d}{d \mu} \overline{\cal S}_{J K} \left(\rho_{ij}, \alpha_s(\mu^2), \epsilon \right) = - \, 
\, \overline{\cal S}_{J I} \left(\rho_{ij}, \alpha_s(\mu^2), \epsilon \right) \, \Gamma^{\overline{\cal S}}_{I K} \left(\rho_{ij}, \alpha_s(\mu^2)
\right) .
\label{renSbar_massless}
\end{equation}
but with an important difference: the anomalous dimension matrix $\Gamma^{\overline{\cal S}}$ is now finite, corresponding to the fact that  $\overline{\cal S}$ has no collinear singularities, just soft ones. 

In ${\overline{\cal S}}$ the rescaling symmetry is restored, and thus it depends exclusively on the following kinematic variables\footnote{The phases $\lambda_{ij}$ are
defined by $\beta_i \cdot \beta_j = - | \beta_i \cdot \beta_j |
{\rm e}^{-{\rm i} \pi \lambda_{ij}}$, where $\lambda_{ij} = 1$ if
$i$ and $j$ are both initial-state partons, or both final-state partons,
and  $\lambda_{ij}=0$ otherwise.}:
\begin{equation}
\rho_{ij} \, \equiv \, \frac{\left(-  \beta_i \cdot \beta_j \right)^2}
{\displaystyle \frac{2(\beta_i \cdot n_i)^2}{n_i^2}
\frac{2(\beta_j \cdot n_j)^2}{n_j^2}} \, = \,
\frac{  
\, \left| \beta_i \cdot \beta_j \right|^2 \, 
{\rm e}^{-2 {\rm i} \pi \lambda_{ij}} }
{\displaystyle \frac{2(\beta_i \cdot n_i)^2}{n_i^2}
\frac{2(\beta_j \cdot n_j)^2}{n_j^2}} \,  ,
\label{rhoij}
\end{equation}
which are inherently scale invariant. Ref. \cite{Gardi:2009qi} constrained $\Gamma^{{\cal S}}$ in (\ref{renS_massless}) by enforcing the cancellation of rescaling violation between the numerator and denominator in (\ref{reduS}). This led to a set of $L$ all-order matrix equations~\cite{Gardi:2009qi}:
\begin{align}
\label{oureq_reformulated}
\sum_{j,\, j \neq i} \frac{\partial}{\partial \ln(\rho_{i j})} \,
\Gamma^{{\overline{\cal S}}}_{IK} \left( 
\rho_{i j}, \alpha_s \right)  = & \frac{1}{4} \, \gamma_K^{(i)} 
\left( \alpha_s \right) \,\delta_{IK} \, ,\qquad\qquad \forall i\,,
\end{align}
which constrain the kinematic and colour dependence of the soft anomalous dimension of any massless scattering amplitude. An equivalent set of constraints was derived in \cite{Becher:2009qa} using the soft-collinear effective theory formulation for the problem.

As mentioned in the introduction, the simplest solution to eqs.~(\ref{oureq_reformulated}) is the sum-over-dipoles formula~\cite{Gardi:2009qi,Becher:2009cu,Becher:2009qa}:
\begin{align}
\begin{split}
\label{ansatz}
\left.\Gamma^{\overline{S}}
\left(\rho_{i j}, \alpha_s \right) \right\vert_{\rm dip.} &= \, - \frac18 \,
\widehat{\gamma}_K\left(\alpha_s \right) \sum_{i=1}^L \sum_{j,\, j\neq i} \,
\ln(\rho_{ij}) \,  \mathrm{T}_i \cdot  \mathrm{T}_j \, 
%\\& \hspace*{40pt}
+ \, \frac12 \, \widehat{\delta}_{{\overline{\cal S}}} ( \alpha_s )  
\sum_{i=1}^L  \mathrm{T}_i \cdot  \mathrm{T}_i \, ,
\end{split}
\end{align} 
where $\widehat{\gamma}_K$ and $\widehat{\delta}_{{\overline{\cal S}}}$ are eikonal anomalous dimensions, depending exclusively on the $d$-dimensional coupling (see~\cite{Gardi:2009qi}). 
The first term in (\ref{ansatz}) correlates directly the colour and kinematic degrees of freedom of pairs of hard partons, while the second term is colour diagonal, and can be absorbed into the jets.
An important aspect of this solution is that the entire matrix structure in colour space is governed by the cusp anomalous dimension to all orders in perturbation theory.

Eq.~(\ref{ansatz}) is indeed a solution to (\ref{oureq_reformulated}) if and only if the cusp anomalous dimension admits Casimir scaling, {\it i.e.} $\gamma_K^{(i)}(\alpha_s)= \mathrm{T}_i \cdot  \mathrm{T}_i \,\, \widehat{\gamma}_K\left(\alpha_s \right)$, where $\widehat{\gamma}_K\left(\alpha_s \right)$ does not depend on the representation of the parton $i$. If the latter condition is violated at some order in perturbation theory -- and, as discussed 
in~\cite{Gardi:2009qi,Dixon:2009ur}, this may happen starting at four loops\footnote{See however \cite{Becher:2009qa} where evidence is provided against such four-loop contributions.} -- there will be additional contributions to $\Gamma^{\overline{S}}$ involving quartic and higher Casimirs.

 In addition, solutions of the homogeneous equations corresponding to (\ref{oureq_reformulated}) may contribute to $\Gamma^{\overline{S}}$ starting at three-loops~\cite{Gardi:2009qi}. 
These are functions of conformally-invariant cross ratios of the form $\rho_{ij}\rho_{kl}/(\rho_{ik}\rho_{jl})$, which simultaneously correlate the colour and kinematic variables of four or more partons. 
A dedicated analysis of such corrections was performed in 
\cite{Dixon:2009ur} taking into account a range of additional constraints, including Bose symmetry, collinear limits~\cite{Becher:2009qa} and considerations of the degree of transcendentality. This analysis showed that although the class of possible functions is strongly constrained, such corrections cannot be altogether excluded even at the three loop order. Moreover, at three-loops there is a unique logarithmic function of conformally-invariant cross ratios that fulfils all constraints, possibly providing a first genuine multi-parton correlation in the massless case. An explicit 3-loop 4-leg calculation would be necessary to determine whether such a correction to (\ref{ansatz}) is indeed present. Such a calculation would greatly benefit from methods for efficiently classifying which diagrams contribute to the exponent, which is the subject of this paper.

\subsection{Anomalous dimensions and webs}

In the previous subsections we have briefly summarised the conventional approach to non-abelian exponentiation in terms of evolution equations. 
The first conclusion from this discussion is that in both the massive and massless cases, the physics of non-abelian exponentiation is encapsulated by the correlator of semi-infinite Wilson lines, defining the soft functions in (\ref{softcorr}) and (\ref{softcorr_massless}) respectively. 
It is this correlator that will be the subject of our diagrammatic analysis in what follows.

Studying soft anomalous dimensions we saw that the kinematic dependence of such Wilson line correlators is tightly linked to their structure in colour space, and that this interesting physics is only visible in the non-abelian theory at finite $N_c$.  

Finally, we learnt that non-collinear soft singularities are always described by finite anomalous dimension matrices, through eqs. (\ref{renS}) and (\ref{renSbar_massless}) in the massive and massless cases, respectively. This corresponds to the simple but crucial fact mentioned above, that at each order in perturbation theory only the
${\cal O}(1/\epsilon^1)$ pole in the exponent encodes genuinely new information, which cannot be deduced from lower orders. 
In terms of Feynman diagrams this observation is highly non-trivial:
after summing up the diagrams corresponding to the exponent of (\ref{softcorr}) or (\ref{softcorr_massless}), 
and accounting for the renormalization of the coupling (and in the massless case, for the eikonal jets) all remining singularities are associated exclusively with the renormalization of the multi-eikonal vertex. The latter comprises at each order, of just one new counter term removing a single ultraviolet divergence. Subdivergences must therefore conspire to cancel.

In the two-parton case (colour singlet hard interaction) 
of~\cite{Sterman:1981jc,Gatheral:1983cz,Frenkel:1984pz}, the situation is simple: diagrams that contribute to the exponent just do not have any subdivergence. So upon considering the massive case, where no collinear singularities are present, and accounting for the renormalization of the coupling, the diagrams that appear in the exponent individually have a single pole each, while others, such as the ladder diagrams, which have subdivergences, do not contribute to the exponent.
For example, in the massive case, at two loops, the diagrams in figure \ref{Non_Abelian_non_connected} (a) and (b) have no subdivergences, as can be seen from the fact that shrinking one of the gluon loops towards the cusp drags in the other attachments, leading to a single overall ultraviolet divergence, while in (c) there is a subdivergence (the internal gluon can shrink to the cusp without affecting the external one) and indeed this diagram does not appear in the exponent.  

In the multi-parton scattering case subdivergences are obviously present in individual diagrams that contribute to the exponent. 
This implies that intricate cancellation must take place between diagrams.
This, as we shall see, goes hand in hand with the need to generalise the concept of webs in the multi-leg case from individual diagrams to particular combinations of diagrams that mix with one another as described by (\ref{setmixintro}). 
We shall return to this fundamental issue in section \ref{sec:subdivergences}, after learning the properties of multi-leg webs. As discussed in the introduction, our analysis will make use of the replica trick, and we introduce this idea in the next section.

\section{Non-abelian exponentiation via webs using the replica trick\label{sec:replica}}

In this section, we introduce the replica trick for multiparton scattering, which we will later use to study the structure of the exponent at higher orders. Our method is inspired by similar methods in statistical physics problems (see e.g.~\cite{Replica}), and was already used in~\cite{Laenen:2008gt} to derive the structure of eikonal and next-to-eikonal webs in two parton scattering. The argument presented in that paper is not immediately generalisable to the $L$ parton case. Thus, here we reformulate the replica trick argument directly for the general $L$-leg case, such that two (and three) parton scattering emerge as special cases. For the moment we focus on the eikonal approximation, and postpone the discussion of next-to-eikonal corrections to section~\ref{sec:next_to_eikonal}. 

This section has two parts. In the first we present the formalism, based on the replica trick, for solving the combinatorial problem of non-abelian exponentiation in the multi-leg case. In the second we present an algorithm for computing ECFs based on this formalism.

\subsection{The replica trick formalism for non-abelian exponentiation}

Our starting point is to consider a hard scattering process $H(x_1,\ldots,x_L)$ which produces $L$ hard outgoing particles with initial positions $\{x_i\}$ and final momenta $\{p_i\}$. 
In the eikonal approximation, one may approximate the spacetime path $z_k(t)$ of the $k^{\text{th}}$ hard external particle by its classical trajectory
\begin{equation}
z_k(t)=x_k+\beta_kt,
\label{class}
\end{equation}
where $\beta_k$ is the velocity factor already introduced in the previous section, and the initial positions all coincide at the hard interaction vertex, so that $x_k=0$. Then the scattering amplitude for this process, dressed by any number of virtual soft gluons can be written as follows: 
\begin{equation}
{\cal M}_{b_1\ldots b_L}(p_1,\ldots,p_L)=\int \left[  {\cal D}{A}^\mu_s \right] H_{a_1\ldots a_L}(0,\ldots, 0)e^{{\mathrm i}S[A^\mu_s]}\prod_k
\left({\cal P}\exp\left[{\mathrm i}g_s\int dt\beta_k\cdot{A}_s\right]\right)_{a_kb_k}.
\label{amppath}
\end{equation}
This expression contains a path integral over the soft gauge field ${A}^\mu_s$ (matrix-valued in colour space), whose action is denoted by $S(A^\mu_s)$. Each external line contributes a path ordered exponential, a Wilson line operator in the direction $\beta_k$, where $a_k$ and $b_k$ represent the initial and final colour indices on line $k$, as illustrated in figure~\ref{fig:soft_hard_factorization}. 

Equation~(\ref{amppath}) generalises the expression given in~\cite{Laenen:2008gt} for the two-parton amplitude to any number of legs. It is a reformulation of a well-known result: outgoing particles emitting soft radiation in physical scattering processes can be treated as Wilson lines (see e.g.~\cite{Korchemsky:1993uz}). Using the same path-integral method Ref.~\cite{Laenen:2008gt} also considered what happens beyond the eikonal approximation, which we return to in section~\ref{sec:next_to_eikonal}. 

One may take the hard interaction outside the path integral and rewrite eq.~(\ref{amppath}) as
\begin{equation}
{\cal M}_{b_1\ldots b_L}(p_1,\ldots,p_L)=H_{a_1\ldots a_L}{\cal Z}_{a_1\ldots a_L,b_1\ldots b_L},
\label{amppath2}
\end{equation}
where
\begin{equation}
{\cal Z}_{a_1\ldots a_L,b_1\ldots b_L}=\int\left[ {\cal D}{A}^\mu_s\right]e^{{\mathrm i}S[A^\mu_s]}\prod_k
\left({\cal P}\exp\left[{\mathrm i}g_s\int dt\,\beta_k\cdot{A}_s\right]\right)_{a_kb_k}
\label{Zdef}
\end{equation}
may be recognised as the generating functional of a quantum field theory for the soft gauge field. The Wilson line factors act as source terms for the soft gauge field, and generate eikonal Feynman rules for gluon emission from the emitting particle legs. Carrying out the path integral thus generates all possible soft gluon subdiagrams which span the external lines. Using the Wilson line notation of eq.~(\ref{Wilson_line_def}) for each line $l=1,2,\ldots L$ one may write eq.~(\ref{Zdef}) more compactly as:
\begin{equation}
{\cal Z}=\int\left[ {\cal D}{A}^\mu_s\right]\,e^{{\mathrm i}S[A^\mu_s]}\,\left[{\Phi}^{(1)}\otimes {\Phi}^{(2)}\otimes\cdots \otimes {\Phi}^{(L)}\right],
\label{Zdef2}
\end{equation}
where the tensor product symbol reminds us that the Wilson lines operate on different parton legs,  involving the respective colour indices in the corresponding representations\footnote{Note that it is straightforward to project ${\cal Z}$ onto a specific colour-flow basis $(c_J)$ as done in (\ref{softcorr}) and (\ref{softcorr_massless}), obtaining:
\begin{equation*}
{\cal Z}_{IJ}=\int\left[ {\cal D}{A}^\mu_s\right]\,e^{{\mathrm i}S[A^\mu_s]}\,\left[{\Phi}^{(1)}\otimes {\Phi}^{(2)}\otimes\cdots \otimes {\Phi}^{(L)}\right]_{IJ},
%\label{Zdef2_IJ}
\end{equation*}
where we have associated the indices $I$ and $J$ with $(a_1\ldots a_L)$ and $(b_1\ldots b_L)$, respectively. }.

To address soft-gluon exponentiation, we essentially have to compute $\ln {\cal Z}$. Here the replica trick comes in handy, which proceeds as follows. 
One considers $N$ identical copies\footnote{The number of replicas $N$ is a new integer parameter (it  should not be confused with the number of colours $N_c$).\, This parameter is only introduced as a tool to handle the combinatorial problem of non-abelian exponentiation, and it need not be thought of as having a physical meaning.} or {\it replicas} of the soft gauge boson field, which all share the same gauge group, say ${\rm SU}(N_c)$. Importantly, the replicas do not interact with each other, so that the soft gauge field actions corresponding to different replicas $S[A_\mu^i]$ combine additively, 
\begin{equation}
S[A_\mu]\longrightarrow \sum_{i=1}^N S[A_\mu^i]\,,
\end{equation}
where $A_\mu^i$ is the gauge field associated with replica number $i$ (we have dropped the subscript $s$ denoting softness). Note that if there is matter in the vacuum, as in QCD, it is replicated as well. Thus, different replicas cannot communicate through matter (e.g. via gluons coupling to a fermion bubble), so they all have the same standard renormalization properties.

By analogy with eq.~(\ref{Zdef2}), the generating functional for this theory is given by
\begin{align}
\begin{split}
{\cal Z}^N=\int \left[{\cal D}{A}_\mu^1\right]
\ldots\left[{\cal D}{A}_\mu^N\right] \,e^{{\mathrm i}\sum_i S[A^{i}_{\mu}]}\,\bigg[&({\Phi}^{(1)}_1{\Phi}^{(1)}_2\ldots{\Phi}^{(1)}_N)\otimes ({\Phi}^{(2)}_1{\Phi}^{(2)}_2\ldots{\Phi}^{(2)}_N )\otimes\cdots\\
&
\cdots \otimes( {\Phi}^{(L)}_1{\Phi}^{(L)}_2\ldots{\Phi}^{(L)}_N)\bigg],
\label{Zdefrep}
\end{split}
\end{align}
where ${\Phi}^{(l)}_i$ is the Wilson line factor associated with parton $l$ (where $l=1\ldots L$) and replica number $i$ (where $i=1\ldots N$).  
Note that ${\Phi}^{(l)}_i$ only provides a source for the field $A_\mu^i$ of the particular replica $i$, however, owing to the non-abelian nature of the theory it does \emph{not} commute with other ${\Phi}^{(l)}_j$ (where $j\neq i$).  

On the left-hand side of (\ref{Zdefrep}), we have recognised that the generating functional for the replicated theory constitutes that of eq.~(\ref{Zdef2}) raised to the $N^{\text{th}}$ power. Each external line now carries a product of Wilson lines, which has the form
\begin{align}
\left[{\Phi}_1^{(l)}{\Phi}_2^{(l)}\ldots{\Phi}_N^{(l)}\right]_{a_1b_1}=&
\left({\cal P}\exp\left[{\mathrm i}g_s\int dt\,\beta_l^\mu{A}^1_\mu\right]\right)_{a_1c_2}
\left({\cal P}\exp\left[{\mathrm i}g_s\int dt\,\beta_l^\mu{A}^2_\mu\right]\right)_{c_2c_3}
\ldots \notag\\&\qquad\times
\left({\cal P}\exp\left[{\mathrm i}g_s\int dt\,\beta_l^\mu{A}^N_\mu\right]\right)_{c_{N}b_1}.
\label{rep3}
\end{align}
In order to derive the Feynman rule for emission of a soft gluon from parton $l$, the product of path-ordered exponentials must be combined into a single path-ordered exponential. This can be accomplished (in analogy with the two eikonal line case of~\cite{Laenen:2008gt}) by introducing an operator ${\cal R}$ that orders the colour generators associated with emissions along external 
line~$l$ according to their replica numbers. That is, one has
\begin{equation}
\left[{\Phi}_1^{(l)}{\Phi}_2^{(l)}\ldots{\Phi}_N^{(l)}\right]_{a_lb_l}=
\left({\cal R}
{\cal P}\exp\left[{\mathrm i}g_s\sum_{i=1}^N\int dt\,\beta_l^\mu {A}_\mu^i\right]\right)_{a_lb_l},
\label{Wprod}
\end{equation}
where the sum in the exponent is over replica numbers\footnote{Note that in the absence of ${\cal R}$ on the {\it r.h.s.} the expansion of the path-ordered exponential would yield terms with all possible orderings of the colour matrices corresponding to the different replicas, not reproducing the {\it l.h.s.} where this order is uniquely fixed. Thus the ${\cal R}$ operation is crucial for this equality to hold.}. Given that we have chosen all momenta to be outgoing, the operator ${\cal R}$ reorders the colour matrices on a given line such that the replica number increases as one moves away from the hard interaction. Equation~(\ref{Zdefrep}) thus becomes
\begin{align}
{\cal Z}^N=\int\left[{\cal D}{A}_\mu^1\right]
\ldots\left[{\cal D}{A}_\mu^N\right]  \,e^{{\mathrm i}\sum_i S[A^{i}_{\mu}]}\,{\cal R}&
\left\{
{\cal P}
\exp\left[{\mathrm i}g_s\sum_{i=1}^N\int dt\,\beta^\mu_1{A}_\mu^i\right]\otimes\ldots
\right.\notag\\
&\left.\qquad\qquad\ldots \,\otimes\,{\cal P}\exp\left[{\mathrm i}g_s\sum_{i=1}^N\int dt\,\beta^\mu_L{A}_\mu^i\right]\right\}\,,
\label{Zdefrep2}
\end{align}
where ${\cal R}$ reorders the colour generators along all the lines.
 
Carrying out the path integrals over the soft gauge fields in (\ref{Zdefrep2}) generates all soft gluon diagrams --- involving all possible replica numbers --- which connect the external lines. Since the actions $S[A_\mu^i]$ are all the same as in the original (unreplicated) theory, the kinematic part of each diagram, ${\cal F}(D)$, is the same as one would obtain in the unreplicated theory for the same configuration of gluons ({\it i.e.} the operator ${\cal R}$ acts only on the colour generators of gluon fields, and leaves their spacetime dependence unchanged). However, the colour factor of each generated diagram is \emph{not} the same as would be obtained in the unreplicated theory ($C(D)$). Rather, the colour factor of a given graph in the replicated theory, which we may denote by $C_N(D)$, is that which results after reordering the replicas according to the ${\cal R}$ operator, and it explicitly depends on the number of replicas~$N$. We will see explicit examples of this in section \ref{sec:special_cases}. 

Next, one may perform a Taylor expansion in $N$ to obtain
\begin{equation}
{\cal Z}^N=1+N\ln{{\cal Z}}+{\cal O}(N^2).
\label{taylorN}
\end{equation}
We now have 
\begin{equation}
\ln {\cal Z}=\sum_D{\cal F}(D)\,\widetilde{C}(D)\,,
\label{lnZ}
\end{equation}
where the sum on the right-hand side is over all diagrams $D$ whose colour factors are ${\cal O}(N)$.
Each diagram contributes the kinematic part ${\cal F}(D)$ as in the original theory, times a modified colour factor $\widetilde{C}(D)$ which is defined as the coefficient of $N^1$ in the Taylor expansion of the colour factor in the replicated theory (note that $\widetilde{C}(D)$ itself carries no $N$ dependence).
The sum in (\ref{lnZ}) is over all diagrams that have a component that is linear in the replica number $N$. Indeed, some diagrams do not have any ${\cal O}(N^1)$ terms in their Taylor expansion, and therefore they do not contribute to the exponent. We shall see such examples in section \ref{sec:special_cases}. 

Given that eq.~(\ref{lnZ}) is manifestly independent of $N$, the final result for the generating functional ${\cal Z}$  
in the original (unreplicated) theory is therefore, 
\begin{equation}
{\cal Z}=\exp\left[\sum_D{\cal F}(D)\,\widetilde{C}(D)\right]\,,
\label{Zexp}
\end{equation}
which is manifestly an exponential form\footnote{
A comment is in order regarding to the nature of the replica trick as used here. In statistical physics applications, the replica number $N$ typically becomes a complex parameter. Recovering the unreplicated from the replicated theory often becomes ambiguous, due to subtleties involving analytic continuation. Here there are no such problems, the colour factor in the replicated theory is a polynomial in $N$, and it is straightforward to take its ${\cal O}(N^1)$ coefficient.}.

\subsection{A replica-trick based algorithm for computing exponentiated colour factors\label{sec:algorithm}}

The formal manipulations leading to (\ref{Zexp}) are useful because they uniquely specify the exponent: they directly translate into a general algorithm by which the ECF associated with any given diagram $D$ can be computed. We may summarise this algorithm by the following sequence of steps:
\begin{enumerate}
\item{} Identify distinct connected pieces in the diagram (see footnote \ref{fotenote:connected} in section \ref{sec:Wilson_lines}). We assume there are $n_c$ such pieces, $f_i$, with $i=1\ldots n_c$.  If $n_c=1$ the diagram as a whole is connected and $\widetilde{C}(D)={C}(D)$.
For $n_c\geq 2$, assign each connected piece $f_i$ in the diagram a replica variable $r_i$, which can take (integer) values from $1$ to $N$.
\item{} Compute the colour factor $C_N(D)$ in the replicated theory. To this end one needs to sum over all possible assignments of the variables $r_i$ in the range $1$ to $N$. In practice, all that matters, based on the definition of the ordering operator ${\cal R}$,  is 
the hierarchy of $r_i$ amongst a set of gluon attachments to a given leg.
Specifically, colour generators associated with two gluon attachments to a given leg will be ordered such that the generator with the higher replica number is further from the hard interaction. 
Given two attachments $i$ and $j$ to the same leg $l$, the ${\cal R}$ operation 
\begin{itemize}
\item{} does nothing if $r_i=r_j$, in which case the colour factor will be determined by the ordering of the gluons in the original diagram $D$. 
\item{} will order them as ${\mathrm T}_i {\mathrm T}_j$ if $r_i<r_j$ and as ${\mathrm T}_j {\mathrm T}_i$ if $r_j<r_i$, independently of the order of the two attachments in the original diagram $D$. 
\end{itemize} 
A convenient way to compute $C_N(D)$ is therefore:
\begin{itemize}
\item[I.]{}
Tabulate all possible hierarchies which may arise when each of the variables $r_i$ takes values 
in the range 1 to $N$. A hierarchy determines whether $r_i>r_j$, $r_i=r_j$ or $r_i<r_j$ for \emph{any} two connected pieces $i$ and $j$.
For a given hierarchy $h$, we denote the number of different $r_i$'s (different active replicas) by $n(h)$. This number can vary from $1$ to the number of connected pieces $n_c$.
An example is provided in table~\ref{17reptab} for the $n_c=3$ case (note that for simplicity the variables $r_i, r_j$ and $r_k$ are denoted there by $i$, $j$ and $k$, respectively).
\item[II.]{} For each hierarchy in the table compute its multiplicity, namely a combinatorial factor counting how many times this hierarchy arises when all $r_i$ go over the range $1\ldots N$.
The multiplicity is equal to the number of ways of choosing $n(h)$ different replica numbers from $N$ (associating them to the $n(h)$ different $r_i$'s), thus is   
 given by the combination function:
\begin{equation}
M_N(h)=\phantom{!}_NC_{n(h)}= \frac{N!}{(N-n(h))!\, n(h)!}\,.
\label{M_N_combination_}
\end{equation}
\item[III.]{} For each hierarchy $h$, determine the colour factor of the diagram after applying the ordering operation ${\cal R}$ to the generators on all legs (${\cal R}$ is only relevant for legs to which there are two or more attachments).  
We denote this colour factor by 
${\cal R} \left[ C(D)\,\vert h \right]$, to be read as ``the ${\cal R}$-ordered colour factor of diagram $D$, given the hierarchy $h$". 
\item[IV.]{} Sum over all possible hierarchies $h$ with their respective multiplicities $M_N(h)$, to obtain the colour factor of the given diagram $D$ in the replicated theory as a function of $N$:
\begin{equation}
\label{C_rep}
C_N(D) =\sum_h M_N(h)\,\,{\cal R} \left[ C(D)\,\vert h \right]
\end{equation}
\end{itemize}
\item{} Expand $C_N(D)$ in powers of $N$ and extract the coefficient of $N^1$, which  is the ECF $\widetilde{C}(D)$. 
\end{enumerate}

One can go further and recast this algorithm as a formula for the ECF.
The crucial observation here is that in our expression for $C_N(D)$ only the multiplicity factor $M_N(h)$ depends on $N$, and  this dependence is through the simple combination function in (\ref{M_N_combination_}).  We can therefore extract the $N^1$ coefficient, obtaining:
\begin{equation}
\label{ECF_in_terms_of_RC}
\widetilde{C}(D) =\sum_h \frac{(-1)^{n(h)-1}}{n(h)}\,\,{\cal R} \left[ C(D)\,\vert h \right]\,.
\end{equation}
We shall continue to develop this in the next section, where we provide a more explicit formula for the ECF.

Implementing the above algorithm requires a notation that identifies connected pieces and simultaneously  specifies the order of attachments of gluons to each of the legs (eikonal lines) in an arbitrary diagram. 
Considering a set of diagrams  
\[
\left\{D\right\}=(n_1,n_2,n_3,\cdots,n_L)
\] 
with $L$ legs and $n_l$ gluon attachments to a given leg $l$, 
we use the following notation for a particular diagram within this set:
\begin{equation}
\label{diag_notation}
D=\left[[s_1^{(1)},s_2^{(1)},\ldots s_{n_1}^{(1)}],\,\, [s_1^{(2)},s_2^{(2)},\ldots s_{n_2}^{(2)}],\,\, \cdots \,\, ,\,\,[s_1^{(L)},s_2^{(L)},\ldots s_{n_L}^{(L)}]\right]\,.
\end{equation}
Each of the square brackets corresponds to a leg (there are $L$ such terms in $D$).
Each of the entries within a given bracket $[s_1^{(l)},s_2^{(l)},\ldots s_{n_l}^{(l)}]$ corresponding to leg $l$, is associated with an individual gluon attachment, where the order of the list indicates the order of gluon attachments to this leg: the list is ordered from the outside inwards toward the hard vertex (cusp).
The variables $s_i^{(l)}$ themselves are assigned values from 1 to $n_c$ (the number of connected pieces in~$D$) associating each of the gluons with a particular connected piece $f_i$ in $D$. This facilitates the assignment of replica numbers. 
Note that this notation does not specify \emph{the way} by which the gluons are connected internally (which may involve additional loops). These details are irrelevant to the combinatorial problem of non-abelian exponentiation. 
The notation (\ref{diag_notation}) is used in section \ref{sec:special_cases} where we consider explicit examples -- see figures \ref{3lsix} through \ref{fourloopdiags}, and also Appendix~\ref{sec:more-3loop-ECF}.

\section{General formula for Exponentiated Colour Factors  \label{sec:formula}}

In the previous section we have set up the replica trick formalism for multiparton eikonal scattering. This allowed us to solve the combinatorial problem of non-abelian exponentiation, constructing directly the exponent of such amplitudes in terms of individual diagrams. 
We have seen that the exponent contains those diagrams which, in the replicated theory, have a term that is linear in the number of replicas $N$. The contribution of each such diagram $D$ to the exponent is determined by an ECF $\widetilde{C}(D)$. We have seen that the replica trick formalism translates into a general algorithm for computing these ECFs.
We will now show that it also leads to an explicit formula for the ECF of a given graph $D$ in terms of a particular linear combination of the products of ordinary colour factors of subgraphs in $D$.
This is the multiparton generalisation of the similar result given in~\cite{Laenen:2008gt} for the case of two eikonal lines. 
Furthermore, in section \ref{sec:inverted_formula} we present an inverted result, expressing the conventional colour factor $C(D)$ of graph $D$ in terms of ECFs. The latter result is of conceptual interest as the multiparton generalisation of Gatheral's formula for webs in two parton scattering~\cite{Gatheral:1983cz}.

\subsection{Exponentiated Colour Factors in terms of conventional ones\label{sec:ECF_formula} }

\begin{figure}[htb]
\begin{center}
\scalebox{0.8}{\includegraphics{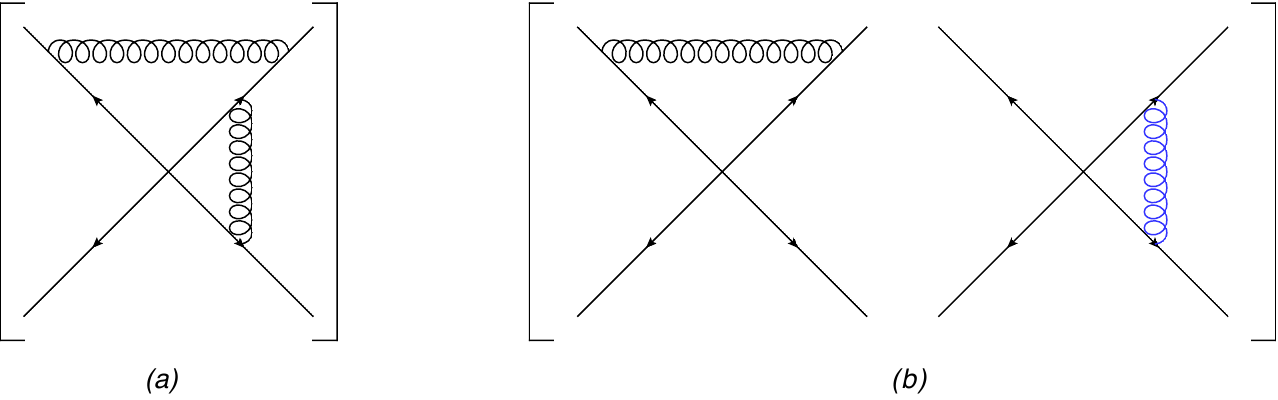}}
\caption{Decompositions of an example two loop diagram. The two decompositions have (a) $n(P)=1$; (b) $n(P)=2$. Different colours represent different replica numbers.}
\label{partex1}
\end{center}
\end{figure}

\begin{figure}[htb]
\begin{center}
\scalebox{0.8}{\includegraphics{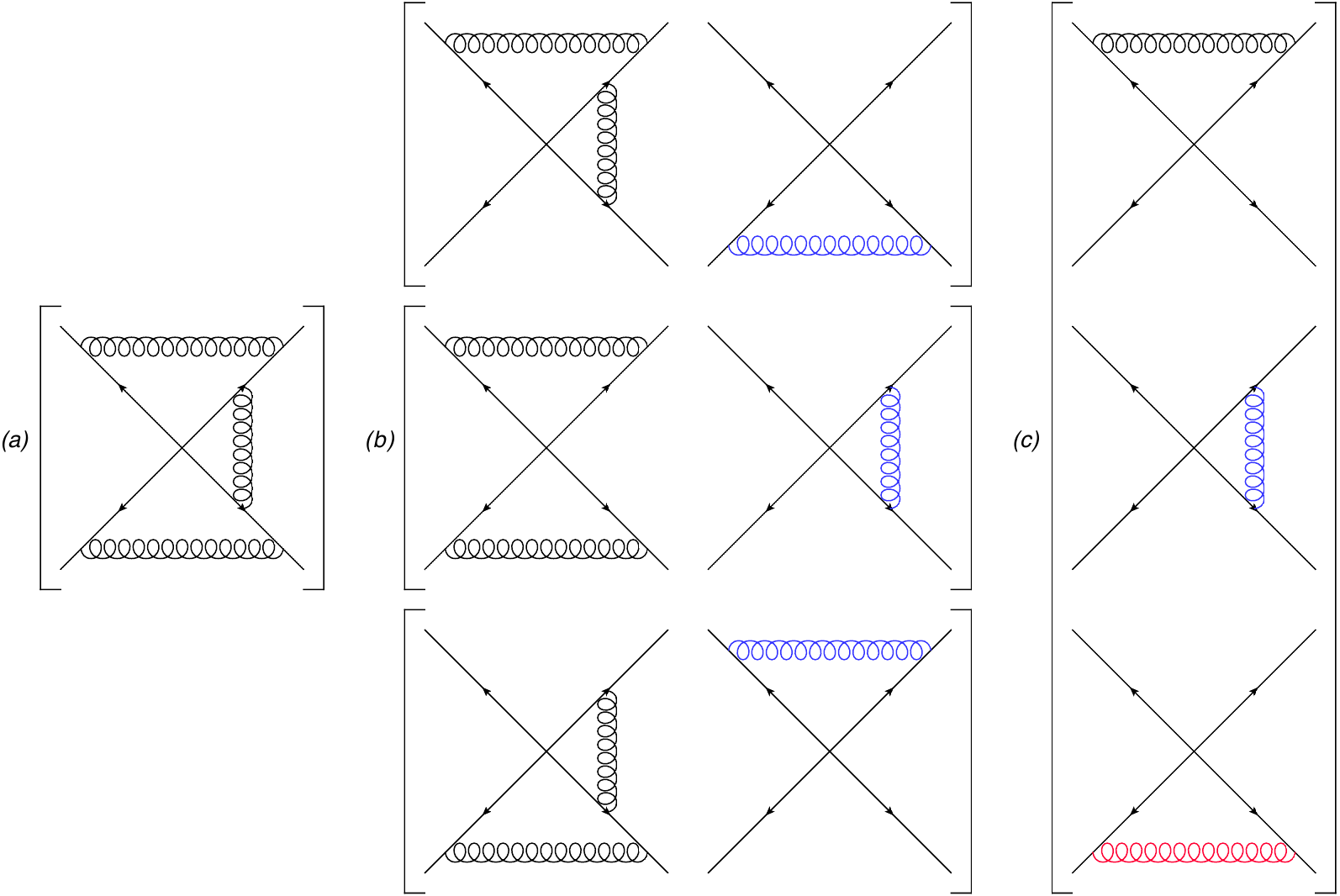}}
\caption{Decompositions of an example three loop diagram, with (a) $n(P)=1$; (b) $n(P)=2$; (c) $n(P)=3$. Different colours represent different replica numbers.}
\label{partex2}
\end{center}
\end{figure}

In order to proceed, let us first introduce the notion of {\it decompositions}\footnote{Note that these were called {\it partitions} in~\cite{Laenen:2008gt}.} of a diagram $D$. A decomposition is a complete set of subgraphs of $D$ -- not necessarily connected -- each containing a single replica number. A given decomposition $P$ contains a number of replicas $n(P)$, which in general is not sufficient to uniquely define the decomposition. A two-loop example is given in figure~\ref{partex1}. Each square bracket represents one decomposition, and there are two in total with $n(P)=1$ and $n(P)=2$, respectively. In this simple case, there is only one decomposition for each choice of the number of replicas. Another example is shown in figure~\ref{partex2}. Here there are 5 decompositions in total.

Each decomposition contains a number of subgraphs $g_1,g_2,\ldots g_{n(P)}$, each of them having a \emph{different} replica number\footnote{The replica numbers of $g_i$ and $g_j$ ($i\neq j$) in a given decomposition $P$ cannot be equal. In this formalism the case where the two corresponding subgraphs are assigned the same replica number is taken into account in some other decomposition $P'$ in which the two belong to one subgraph $g_k$.}. 
We now wish to express the ECF $\widetilde{C}(D)$ in terms of products of ordinary colour factors of these subgraphs $C(g_i)$. To this end we need to determine the colour factor in the replicated theory,  $C_N(D)$. We shall see that owing to the replica ordering operator ${\cal R}$, the latter contains a sum of products of $C(g_i)$ in all possible orders.

The simplest way to determine this sum is to use eq.~(\ref{C_rep}) and split the sum over hierarchies $h$ in this equation into a sum over decompositions $P$, where for each decomposition there is an internal sum over possible hierarchies. Crucially, this internal sum becomes trivial, since given the decomposition $P$ all the possible hierarchies have $n(h)=n(P)$, and therefore they have identical multiplicities $M_N(h)$, which we denote $M_N(P)$. Given $P$, for any particular $h$ the ${\cal R}$-ordered colour factor would be a particular product 
\begin{equation}
{\cal R} \left[ C(D)\,\vert h \right] = C(g_{\pi(h)_1})\ldots C(g_{\pi(h)_{n(P)}}),
\end{equation}
where $\pi(h)$ is the permutation of $(1,2,\ldots, n(P))$ corresponding to the hierarchy $h$. 
But since all $M_N(h)$ given $P$ are the same, \emph{all} the permutations contribute symmetrically to $C_N(D)$. One therefore obtains:
\begin{equation}
\label{modcol}
C_N(D) =\sum_P M_N(P) \sum_{\pi} \,\,C(g_{\pi_1})\ldots C(g_{\pi_{n(P)}}),
\end{equation}
where the internal sum goes over all permutations of $(1,2,\ldots, n(P))$ and $M_N(P)$ is given by the combination function in (\ref{M_N_combination_}) with $n(h)=n(P)$.

We can now proceed, as in eq. (\ref{ECF_in_terms_of_RC}), to extract the ${\cal O}(N^1)$ part of eq.~(\ref{modcol}). We obtain:
\begin{equation}
\widetilde{C}(D)=\sum_P\frac{(-1)^{n(P)-1}}{n(P)}\sum_\pi C(g_{\pi_1})\ldots C(g_{\pi_{n(P)}}),
\label{modcol2}
\end{equation}
which thus represents the ECF of graph $D$, written explicitly in terms of ordinary colour factors of its subgraphs. 
A couple of examples for the application of eq.~(\ref{modcol2}) are provided in appendix \ref{app-col2}; the same examples will be analysed in section \ref{sec:special_cases} by applying the replica trick algorithm according to section \ref{sec:algorithm}, as well as by direct exponentiation.

It is interesting to note a further simplification of this result when the colour factors $C(X)$ of subgraphs commute with each other. When this is the case, one may write
\begin{equation}
\sum_\pi C(g_{\pi_1})\ldots C(g_{\pi_{n(P)}})=n(P)!\prod_{g\in P}C(g),
\label{perms}
\end{equation}
so that eq.~(\ref{modcol}) becomes
\begin{equation}
C_N(D)=\sum_P\phantom{!}_NP_{n(P)}\prod_{g\in P} C(g),
\label{modcol3_N}
\end{equation}
where 
\begin{equation}
\phantom{!}_NP_{n(P)}=\frac{N!}{(N-n(P))!} 
\end{equation}
is the permutation function. 
Expanding in powers of $N$,
\begin{equation}
\phantom{!}_NP_{n(P)}= (-1)^{n(P)-1}\, (n(P)-1)! \,\, N+{\cal O}(N^2)
\end{equation}
we extract the ECF for the case of subgraphs with commuting colour factors:
\begin{equation}
\widetilde{C}(D)=\sum_P (-1)^{n(P)-1}\, (n(P)-1)!\,\prod_{g\in P} C(g)\,.
\label{modcol3}
\end{equation}
This is precisely the result that was given in~\cite{Laenen:2008gt} for the two eikonal line case, when the colour factors do indeed commute.

\subsection{Conventional colour factors in terms of exponentiated ones \label{sec:inverted_formula}}

The above formula expresses the exponentiated colour factors in terms of the conventional colour factors. It is also instructive to write an inverted formula which expresses things the other way around. Such a formula is the generalisation of Gatheral's original definition of exponentiated colour factors~\cite{Gatheral:1983cz} to the multileg case. 

The argument here is similar to that presented in~\cite{Laenen:2008gt} for two eikonal lines, and runs as follows. 
One again uses the notion of decompositions of a diagram $D$ into subdiagrams, as described above. A given decomposition is labelled by a set of numbers $\{m_H\}$, where $m_H$ counts how many times a given subdiagram $H$ appears in the decomposition.
The index $H$ runs over all possible subdiagrams. For example, the decomposition of figure~\ref{partex2}(a) has $m_{3a}=1$ and all other $m_H=0$; figure~\ref{partex2}(c) has $m_{1a}=m_{1b}=m_{1c}=1$ and all other $m_H=0$ (n.b. we have used labels introduced in figures~\ref{17-20}, \ref{1a-c}, \ref{2a-d} and \ref{2e}). One may write the exponentiated amplitude as
\begin{equation}
\exp\left\{\sum_H{\cal F}(H)\, \widetilde{C}(H)\right\}=\sum_{n=0}^\infty\frac{1}{n!}\left(\sum_H {\cal F}(H)\, \widetilde{C}(H)\right)^n,
\label{expamp}
\end{equation}
where as usual ${\cal F}(H)$ represents the kinematic part of diagram $H$. The expansion of the exponential on the right-hand side contains a product of sums. One may rewrite this as a sum over all possible products of subdiagrams by using the notation for decompositions introduced above:
\begin{equation}
\left(\sum_H{\cal F}(H)\, \widetilde{C}(H)\right)^n=\sum_{\{m_H\}}\left(\prod_H \frac{{\cal F}(H)^{m_H}}{m_H!}\right)\left[\widetilde{C}(H_1)^{m_1}\widetilde{C}(H_2)^{m_2}\ldots +\text{perms}\right].
\label{expamp2}
\end{equation}
Some explanatory remarks are in order. Firstly, the sum over all possible decompositions constitutes a sum over all possible products of subdiagrams. The kinematic factors may be combined into a single term, given that the ${\cal F}(H)$ factors commute. Each such product also contains a string of exponentiated colour factors, one for each subgraph ({\it i.e.} $m_H=0$ if graph $H$ is not present). All permutations of the exponentiated colour factors are present, however one should not multiply count permutations obtained by interchanging colour factors corresponding to the same subdiagram. Thus, one may write the colour part as a sum over all permutations, but with inverse factorial factors $m_H!$ to compensate for the overcounting. 

Any product of kinematic factors can always be expressed as a sum of kinematic factors of graphs $D$, whose topology is consistent with the diagrams which enter the product. More formally, one may write
\begin{equation}
\prod_H{\cal F}(H)^{m_H}=\sum_D{\cal F}(D)N_{D|\{m_H\}},
\label{momprod}
\end{equation}
where $N_{D|\{m_H\}}$ is the number of ways in which the diagram $D$ can be obtained from the product of kinematic factors in the decomposition given by $\{m_H\}$ (note that this may be zero, for the cases in which $D$ cannot be formed from the decomposition). Substituting this into eq.~(\ref{expamp2}), the exponentiated amplitude becomes
\begin{align}
\exp\left\{\sum_H{\cal F}(H)\,\widetilde{C}(H)\right\}&=\sum_{n=0}^\infty\frac{1}{n!}\sum_{\{m_H\}}\sum_D {\cal F}(D)N_{D|\{m_H\}}\left(\prod_Hm_H!\right)^{-1}\notag\\
&\quad\times\left[\widetilde{C}(H_1)^{m_1}\widetilde{C}(H_2)^{m_2}\ldots +\text{perms}\right].
\label{expamp3}
\end{align}
Currently in the sum over decompositions, it is understood that the sum of all $m_H$ is equal to $n$. One may thus replace the multiple sums over $n$ and $\{m_H\}$ with a single sum over all decompositions to obtain 
\begin{align}
\exp\left\{\sum_H{\cal F}(H)\,\widetilde{C}(H)\right\}&=\sum_{D}{\cal F}(D)\sum_{\{m_H\}}\frac{N_{D|\{m_H\}}}{n!}\left(\prod_Hm_H!\right)^{-1}\notag\\
&\quad\times\left[\widetilde{C}(H_1)^{m_1}\widetilde{C}(H_2)^{m_2}\ldots +\text{perms}\right],
\label{expamp4}
\end{align}
where we have also pulled the sum over $D$ to the front. 

Eq.~(\ref{expamp4}) must be equal to the original amplitude, which we may express as a sum over all diagrams:
\begin{equation}
\sum_D{\cal F}(D)\,C(D)\,,
\end{equation}
where $C(D)$ is the conventional colour factor of graph $D$. Equating coefficients of ${\cal F}(D)$ with eq.~(\ref{expamp4}), one finally finds
\begin{equation}
C(D)=\sum_{\{m_H\}}\frac{N_{D|\{m_H\}}}{n!}\left(\prod_Hm_H!\right)^{-1}\left[\widetilde{C}(H_1)^{m_1}\widetilde{C}(H_2)^{m_2}\ldots +\text{perms}\right],
\label{colfacfin}
\end{equation}
with $n=\sum_H m_H$. This formula expresses the conventional colour factors in terms of the exponentiated colour factors, and thus is the generalisation of Gatheral's formula to the multiparton case.  A couple of examples for the application of eq.~(\ref{colfacfin}) are provided in appendix \ref{app-gatheral}. The same examples will be analysed in the next section using the replica trick method, as well as by direct exponentiation.

Note that if the colour factors $\widetilde{C}(D)$ commute with each other (as happens for two eikonal lines), the colour factor of eq.~(\ref{colfacfin}) becomes
\begin{equation}
\widetilde{C}(H_1)^{m_1}\widetilde{C}(H_2)^{m_2}\ldots +\text{perms}=n!\prod_H\widetilde{C}(H)^{m_H},
\label{colfaccom}
\end{equation}
so that one has
\begin{equation}
C(D)=\sum_{\{m_H\}}N_{D|\{m_H\}}\prod_H\frac{\widetilde{C}(H)^{m_H}}{m_H!}.
\label{gatheral}
\end{equation}
This is precisely the result of Gatheral's formula, albeit expressed in the notation of~\cite{Laenen:2008gt}.  

\newpage
\section{Exponentiated Colour Factors in special cases\label{sec:special_cases}}

In the previous sections, we have set up the replica trick for multiparton scattering, and presented a general algorithm for calculating the exponentiated colour factor of any graph. Furthermore, we have used the algorithm to derive explicit combinatoric formulae for exponentiated colour factors, which can in principle be applied to arbitrary order in perturbation theory. In this section, we present particular cases of applying these results, and our aim is twofold. Firstly, we wish to illustrate and thus clarify the somewhat technical discussion of the previous sections. Secondly, we will see structures emerging in the exponent which allow us to meaningfully generalise the notion of webs from two eikonal line scattering to the $L$-parton case.

\begin{figure}[htb]
\begin{center}
\scalebox{1.0}{\includegraphics{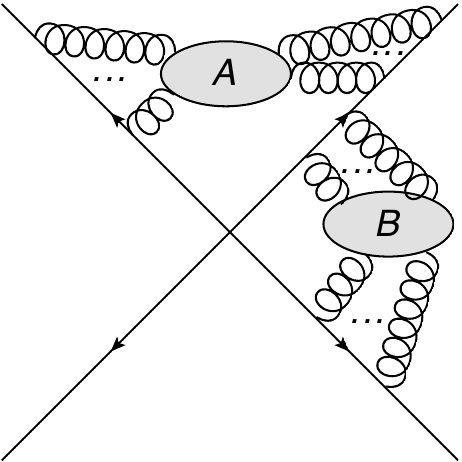}}
\caption{Example of a reducible topology in which three partons are linked by soft gluon exchanges.
Such diagrams have subdivergences, and yet they do contribute to the exponent (as part of a multi-diagram web). }
\label{fig-reducible}
\end{center}
\end{figure}

As already stated in the introduction, in the $L$-parton case one is led to abandon the assumption that webs must be single diagrams. Rather, one finds closed sets of diagrams in the exponent, whose members are related by permutations of gluons on the eikonal lines, and mix only with each other. These diagrams include, in general, ones with reducible topologies, with subgraphs that are separable by cutting the eikonal lines, as shown for example in figure \ref{fig-reducible}.
Such diagrams have subdivergences, in contrast with the familiar two-eikonal line irreducible webs.

Before considering this general structure, it is worth noting that there is one way by which 
the irreducibility criterion of webs remains relevant to the multiparton case. 
This is the fact that diagrams that are reducible by mere separation at the hard vertex -- 
namely those containing separate clusters of partons that do not communicate by gluon exchanges -- 
never contribute to the exponent\footnote{Another case in which only irreducible diagrams contribute to the exponent is the planar (large 't Hooft coupling) limit, as discussed already, albeit from a different point of view, in~\cite{Bern:2005iz}.}. We examine this in the following subsection.

\subsection{Diagrams with distinct parton clusters}

\begin{figure}[htb]
\begin{center}
\scalebox{1.0}{\includegraphics{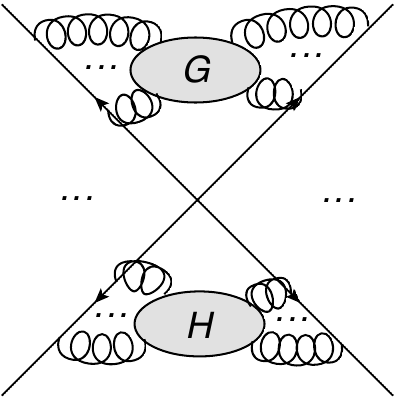}}
\caption{Example of a diagram in which distinct clusters of partons are linked by soft gluon exchanges.}
\label{clusform}
\end{center}
\end{figure}

As a first demonstration of the replica trick, we consider diagrams in which distinct clusters of partons occur, each of which is internally linked by soft gluons, but between which there are no exchanges. 
As an example, we consider explicitly the form shown in figure~\ref{clusform}, and demonstrate that such diagrams do not contribute to the exponent, independently of the internal structure of the subdiagrams $G$ and $H$. The proof clearly generalises to more than two clusters. 

The contribution to the exponentiated colour factor from a given assignment of replica numbers in figure~\ref{clusform} is
\begin{equation}
\prod_{i=1}^N C(G_i)C(H_i),
\label{cluscol1}
\end{equation}
where $G_i$ is the part of subdiagram $G$ composed of replica number $i$, and $C(G_i)$ its colour factor. The matrices $C(G_i)$ and $C(H_j)$ (for any $i$ and $j$) commute with each other because they act on different parton indices. Thus one may rewrite eq.~(\ref{cluscol1}) as
\begin{equation}
\left[\prod_{i=1}^NC(G_i)\right]\left[\prod_{j=1}^NC(H_j)\right],
\label{cluscol2}
\end{equation}
and one sees that the colour factors for subdiagrams $G$ and $H$ are completely independent of each other. The number of ways of forming the subdiagram $GH$ is then given by the product of the numbers of ways of forming diagrams $G$ and $H$ separately. Given that each of the latter is at least ${\cal O}(N)$, it follows that diagrams of the form shown in figure~\ref{clusform} 
are~${\cal O}(N^2)$, that is their colour factor in the replicated theory, $C_N(GH)$, has no linear component in $N$, so they do not contribute to the exponent.

\subsection{Webs connecting three lines at two loops}
\label{W3l2l}
In the previous subsection we have seen an example where reducible diagrams do not contribute to the exponent. In general, in the multiparton case, things are more complicated than this, and in contrast to the two-line case certain reducible diagrams, such as those of figure \ref{fig-reducible}, do contribute. 
We start by examining two-loop diagrams, in which at most three partons may be connected. These diagrams have already been studied in the context of soft anomalous dimensions~\cite{Aybat:2006mzy,Mitov:2009sv,Mitov:2010xw}, thus we will be able to demonstrate the replica trick in a familiar context, and also reinterpret the previously known results in terms of webs.

\begin{figure}[htb]
\begin{center}
\scalebox{0.9}{\includegraphics{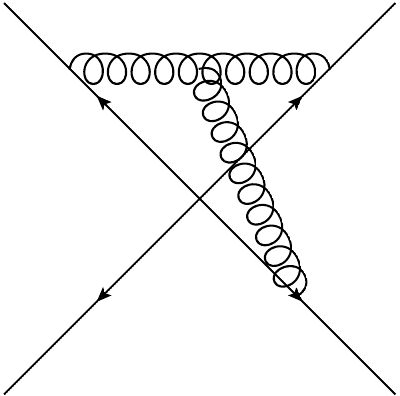}}
\caption{Example two-loop diagram in which three partons are connected via a three gluon vertex.}
\label{3gfig}
\end{center}
\end{figure}

First, there is the diagram in which three gluons are connected by a three-gluon vertex, shown in figure~\ref{3gfig}. It is clear that this diagram is a web from the replica trick, as it is ${\cal O}(N)$ in the replicated theory (only one replica can be involved due to the lack of self-interactions between the replicas). It was indeed shown to be present in the soft anomalous dimension matrix for massive eikonal lines~\cite{Mitov:2009sv} (it in fact vanishes for massless eikonal lines, once kinematic information is included~\cite{Aybat:2006mzy}).

\begin{figure}[htb]
\begin{center}
\scalebox{0.9}{\includegraphics{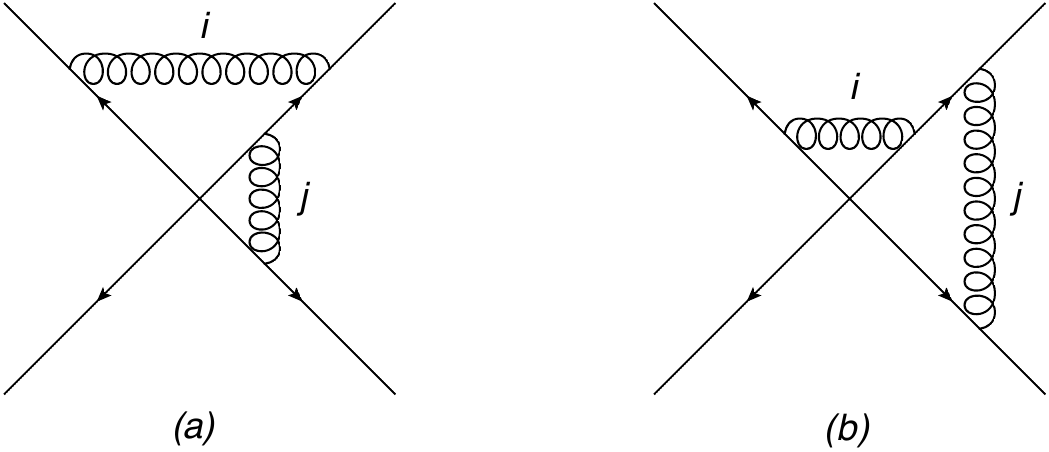}}
\caption{Example two-loop diagrams involving two gluon exchanges between three lines.}
\label{twoemfig}
\end{center}
\end{figure}

There are also the diagrams in which three lines are connected by two single gluon exchanges, shown here in figure~\ref{twoemfig}. Using a similar notation to before, we denote by $C(a)$ and ${\cal F}(a)$ the colour and kinematic parts respectively of diagram $a$ (and similarly for $b$). We may then evaluate the exponentiated colour factors using the algorithm defined in section~\ref{sec:algorithm}. First one identifies distinct connected pieces, and assigns a replica number to each. This has already been carried out in figure~\ref{twoemfig}, where replica numbers $i$ and $j$ are introduced. Next, one considers all possible hierarchies $h$ of the replica numbers. There are three such hierarchies: $i=j$, $i<j$ and $i>j$. For each $h$, one must evaluate the colour factor in the replicated theory (${\cal R}[C(D)|h]$) for each diagram $D$.

Considering $i=j$, the exponentiated colour contributions from figure~\ref{twoemfig}(a) and (b) are 
\begin{equation}
M_N(i=j){\cal R}[C(a)|i=j]=NC(a)
\label{twoemcol1}
\end{equation}
and
\begin{equation}
M_N(i=j){\cal R}[C(b)|i=j]=NC(b).
\label{twoemcol2}
\end{equation}
That is, these are the conventional colour factors for these diagrams, weighted by $M_N(i=j)$, the number of possible assignments of replica numbers $i$ and $j$ such that $i=j$, which is equal to the number of replicas $N$. 

Next there is the case $i<j$. Given the replica ordering in the replicated theory, the gluons on the middle line in diagram (a) get interchanged, so that the colour factor corresponds to that of diagram (b). However, no such reordering occurs in diagram (b). Thus, the contributions from diagrams (a) and (b) are
\begin{equation}
M_N(i<j){\cal R}[C(a)|i<j]=\frac{N(N-1)}{2}C(b)
\label{twoemcol3}
\end{equation}
and
\begin{equation}
M_N(i<j){\cal R}[C(b)|i<j]=\frac{N(N-1)}{2}C(b)
\label{twoemcol4}
\end{equation}
respectively, where the prefactor arises from the number of ways of choosing $i<j$. 

Finally, one has the case $i>j$. This is similar to the case $i<j$, except that reordering of gluons occurs in diagram (b) but not in (a), and the two diagrams give contributions
\begin{equation}
M_N(i>j){\cal R}[C(a)|i>j]=\frac{N(N-1)}{2}C(a)
\label{twoemcol5}
\end{equation}
and
\begin{equation}
M_N(i>j){\cal R}[C(b)|i>j]=\frac{N(N-1)}{2}C(a).
\label{twoemcol6}
\end{equation}
Summing over all hierarchies $h$ (as in eq.~(\ref{C_rep})), the total colour factors of diagrams (a) and (b) in the replicated theory are
\begin{align}
C_N(a)&=\frac{N}{2}\big[C(a)-C(b)\big]+\frac{N^2}{2}\big[C(a)+C(b)\big].\label{arepres}\\
C_N(b)&=\frac{N}{2}\big[C(b)-C(a)\big]+\frac{N^2}{2}\big[C(a)+C(b)\big].\label{brepres}
\end{align}
The exponentiated colour factors $\widetilde{C}(a)$ and $\widetilde{C}(b)$ are given by the coefficient of the ${\cal O}(N)$ parts of these results (eq.~(\ref{ECF_in_terms_of_RC})). Combining with the kinematic factors for each graph, the total contribution to the exponent from the diagrams of figure~\ref{twoemfig} is
\begin{equation}
\frac{1}{2}\Big(C(a)-C(b)\Big)\,\Big({\cal F}(a)-{\cal F}(b)\Big).
\label{twoemcol7}
\end{equation}
Thus, the only surviving part which exponentiates has the form of a function which is antisymmetric in both colour and kinematics. This agrees exactly with the results of~\cite{Aybat:2006mzy}, which state that the sum of diagrams (a) and (b) can be decomposed into a term which is symmetric in both colour and kinematics, and a part which is antisymmetric in both. The former can be shown to be a product of lower order diagrams, thus does not appear in the exponent of the eikonal scattering amplitude. Here we see this explicitly in eqs.~(\ref{arepres}) and~(\ref{brepres}), where the symmetric combination is ${\cal O}(N^2)$. The antisymmetric combination indeed survives (is ${\cal O}(N)$)~\footnote{Note that the antisymmetric combination of kinematic factors in eq.~(\ref{twoemcol7}) actually vanishes in the case of massless eikonal lines~\cite{Aybat:2006mzy}, although not in the case of massive lines~\cite{Mitov:2009sv}.}. This is our first example of the fact that reducible diagrams can appear in the exponent. Furthermore, the diagrams of figure~\ref{twoemfig} form a closed set, mixing only with each other in colour space. We will see more examples in the following section, where we consider diagrams at three loop order.

\subsection{Webs in multi-parton scattering: three loops\label{sec:ECF-3loops}}

Having shown examples of applying the replica trick to diagrams that have been previously studied using other methods, we now turn our attention to cases that have not been previously explored, at three-loop order. 
Although we shall consider here the specific case of three loops (and, in the next section, one set of four-loop diagrams) we emphasise that the replica trick algorithm, as set up in section~\ref{sec:replica}, can be directly applied at any order in the exponent: it does not require prior analysis of lower-order graphs. 

Further to providing examples for using the replica trick algorithm, we will explicitly check in each case that the linear combinations of colour factors found from the replica trick agree precisely with what is obtained upon calculating the exponent by subtracting 
the exponentiation of lower-order diagrams from the full result. The latter calculation is rather long, and it requires manipulating products of colour factors well as kinematic parts of lower-order diagrams. This comparison therefore illustrates how effective the replica trick algorithm is in determining directly the structure of the exponent.

\subsubsection*{Replica trick computation of $W_{(1,2,2,1)}$}

\begin{figure}[htb]
\begin{center}
\scalebox{.93}{\includegraphics{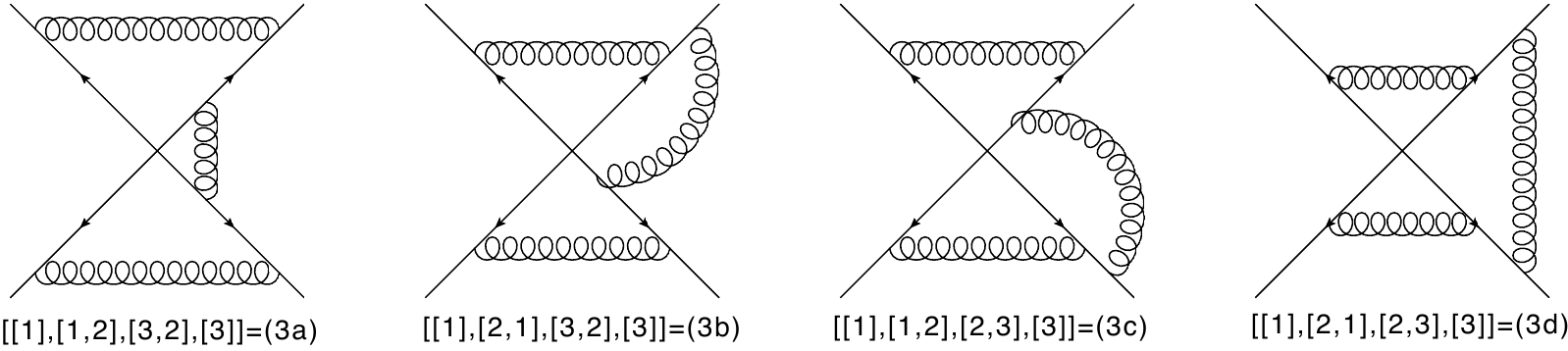}}
\caption{Example three loop diagrams.}
\label{17-20}
\end{center}
\end{figure}

We first focus on the set of diagrams shown in figure~\ref{17-20}, where we have labelled each diagram as shown in the figure.
Considering first diagram (3a), one must again assign replica numbers to the distinct connected pieces, and consider all possible hierarchies $h$. Labelling the upper, right-hand and lower gluons by $i$, $j$ and $k$ respectively, the list of hierarchies is given in table~\ref{17reptab}. The second column shows, for each $h$, the replica ordered colour factor (corresponding to the conventional colour factor of one of the graphs in figure~\ref{17-20}). The third column shows the multiplicity factor $M_N(h)$ {\it i.e.} the number of ways of choosing the number of replica numbers $i$, $j$ and $k$ out of $N$, given $h$. Finally, the fourth column shows the ${\cal O}(N)$ part of this multiplicity factor, which enters the expression for the exponentiated colour factor (eq.~(\ref{ECF_in_terms_of_RC})). 

\begin{table}
\begin{center}
\begin{tabular}{c|c|c|c}
$h$ &${\cal R}[(3a)|h]$&$M_N(h)$&${\cal O}(N^1)$ part of $M_N(h)$\\
\hline
$i=j=k$&$C(3a)$ & $N$ &1\\
$i=j,k>i$&$C(3a)$ & $N(N-1)/2$ &$-\frac{1}{2}$\\
$i=j,k<i$&$C(3c)$ & $N(N-1)/2$ &$-\frac{1}{2}$\\
$i=k,j>i$&$C(3d)$& $N(N-1)/2$ &$-\frac{1}{2}$\\
$i=k,j<i$&$C(3a)$& $N(N-1)/2$ &$-\frac{1}{2}$\\
$j=k,i>j$&$C(3a)$& $N(N-1)/2$ &$-\frac{1}{2}$\\
$j=k,i<j$&$C(3b)$& $N(N-1)/2$ &$-\frac{1}{2}$\\
$i<j<k$&$C(3b)$& $N(N-1)(N-2)/6$ &$\frac{1}{3}$\\
$i<k<j$&$C(3d)$& $N(N-1)(N-2)/6$ &$\frac{1}{3}$\\
$j<i<k$&$C(3a)$& $N(N-1)(N-2)/6$ &$\frac{1}{3}$\\
$j<k<i$&$C(3a)$& $N(N-1)(N-2)/6$ &$\frac{1}{3}$\\
$k<i<j$&$C(3d)$& $N(N-1)(N-2)/6$ &$\frac{1}{3}$\\
$k<j<i$&$C(3c)$& $N(N-1)(N-2)/6$ &$\frac{1}{3}$
\end{tabular}
\caption{Replica trick analysis for diagram (3a), as shown in figure~\ref{17-20}. The replica indices $i$, $j$ and $k$ label the gluons on the upper, right-hand and lower parts of the diagram respectively.}
\label{17reptab}
\end{center}
\end{table}
Adding things together, the ${\cal O}(N)$ part of the colour factor for this diagram in the replicated theory ({\it i.e.} the exponentiated colour factor in the unreplicated theory) is
\begin{equation}
\widetilde{C}(3a)=\frac{1}{6}\left[C(3a)-C(3b)-C(3c)+C(3d)\right].
\label{17rep}
\end{equation}
One may evaluate the exponentiated colour factors of the remaining diagrams in figure~\ref{17-20} in a similar fashion to obtain
\begin{align}
\widetilde{C}(3b)&=\frac{1}{3}\left[-C(3a)+C(3b)+C(3c)-C(3d)\right]\label{18rep}\\
\widetilde{C}(3c)&=\frac{1}{3}\left[-C(3a)+C(3b)+C(3c)-C(3d)\right]\label{19rep}\\
\widetilde{C}(3d)&=\frac{1}{6}\left[C(3a)-C(3b)-C(3c)+C(3d)\right]\label{20rep},
\end{align}
so that the total contribution to the exponent from these diagrams is
\begin{equation}
W_{(1,2,2,1)}=\frac{1}{6}\Big(C(3a)-C(3b)-C(3c)+C(3d)\Big)
\Big({\cal F}(3a)-2{\cal F}(3b)-2{\cal F}(3c)+{\cal F}(3d)\Big),
\label{17-20rep}
\end{equation}
where, as usual, ${\cal F}(D)$ denotes the kinematic part of diagram $D$, and on the left-hand side we have used the notation of eq.~(\ref{setmixintro}) denoting the number of gluon attachments on each parton line. 

It is also instructive to see the above results derived from the closed form solution for exponentiated colour factors of eq.~(\ref{modcol2}). We carry out this calculation in  Appendix~\ref{app-col2}. 
Similarly to the two-loop case considered in the previous section, we see that the diagrams of figure~\ref{17-20} form a ``closed set'' of diagrams which mix with each other in the exponent. Equation~(\ref{17-20rep}) is thus the result for the ``web'' formed from these diagrams. Furthermore, a non-trivial degree of combinatorics enters eq.~(\ref{17-20rep}). One might have not anticipated, for example, the factors of two in this expression. We shall return to this point below.

\subsubsection*{Replica trick computation of $W_{(2,3,1)}$}

\begin{figure}[htb]
\begin{center}
\scalebox{1}{\includegraphics{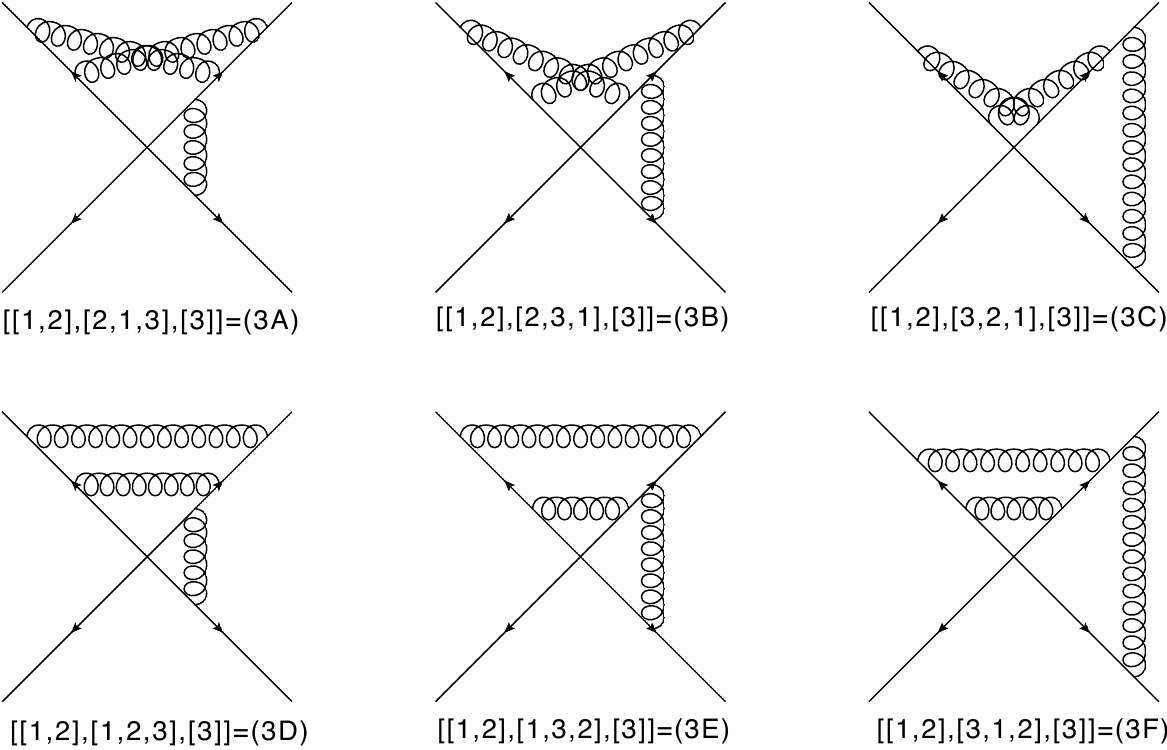}}
\caption{Example diagrams in which three parton lines are linked by three gluon exchanges.}
\label{3lsix}
\end{center}
\end{figure}

First let us consider another example of a closed set of diagrams, namely the set $(2,3,1)$ in the notation of eq.~(\ref{setmixintro}). These are shown in figure~\ref{3lsix}, where we also introduce some convenient labels.

Using the replica trick as outlined above, one may show that the exponentiated colour factors are
\begin{align}
\widetilde{C}(3A)&=\frac{1}{6}\Big[3C(3A)-2C(3D)-2C(3E)-3C(3C)+4C(3F)\Big]\label{resA}\\
\widetilde{C}(3B)&=\frac{1}{6}\Big[-3C(3A)+6C(3B)+C(3D)-2C(3E)+C(3F)-3C(3C)\Big]\label{resB}\\
\widetilde{C}(3C)&=\frac{1}{6}\Big[3C(3C)-2C(3E)-2C(3F)-3C(3A)+4C(3D)\Big]\label{resC}\\
\widetilde{C}(3D)&=\frac{1}{6}\Big[C(3D)-2C(3E)+C(3F)\Big]\label{resD}\\
\widetilde{C}(3E)&=\frac{1}{3}\Big[-C(3D)+2C(3E)-C(3F)\Big]\label{resE}\\
\widetilde{C}(3F)&=\frac{1}{6}\Big[C(3D)-2C(3E)+C(3F)\Big],\label{resF}
\end{align}
such that the total contribution to the exponent from these diagrams is
\begin{align}
\label{1-6rep}
\begin{split}
W_{(2,3,1)}
  &=\frac{1}{6}\Big[3C(3A)-3C(3C)-2C(3D)-2C(3E)+4C(3F)\Big]{\cal F}(3A)
\\&+\frac{1}{6}\Big[-3C(3A)+6C(3B)-3C(3C)+C(3D)-2C(3E)+C(3F)\Big]{\cal F}(3B)
\\&+\frac{1}{6}\Big[-3C(3A)+3C(3C)+4C(3D)-2C(3E)-2C(3F)\Big]{\cal F}(3C)
\\&+\frac{1}{6}\Big[C(3D)-2C(3E)+C(3F)\Big]{\cal F}(3D)
\\&+\frac{1}{3}\Big[-C(3D)+2C(3E)-C(3F)\Big]{\cal F}(3E)
+\frac{1}{6}\Big[C(3D)-2C(3E)+C(3F)\Big]{\cal F}(3F).
\end{split}
\end{align}
Once again, it is instructive to rederive the above results using eq.~(\ref{modcol2}) -- see appendix~\ref{app-col2}. Here the combinatorics of how each coefficient is obtained are even 
more complicated, and it is useful to reproduce the above results by explicitly exponentiating  lower-order diagrams. 

\subsubsection*{Calculation of $W_{(1,2,2,1)}$ by direct exponentiation of lower orders\label{exp1720}}

Consider again the diagrams of figure~\ref{17-20}, whose contribution to the exponent, as found from the replica trick, is given by eq.~(\ref{17-20rep}). One may verify this result explicitly as follows. One exponentiates all one and two loop diagrams (with the appropriate exponentiated colour factors) that lead to the diagrams whose form matches those of figure~\ref{17-20}. Then, one subtracts the result from the appropriate contribution to the unexponentiated amplitude. The result is, by definition, the contribution to the exponent from these three loop diagrams. 

The aim of this exercise is both to demonstrate how exponentiation of lower-order diagrams proceeds in the multileg case, and to highlight the drastic simplification achieved by the replica trick. We will see that the results we have obtained above, even though they relate to three-loop diagrams of relatively simple topology, are much more complicated to derive using explicit exponentiation, which involves manipulating both kinematic and colour factors of all relevant 
lower-order diagrams. The reader who is not interested in seeing such a rederivation of the above results may skip to the end of this section, where we discuss the general structure of our results in more detail.

\begin{figure}[htb]
\begin{center}
\scalebox{.93}{\includegraphics{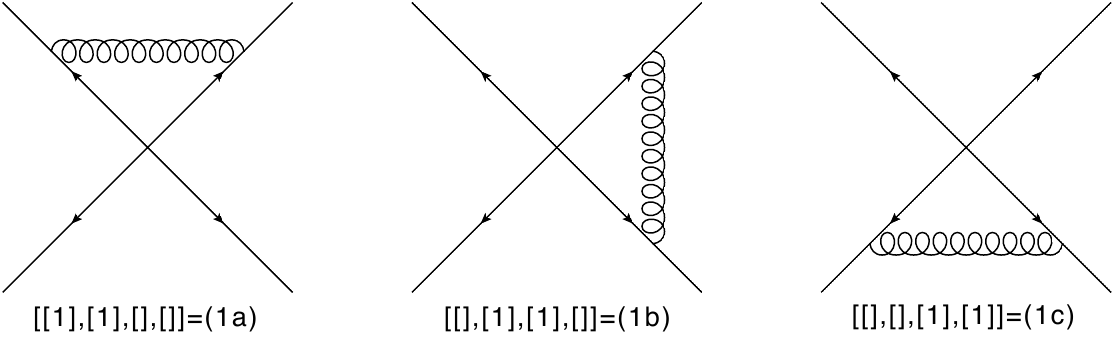}}
\caption{One loop diagrams which, when exponentiated, contribute to eq.~(\ref{17-20rep}).}
\label{1a-c}
\end{center}
\end{figure}

\begin{figure}[htb]
\begin{center}
\scalebox{.93}{\includegraphics{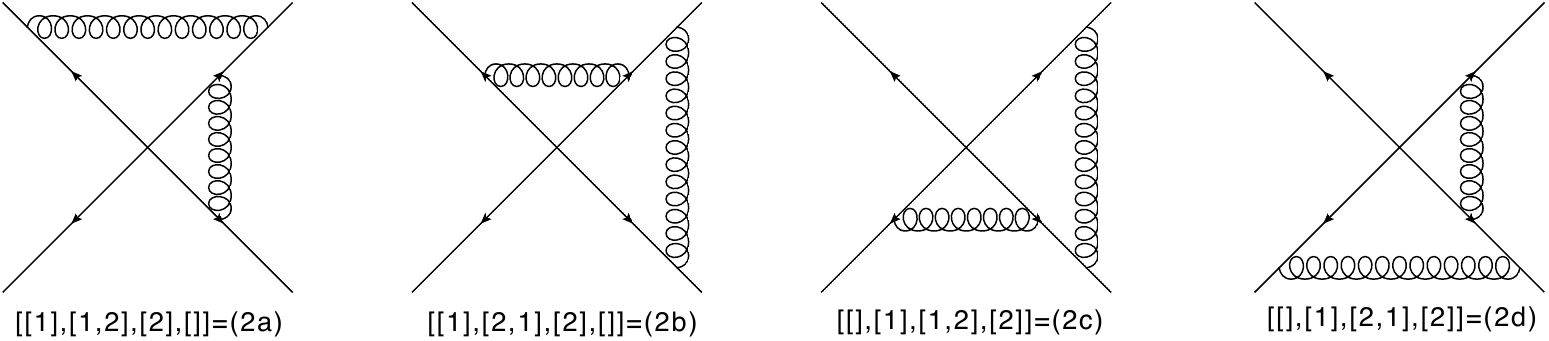}}
\caption{Two loop diagrams which, when exponentiated, contribute to eq.~(\ref{17-20rep}).}
\label{2a-d}
\end{center}
\end{figure}

To be more specific, one may write the eikonal scattering amplitude in either conventional or exponentiated form as
\begin{equation}
{\cal A}=\sum_{n}A_n\alpha_s^n\,=\,\exp\left(\sum_mB_m\alpha_s^m\right),
\label{ampexp}
\end{equation}
where the coefficients $A_i$ and $B_i$ are matrix-valued in the space of possible colour flows, thus do not commute in general. In the notation introduced above, the $\{A_i\}$ are given by
\begin{equation}
A_n=\sum_GC(G){\cal F}(G),
\label{Aidef}
\end{equation}
where the sum is over all Feynman diagrams $G$ contributing at ${\cal O}(\alpha_s^n)$. By expanding the exponential on the right-hand side, the $\{B_i\}$ and $\{A_i\}$ can be related to each other. Up to three-loop order, one has
\begin{align}
A_1&=B_1\notag\\
A_2&=B_2+\frac{1}{2}B_1^2\notag\\
A_3&=B_3+\frac{1}{2}\{B_1,B_2\}+\frac{1}{6}B_1^3.
\label{ABrel}
\end{align}
In order to rederive the replica trick result ({\it i.e.} that $B_3$ is given by eq.~(\ref{17-20rep})) using (\ref{ABrel}), one thus needs to evaluate the one and two-loop coefficients $B_1$ and $B_2$, and subtract the appropriate combination from $A_3$. In calculating $B_1$ and $B_2$, one need only consider diagrams that, when exponentiated, will give rise to three loop diagrams of the form shown in figure~\ref{17-20}. 
The relevant diagrams are shown in figures~\ref{1a-c} and~\ref{2a-d}. We have seen some of these diagrams already. However, we redraw them all together here for convenience, and also to introduce labels that will be used in what follows. From previous results we know that 
\begin{align}
B_1&=C(1a){\cal F}(1a)+C(1b){\cal F}(1b)+C(1c){\cal F}(1c)\notag\\
B_2&=\frac{1}{2}\Big(C(2a)-C(2b)\Big)\Big({\cal F}(2a)-{\cal F}(2b)\Big)
+\frac{1}{2}\Big(C(2c)-C(2d)\Big)\Big({\cal F}(2c)-{\cal F}(2d)\Big),
\label{onetwoloop}
\end{align}
where we have kept only those terms which contribute to eq.~(\ref{17-20rep}). The first result follows immediately from $B_1=A_1$, and the second result was first derived in~\cite{Aybat:2006mzy} (we have also seen part of this result in eq.~(\ref{twoemcol7})). In combining these results to explicitly derive $B_3$, one needs to know how to deal with products of eikonal diagrams (e.g. $C(1a)C(2a){\cal F}(1a){\cal F}(2a)$). To illustrate the method, it is simpler to first rederive $B_2$ by exponentiating the one-loop result. 


\begin{figure}[htb]
\begin{center}
\scalebox{1.0}{\includegraphics{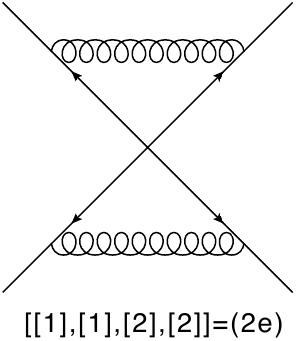}}
\caption{Two-loop diagram present in the amplitude, but absent from the exponent.}
\label{2e}
\end{center}
\end{figure}

At two loops (\ref{ABrel}) yields 
\begin{equation}
B_2=A_2-\frac{1}{2}{B_1^2},
\label{B2def}
\end{equation}
where
\begin{equation}
A_2=C(2a){\cal F}(2a)+C(2b){\cal F}(2b)+C(2c){\cal F}(2c)+C(2d){\cal F}(2d)+C(2e){\cal F}(2e).
\label{A2def}
\end{equation}
The final term contains the diagram shown in figure~\ref{2e} which, as we will see, does not appear in the exponent. 

Evaluating $B_1^2$ from eq.~(\ref{onetwoloop}) gives
\begin{align}
\frac{1}{2}B_1^2&=\frac{1}{2}\Big[
{\cal F}(1a){\cal F}(1b)\left\{C(1a),C(1b)\right\}
+{\cal F}(1a){\cal F}(1c)\left\{C(1a),C(1c)\right\}\notag\\
&\quad+{\cal F}(1b){\cal F}(1c)\left\{C(1b),C(1c)\right\}+\ldots\Big].
\label{B1sq}
\end{align}
On the right-hand side, we have kept only products of diagrams which give rise to graphs of the same form as those shown in figure~\ref{2a-d}. Furthermore, we have used the fact that the kinematic factors commute ($[{\cal F}(G),{\cal F}(H)]=0$), but the colour factors do not in general. Considering the first term, the product of kinematic factors gives
\begin{align}
{\cal F}(1a){\cal F}(1b)&=v_{12}\,v_{23}\,\int_0^\infty dt_1\int_0^\infty dt_2\int_0^\infty ds_2\int_0^\infty ds_3
% \notag\\&\quad\times 
D(t_1v_1-t_2v_2)D(s_2v_2-s_3v_3),
\label{M1a1bprod}
\end{align}
where $v_i$ is the 4-velocity (or rescaled 4-momentum) of the $i^{\text{th}}$ eikonal line, and where we introduce the notation $v_{ij}\equiv v_i\cdot v_j$ for the scalar product of $v_i$ and $v_j$. Here $D(x-y)$ is the position-space propagator for a soft gluon propagating between $x$ and $y$. We have ignored coupling factors etc. for brevity. The colour factors may be given in Catani-Seymour notation~\cite{Catani:1997vz} as
\begin{equation}
C(1a)={\mathrm T}_1^a{\mathrm T}_2^a,\quad C(1b)={\mathrm T}_2^b{\mathrm T}_3^b,
\label{C1a1b}
\end{equation}
(with ${\mathrm T}_i^a$ a colour generator in the appropriate representation of parton $i$) so that
\begin{equation}
\{C(1a),C(1b)\}={\mathrm T}_1^a\{{\mathrm T}_2^a,{\mathrm T}_2^b\}{\mathrm T}_3^b.
\label{C1a1bprod}
\end{equation}
Then the first term on the right-hand side of eq.~(\ref{B1sq}) gives
\begin{equation}
\frac{1}{2}\,v_{12}\,v_{23}\,{\mathrm T}_1^a\{{\mathrm T}_2^a,{\mathrm T}_2^b\}{\mathrm T}_3^b\int_0^\infty dt_1\int_0^\infty dt_2\int_0^\infty ds_2\int_0^\infty ds_3 D(t_1v_1-t_2v_2)D(s_2v_2-s_3v_3).
\label{B1sq1}
\end{equation}
Similarly, the second and third terms give
\begin{equation}
v_{12}\,v_{23}\,{\mathrm T}_1^a{\mathrm T}_2^a{\mathrm T}_3^c{\mathrm T}_4^c\int_0^\infty dt_1\int_0^\infty dt_2\int_0^\infty dt_3\int_0^\infty dt_4D(t_1v_1-t_2v_2)D(t_3v_3-t_4v_4)
\label{B1sq2}
\end{equation}
and 
\begin{equation}
\frac{1}{2}\,v_{12}\,v_{23}\,{\mathrm T}_2^b\{{\mathrm T}_3^b,{\mathrm T}_3^c\}{\mathrm T}_4^c\int_0^\infty ds_2\int_0^\infty ds_3\int_0^\infty dt_3\int_0^\infty dt_4D(s_2v_2-s_3v_3)D(t_3v_3-t_4v_4)
\label{B1sq3}
\end{equation}
respectively. From eq.~(\ref{A2def}), one has
\begin{align}
A_2&=v_{12}\,v_{23}\,\bigg[{\mathrm T}_1^a{\mathrm T}_2^a{\mathrm T}_2^b{\mathrm T}_3^b\int_0^\infty dt_1\int_0^\infty dt_2\int_0^{t_2}ds_2\int_0^\infty ds_3D(t_1v_1-t_2v_2) D(s_2v_2-s_3v_3)\notag\\
&\quad+{\mathrm T}_1^a{\mathrm T}_2^b{\mathrm T}_2^a{\mathrm T}_3^b\int_0^\infty dt_1\int_0^\infty dt_2\int_{t_2}^\infty ds_2\int_0^\infty ds_3 D(t_1v_1-t_2v_2)D(s_2v_2-s_3v_3)\notag\\
&\quad+{\mathrm T}_2^b{\mathrm T}_3^b{\mathrm T}_3^c{\mathrm T}_4^c\int_0^\infty ds_2\int_0^\infty dt_3\int_{t_3}^\infty ds_3\int_0^\infty dt_4 D(s_2v_2-s_3v_3)D(t_3v_3-t_4v_4)\notag\\
&\quad+{\mathrm T}_2^b{\mathrm T}_3^c{\mathrm T}_3^b{\mathrm T}_4^c\int_0^\infty ds_2\int_0^\infty dt_3\int_0^{t_3}ds_3\int_0^\infty dt_4D(s_2v_2-s_3v_3)D(t_3v_3-t_4v_4)\notag\\
&\quad+{\mathrm T}_1^a{\mathrm T}_2^a{\mathrm T}_3^c{\mathrm T}_4^c\int_0^\infty dt_1\int_0^\infty dt_2\int_0^\infty dt_3\int_0^\infty dt_4D(t_1v_1-t_2v_2)D(t_3v_3-t_4v_4)\bigg],
\label{A2calc}
\end{align}
where the five terms correspond to diagrams (2a)-(2e) respectively. Using eqs.~(\ref{B2def},\ref{B1sq1}-\ref{B1sq3}) one finds
\begin{align}
B_2&= v_{12}\,v_{23}\,
\left\{\frac{1}{2}{\mathrm T}_1^a[{\mathrm T}_2^a,{\mathrm T}_2^b]{\mathrm T}_3^b\int_0^\infty dt_1\int_0^\infty dt_2\int_0^\infty ds_3\left[\int_0^{t_2}ds_2-\int_{t_2}^\infty ds_2\right]\right.\notag\\
&\quad\times D(t_1v_1-t_2v_2)D(s_2v_2-s_3v_3)+\frac{1}{2}{\mathrm T}_2^b[{\mathrm T}_3^b,{\mathrm T}_3^c]{\mathrm T}_4^c\int_0^\infty ds_2\int_0^\infty dt_3\int_0^\infty dt_4\notag\\
&\left.\quad\times\left[\int_{t_3}^{\infty}ds_3-\int_0^{t_3}ds_3\right]D(s_2v_2-s_3v_3)D(t_3v_3-t_4v_4)\right\}\notag\\
&=\frac{1}{2}\left(C(2a)-C(2b)\right)\left({\cal F}(2a)-{\cal F}(2b)\right)+\frac{1}{2}\left(C(2c)-C(2d)\right)\left({\cal F}(2c)-{\cal F}(2d)\right),
\label{B2res}
\end{align}
which indeed agrees with eq.~(\ref{onetwoloop}). 

Having seen explicitly how exponentiation of a subset of diagrams works at two loops, we now consider the three loop example of figure~\ref{17-20} by evaluating $B_3$, given by eq.~(\ref{ABrel}) as
\begin{equation}
B_3=A_3-\frac{1}{6}B_1^3-\frac{1}{2}\{B_1,B_2\},
\label{B3def}
\end{equation}
where
\begin{equation}
A_3=C(3a){\cal F}(3a)+C(3b){\cal F}(3b)+C(3c){\cal F}(3c)+C(3d){\cal F}(3d).
\label{A3calc}
\end{equation}
Firstly, one has
\begin{align}
\frac{1}{6}B_1^3&=\frac{1}{6}\,\Big(C(1a){\cal F}(1a)+C(1b){\cal F}(1b)+C(1c){\cal F}(1c)\Big)^3\notag\\
&=\frac{1}{6}\,{\cal F}(1a){\cal F}(1b){\cal F}(1c)
\,\Big(C(1a)C(1b)C(1c)+C(1a)C(1c)C(1b)+C(1b)C(1a)C(1c) \notag\\
&\quad+C(1b)C(1c)C(1a)+C(1c)C(1a)C(1b)+C(1c)C(1b)C(1a)\Big)\,+\cdots\,,
\label{B1cubed}
\end{align}
where we have kept only combinations in the cubic product that lead to diagrams of the form shown in figure~\ref{17-20}, and have also used the fact that the kinematic factors commute. Considering the first term in the sum over products of colour factors, this is
\begin{align}
C(1a)C(1b)C(1c)&=({\mathrm T}_1^a{\mathrm T}_2^a)({\mathrm T}_2^b{\mathrm T}_3^b)({\mathrm T}_3^c{\mathrm T}_4^c)={\mathrm T}_1^a{\mathrm T}_2^a{\mathrm T}_2^b{\mathrm T}_3^b{\mathrm T}_3^c{\mathrm T}_4^c\notag\\
&=C(3c).
\label{B13col1}
\end{align}
Note we have used the rule in combining the colour factors that in each product of matrices associated with the same parton line (e.g. ${\mathrm T}_2^b{\mathrm T}_2^c$), the matrices are ordered from the external line towards the hard interaction. Evaluating the remaining terms in eq.~(\ref{B1cubed}), one finds 
\begin{equation}
\frac{1}{6}B_1^3=\frac{1}{6}\,{\cal F}(1a){\cal F}(1b){\cal F}(1c)\,\Big(C(3c)+2C(3a)+2C(3d)+C(3b)\Big).
\label{B1cubed2}
\end{equation}
The kinematic prefactor has the explicit form
\begin{align}
{\cal F}(1a){\cal F}(1b){\cal F}(1c)&=v_{12}v_{23}v_{34}\int_0^\infty dt_1\int_0^\infty dt_2\int_0^\infty dt_3\int_0^\infty ds_2\int_0^\infty ds_3\notag\\
&\quad\times\int_0^\infty dt_4D(t_1v_1-t_2v_2)D(s_2v_2-s_3v_3)D(t_3v_3-t_4v_4).
\label{M1a1b1cprod}
\end{align}
All of the kinematic factors from now on will involve the same integrals over $\{t_i\}$, and the same product of propagators and $v_{ij}$ factors. Thus we may shorten notation by writing each kinematic factor according to its $s_2$ and $s_3$ integrals {\it i.e.}
\begin{equation}
{\cal F}(1a){\cal F}(1b){\cal F}(1c)\equiv \int_0^\infty ds_2\int_0^\infty ds_3.
\label{M1a1b1cprod2}
\end{equation}
We may express this in terms of the kinematic factors of the individual diagrams in figure~\ref{17-20}. After writing
\begin{equation}
\int_0^\infty ds_i=\int_0^{t_i}ds_i+\int_{t_i}^\infty ds_i,
\label{momsep}
\end{equation}
eq.~(\ref{M1a1b1cprod2}) becomes
\begin{equation}
{\cal F}(1a){\cal F}(1b){\cal F}(1c)= {\cal F}(3a)+{\cal F}(3b)+{\cal F}(3c)+{\cal F}(3d),
\label{M1a1b1cprod3}
\end{equation}
where we have identified
\begin{align}
&{\cal F}(3a)\equiv \int_0^{t_2}ds_2\int_0^{t_3}ds_3,\quad {\cal F}(3b)\equiv\int_{t_2}^\infty ds_2\int_0^{t_3}ds_3\notag\\
&{\cal F}(3c)\equiv \int_0^{t_2}ds_2\int_{t_3}^{\infty}ds_3,\quad {\cal F}(3d)\equiv\int_{t_2}^\infty ds_2\int_{t_3}^{\infty}ds_3.
\label{Mdefs}
\end{align}
Thus one has
\begin{equation}
\frac{1}{6}B_1^3=\frac{1}{6}\,\Big({\cal F}(3a)+{\cal F}(3b)+{\cal F}(3c)+{\cal F}(3d)\Big)\,\Big(2C(3a)+C(3b)+C(3c)+2C(3d)\Big).
\label{B1cubed3}
\end{equation}
This result is already interesting, in that some nontrivial combinatoric factors appear. The three-loop diagrams do not occur equally, with a separation between diagrams (3a, 3d) and (3b, 3c). We will see that the same separation persists in the final result for $B_3$ (as has already been found using the replica trick). 

The last ingredient in (\ref{B3def}) is the term
\begin{align}
\frac{1}{2}\{B_1,B_2\}=&\,\frac{1}{4}{\cal F}(1a)
\Big({\cal F}(2c)-{\cal F}(2d)\Big)\Big[C(1a)\Big(C(2c)-C(2d)\Big)
+\Big(C(2c)-C(2d)\Big)C(1a)\Big]\notag\\
\,+&\,\frac{1}{4}{\cal F}(1c)\Big({\cal F}(2a)-{\cal F}(2b)\Big)
\Big[C(1c)\Big(C(2a)-C(2b)\Big)+\Big(C(2a)-C(2b)\Big)C(1c)\Big]\notag\\
\,+&\,\ldots
\label{B1B2}
\end{align}
Again we have only kept terms leading to diagrams having the forms shown in figure~\ref{17-20}. The kinematic factors may be simplified as follows:
\begin{align}
{\cal F}(1a)\left({\cal F}(2c)-{\cal F}(2d)\right)&\equiv \int_0^\infty ds_2\int_{t_3}^\infty ds_3-\int_0^\infty ds_2\int_0^{t_3}ds_3\notag\\
&=\int_0^{t_2}ds_2\int_{t_3}^\infty ds_3+\int_{t_2}^\infty ds_2\int_{t_3}^\infty ds_3-\int_0^{t_2}ds_2\int_0^{t_3}ds_3\notag\\
&\quad-\int_{t_2}^\infty ds_2\int_0^{t_3}ds_3\notag\\
&={\cal F}(3c)+{\cal F}(3d)-{\cal F}(3a)-{\cal F}(3b);
\label{mom1}
\end{align}
\begin{align}
{\cal F}(1c)\left({\cal F}(2a)-{\cal F}(2b)\right)&\equiv \int_0^{t_2} ds_2\int_0^\infty ds_3-\int_{t_2}^\infty ds_2\int_0^\infty ds_3\notag\\
&=\int_0^{t_2} ds_2\int_0^{t_3}ds_3+\int_0^{t_2}ds_2\int_{t_3}^\infty ds_3-\int_{t_2}^\infty ds_2\int_0^{t_3}ds_3\notag\\
&\quad-\int_{t_2}^\infty ds_2\int_{t_3}^\infty ds_3\notag\\
&={\cal F}(3a)+{\cal F}(3c)-{\cal F}(3b)-{\cal F}(3d).
\label{mom2}
\end{align}
Furthermore, the colour factors can be evaluated to give
\begin{equation}
C(1a)\Big(C(2c)-C(2d)\Big)+\Big(C(2c)-C(2d)\Big)C(1a)\,=\,
C(3c)-C(3a)+C(3d)-C(3b).
\label{col1}
\end{equation}
and
\begin{equation}
C(1c)\Big(C(2a)-C(2b)\Big)+\Big(C(2a)-C(2b)\Big)C(1c)\,=\,
C(3c)-C(3d)+C(3a)-C(3b).
\end{equation}
Putting things together, one finds
\begin{align}
\frac{1}{2}\{B_1,B_2\}&=\frac{1}{2}\Big[C(3a)\Big({\cal F}(3a)-{\cal F}(3d)\Big)
+C(3b)\Big({\cal F}(3b)-{\cal F}(3c)\Big)
+C(3c)\Big({\cal F}(3c)-{\cal F}(3b)\Big)
\notag\\
&+C(3d)\Big({\cal F}(3d)-{\cal F}(3a)\Big)\Big].
\label{B1B2tot}
\end{align}
Finally, combining eqs.~(\ref{B3def}, \ref{A3calc}, \ref{B1cubed3}, \ref{B1B2tot}) one finds
\begin{equation}
B_3=\frac{1}{6}\Big(C(3a)-C(3b)-C(3c)+C(3d)\Big)
\Big({\cal F}(3a)-2{\cal F}(3b)-2{\cal F}(3c)+{\cal F}(3d)\Big),
\label{B3res}
\end{equation}
which agrees precisely with the replica trick result of eq.~(\ref{17-20rep}). Furthermore, we have learnt the following lessons:
\begin{itemize}
\item The exponentiation of lower-order diagrams results in products of colour factors, as well as products of kinematic factors, which can be recast as linear combinations of the colour and kinematic factors of recognisable diagrams. These manipulations bring about non-trivial combinatorics.
\item The replica trick greatly simplifies the calculation. The degree of simplification becomes yet more apparent at higher orders.
\end{itemize}

\subsubsection*{Calculation of $W_{(2,3,1)}$ by direct exponentiation of lower orders}

\begin{figure}[htb]
\begin{center}
\scalebox{1.2}{\includegraphics{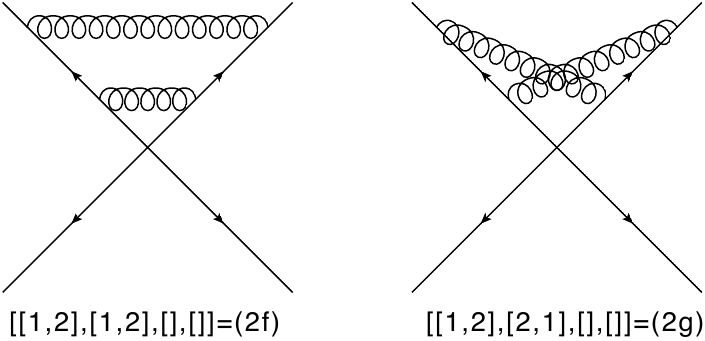}}
\caption{Two loop diagrams which, when exponentiated, contribute to the diagrams shown in figure~\ref{3lsix}.}
\label{2f2g}
\end{center}
\end{figure}

As a further demonstration, one may consider the diagrams of figure~\ref{3lsix}. The combination of these diagrams which appears in the exponent is given in eq.~(\ref{1-6rep}), and one may check that the same result is obtained on subtracting the relevant exponentiated one and two loop graphs from the three loop amplitude. In this case, the terms $B_i$ in the exponent up to three loops are given by eq.~(\ref{ABrel}) with
\begin{align}
A_1&=C(1a){\cal F}(1a)+C(1b){\cal F}(1b)\notag\\
A_2&=C(2a){\cal F}(2a)+C(2b){\cal F}(2b)+C(2f){\cal F}(2f)+C(2g){\cal F}(2g)\notag\\
A_3&=C(3A){\cal F}(3A)+C(3B){\cal F}(3B)+C(3C){\cal F}(3C)+C(3D){\cal F}(3D)\notag\\
&\quad+C(3E){\cal F}(3E)+C(3F){\cal F}(3F),
\label{A1-6}
\end{align}
where the second line contains the diagrams shown in figure~\ref{2f2g}.
As before, we first calculate the two loop contribution to the exponent, before extending the calculation to three loops.

For this set of diagrams, one has 
\begin{align}
\frac{1}{2}\,B_1^2&=\frac{1}{2}\Big(C(1a){\cal F}(1a)+C(1b){\cal F}(1b)\Big)^2\notag\\
&=\frac{1}{2}\,C^2(1a){\cal F}^2(1a)+\frac{1}{2}\{C(1a),C(1b)\}{\cal F}(1a){\cal F}(1b)+\ldots,
\label{b12a}
\end{align}
where the ellipsis denotes terms which do not contribute to diagrams of the form shown in figure~\ref{3lsix}. The kinematic factor in the first term is
\begin{equation}
{\cal F}^2({1a})=v_{12}^2\int_0^{\infty}dt_1\int_0^{\infty}dt_2 \int_0^\infty ds_1\int_0^\infty ds_2D(t_1v_1-t_2v_2)D(s_1v_1-s_2v_2),
\label{M1a2}
\end{equation}
which must be rewritten in terms of the kinematic parts of the diagrams shown in figure~\ref{2f2g}. 
We parametrize the integrals along $v_1$ by $t_1$ and $s_1$, the ones along $v_2$ by $t_2$, $s_2$ and $s_3$, and the one along $v_3$ by $t_3$.
Given these notations, we note that the integrals considered share the same $t_i$ integrals as well as the same propagator factors, while the limits on the $s_1$, $s_2$ and $s_3$ integrations differ between them. We can therefore represent the kinematic part of each diagram by its $s_i$ integrals\footnote{Similar notation will be used below at three loops.} {\it i.e.} for the two-loop integrals in figure~\ref{2f2g},
\begin{align}
{\cal F}(2f)\equiv \int_{t_1}^\infty ds_1\int_{t_2}^\infty ds_2\quad\text{or}\quad \int_{0}^{t_1}ds_1\int_0^{t_2}ds_2;\notag\\
{\cal F}(2g)\equiv \int_{t_1}^\infty ds_1\int_0^{t_2}ds_2\quad\text{or}\quad \int_0^{t_1}ds_1\int_{t_2}^\infty ds_2.
\label{Mparts2f2g}
\end{align}
Note that each diagram can be rewritten in two ways, which correspond to different labellings of gluons in terms of the parameters $s_i$ and $t_i$. One then has
\begin{align}
{\cal F}^2({1a})&\equiv\left[\int_0^{t_1}ds_1+\int_{t_1}^\infty ds_1\right]\left[\int_0^{t_2}ds_2+\int_{t_2}^\infty ds_2\right]\notag\\
&=2\Big({\cal F}(2f)+{\cal F}(2g)\Big).
\label{M1a22}
\end{align}
One also needs the kinematic combination
\begin{align}
{\cal F}(1a){\cal F}(1b)&=
v_{12}\,v_{23}
\int_0^{\infty}dt_1\int_0^{\infty}dt_2\int_0^{\infty}ds_3\int_0^{\infty}dt_3 D(t_1v_1-t_2v_2)\,D(s_3v_2-t_3v_3)\notag\\
&=
v_{12}\,v_{23}\,
\int_0^\infty dt_1\int_0^{\infty}dt_2\int_0^\infty dt_3\notag\\ &\qquad\quad\left[\int_0^{t_2}ds_3+\int_{t_2}^\infty ds_3\right]D(t_1v_1-t_2v_2)\, D(s_3v_2-t_3v_3)\notag\\
&={\cal F}(2a)+{\cal F}(2b).
\label{M1a1b}
\end{align}
The colour factors appearing in eq.~(\ref{b12a}) combine to give
\begin{align}
C^2(1a)&=C(2f)\notag\\
\{C(1a),C(1b)\}&=C(2a)+C(2b).
\label{colfacsb12}
\end{align}
Putting things together, one finds
\begin{equation}
\frac{B_1^2}{2}=C(2f)\Big({\cal F}(2f)+{\cal F}(2g)\Big)+\frac{1}{2}\Big(C(2a)+C(2b)\Big)\Big({\cal F}(2a)+{\cal F}(2b)\Big).
\label{B12res}
\end{equation}
Combining this with eqs.~(\ref{ABrel}, \ref{A1-6}), the two-loop contribution to the exponent for the diagrams of figure~\ref{3lsix} is
\begin{equation}
B_2=\Big(C(2g)-C(2f)\Big){\cal F}(2g)+\frac{1}{2}\Big(C(2a)-C(2b)\Big)\Big({\cal F}(2a)-{\cal F}(2b)\Big),
\label{B2res2}
\end{equation}
which agrees with the results of the replica trick, and also those of~\cite{Aybat:2006mzy}.

Next, we calculate the three loop contribution to the exponent from the diagrams of figure~\ref{3lsix}. From eq.~(\ref{ABrel}), one first has the cube of the one loop exponent
\begin{align}
\frac{B_1^3}{6}&=\frac{1}{6}\Big(C(1a){\cal F}(1a)+C(1b){\cal F}(1b)\Big)^3\notag\\
&=\frac{1}{6}{\cal F}^2({1a}){\cal F}(1b)\Big(C^2(1a)C(1b)+C(1a)C(1b)C(1a)+C(1b)C^2(1a)\Big)+\ldots,
\label{oneloopcube}
\end{align}
where we again neglect terms corresponding to diagrams which do not have the form of figure~\ref{3lsix}. The kinematic factor must be rewritten in terms of the kinematic factors of each diagram. The kinematic factor for diagram (3A) is
\begin{align}
{\cal F}(3A)&=v_{12}^2v_{23}\int_0^\infty dt_1\int_0^\infty dt_2\int_0^\infty dt_3\int_0^{t_1}ds_1\int_{t_2}^\infty ds_2\int_0^{t_2}ds_3\notag\\
&\quad\times D(t_1v_1-t_2v_2)D(s_1v_1-s_2v_2)D(s_3v_2-t_3v_3.)
\label{mom3A}
\end{align}
Again we may neglect the propagator factors and $t_i$ integrals, and represent each diagram by its $s_i$ integrals. The results are
\begin{align}
{\cal F}(3A)&\equiv\int_0^{t_1}ds_1\int_{t_2}^\infty ds_2\int_0^{t_2}ds_3\quad\text{or}\quad \int_{t_1}^\infty ds_1\int_0^{t_2}ds_2\int_0^{s_2}ds_3;\notag\\
{\cal F}(3B)&\equiv\int_0^{t_1}ds_1\int_{t_2}^\infty ds_2\int_{t_2}^{s_2}ds_3\quad\text{or}\quad\int_{t_1}^\infty ds_1\int_0^{t_2}ds_2\int_{s_2}^{t_2}ds_3;\notag\\
{\cal F}(3C)&\equiv\int_0^{t_1}ds_1\int_{t_2}^\infty ds_2\int_{s_2}^\infty ds_3\quad\text{or}\quad\int_{t_1}^\infty ds_1\int_0^{t_2}ds_2\int_{t_2}^\infty ds_3\notag\\
{\cal F}(3D)&\equiv\int_0^{t_1}ds_1\int_0^{t_2}ds_2\int_0^{s_2}ds_3\quad\text{or}\quad\int_{t_1}^\infty ds_1\int_{t_2}^\infty ds_2\int_0^{t_2}ds_3;\notag\\
{\cal F}(3E)&\equiv\int_0^{t_1}ds_1\int_0^{t_2}ds_2\int_{s_2}^{t_2}ds_3\quad\text{or}\quad\int_{t_1}^\infty ds_1\int_{t_2}^\infty ds_2\int_{t_2}^{s_2}ds_3\notag\\
{\cal F}(3F)&\equiv\int_0^{t_1}ds_1\int_0^{t_2}ds_2\int_{t_2}^\infty ds_3\quad\text{or}\quad\int_{t_1}^\infty ds_1\int_{t_2}^\infty ds_2\int_{s_2}^\infty ds_3.
\label{momfacs}
\end{align}
where we have used the fact that each diagram may be written in different, but equivalent ways. Then one can write
\begin{align}
{\cal F}^2({1a}){\cal F}(1b)&=\int_0^\infty ds_1\int_0^\infty ds_2\int_0^\infty ds_3\notag\\
&=\int_0^{t_1}ds_1\int_0^{t_2}ds_2\left[\int_0^{s_2}ds_3+\int_{s_2}^{t_2}ds_3+\int_{t_2}^\infty ds_3\right]+\int_{t_1}^\infty ds_1\int_0^{t_2}ds_2\left[\int_0^{s_2}ds_3\right.\notag\\
&\left.+\int_{s_2}^{t_2}ds_3+\int_{t_2}^\infty ds_3\right]+\int_0^{t_1}ds_1\int_{t_2}^\infty ds_2\left[\int_0^{t_2}ds_3+\int_{t_2}^{s_2}ds_3+\int_{s_2}^\infty ds_3\right]\notag\\
&+\int_{t_1}^\infty ds_1\int_{t_2}^\infty ds_2\left[\int_0^{t_2}ds_3+\int_{t_2}^{s_2}ds_3+\int_{s_2}^\infty ds_3\right]\notag\\
&=2({\cal F}(3A)+{\cal F}(3B)+{\cal F}(3C)+{\cal F}(3D)+{\cal F}(3E)+{\cal F}(3F)).
\label{oneloopcube2}
\end{align}
The colour factors in eq.~(\ref{oneloopcube}) give
\begin{align}
C^2(1a)C(1b)&=C(3D);\notag\\
C(1a)C(1b)C(1a)&=C(3E);\notag\\
C(1b)C^2(1a)&=C(3F),
\label{oneloopcubecol}
\end{align}
so that one has
\begin{equation}
\frac{B_1^3}{6}=\frac{1}{3}(C(3D)+C(3E)+C(3F))({\cal F}(3A)+{\cal F}(3B)+{\cal F}(3C)+{\cal F}(3D)+{\cal F}(3E)+{\cal F}(3F)).
\label{oneloopcuberes}
\end{equation}
Next, one has the contribution of one and two loop webs from the exponentiation {\it i.e.} the third term in eq.~(\ref{B3def}), which in the present case gives
\begin{align}
\frac{1}{2}\{B_1,B_2\}&=\frac{1}{4}\{C(1a),C(2a)-C(2b)\}{\cal F}(1a)({\cal F}(2a)-{\cal F}(2b))\notag\\
&\quad+\frac{1}{2}\{C(1b),C(2g)-C(2f)\}{\cal F}(1b){\cal F}(2g).
\label{B31-6}
\end{align}
The first kinematic factor on the right-hand side is given by
\begin{align}
{\cal F}(1a)({\cal F}(2a)-{\cal F}(2b))&=\int_0^\infty ds_1\int_0^\infty ds_2\int_0^{s_2}ds_3-\int_0^\infty ds_1\int_0^\infty ds_2\int_{s_2}^\infty ds_3\notag\\
&=\left[\int_0^{t_1}ds_1\int_0^{t_2}ds_2+\int_{t_1}^\infty ds_1\int_0^{t_2}ds_2+\int_0^{t_1}ds_1\int_{t_2}^\infty ds_2\right.\notag\\
&\left.\quad+\int_{t_1}^\infty ds_1\int_{t_2}^\infty ds_2\right]\left[\int_0^{s_2}ds_3-\int_{s_2}^\infty ds_3\right]\notag\\
&=2({\cal F}(3A)-{\cal F}(3C)+{\cal F}(3D)-{\cal F}(3F)).
\label{mom1eq}
\end{align}
Also, one has
\begin{align}
{\cal F}(1b){\cal F}(2g)&=\int_0^{t_1}ds_1\int_{t_2}^\infty ds_2\int_0^\infty ds_3\notag\\
&=\int_0^{t_1}ds_1\int_{t_2}^\infty ds_2\left[\int_0^{t_2}ds_3+\int_{t_2}^{s_2}ds_3+\int_{s_2}^\infty ds_3\right]\notag\\
&={\cal F}(3A)+{\cal F}(3B)+{\cal F}(3C).
\label{mom2eq}
\end{align}
The colour factors in eq.~(\ref{B31-6}) give
\begin{align}
\{C(1a),C(2a)-C(2b)\}&=C(3D)-C(3F)\notag\\
\{C(1b),C(2g)-C(2f)\}&=C(3A)+C(3C)-C(3D)-C(3F).
\label{colfaceq}
\end{align}
Putting things together, one has
\begin{align}
\frac{1}{2}\{B_1,B_2\}&=\frac{1}{2}\Big(C(3D)-C(3F)\Big)
\Big({\cal F}(3A)-{\cal F}(3C)+{\cal F}(3D)-{\cal F}(3F)\Big)\notag\\
&\quad+\frac{1}{2}\Big(C(3A)+C(3C)-C(3D)-C(3F)\Big)\Big({\cal F}(3A)+{\cal F}(3B)+{\cal F}(3C)\Big).
\label{B1B21-6}
\end{align}
Finally, combining eqs.~(\ref{B3def}, \ref{A1-6}, \ref{oneloopcuberes}, \ref{B1B21-6}), one finds that $B_3$ is given by exactly the same expression as eq.~(\ref{1-6rep}), thus agreeing with what has already been found using the replica trick.

\subsubsection*{Discussion}

We have now seen for two non-trivial examples that the replica trick produces the same results as can be obtained from the explicit exponentiation of lower order diagrams, as it must do. We see that in general, the exponent contains distinct sets of diagrams, related by permutations of the gluons on each eikonal line. That such diagrams mix is an expression of the fact that they are colour linked. We may then introduce a matrix notation, and write the results of eqs.~(\ref{17-20rep}) and~(\ref{1-6rep}) as
\begin{equation}
W_{(1,2,2,1)}=\left(\begin{array}{c}{\cal F}(3a)\\{\cal F}(3b)\\{\cal F}(3c)\\{\cal F}(3d)\end{array}\right)^T\frac{1}{6}\left(\begin{array}{rrrr}1&-1&-1&1\\-2&2&2&-2\\-2&2&2&-2\\1&-1&-1&1\end{array}\right)\left(\begin{array}{c}C(3a)\\C(3b)\\C(3c)\\C(3d)\end{array}\right)
\label{17-20mat}
\end{equation}
and
\begin{equation}
W_{(2,3,1)}=\left(\begin{array}{c}{\cal F}(3A)\\{\cal F}(3B)\\{\cal F}(3C)\\{\cal F}(3D)\\{\cal F}(3E)\\{\cal F}(3F)\end{array}\right)^T\frac{1}{6}\left(\begin{array}{rrrrrr}3&0&-3&-2&-2&4\\-3&6&-3&1&-2&1\\-3&0&3&4&-2&-2\\0&0&0&1&-2&1\\0&0&0&-2&4&-2\\0&0&0&1&-2&1\end{array}\right)\left(\begin{array}{c}C(3A)\\C(3B)\\C(3C)\\C(3D)\\C(3E)\\C(3F)\end{array}\right)
\label{1-6mat}
\end{equation}
respectively. In each case a square matrix appears which compactly encodes the mixing of each set of diagrams. 
The general form of such a set is thus
\begin{equation}
W_{(n_1,n_2,\ldots,n_L)}\,\equiv\, \sum_{D}{\cal F}(D)\,\widetilde{C}(D)=\sum_{D,D'}{\cal F}(D) \,R_{DD'}\,
C(D'),
\label{setmix}
\end{equation}
where $D$ and $D'$ label diagrams in a given set, which are mutually related by permutation of the gluon attachments to any of the eikonal lines. The set itself is characterized by the number of attachments to each leg $(n_1,n_2,\ldots,n_L)$, as well as by the way subsets of these gluons are connected away from the eikonal lines. 
As usual, $C(D)$ and ${\cal F}(D)$ denote colour and kinematic parts.

The mixing matrix $R_{DD'}$ encapsulates the explicit results for the exponentiated colour factors in terms of the conventional colour factors. Note that the form of eq.~(\ref{setmix}) also applies in the trivial case in which a set contains only one diagram. In such a case, the mixing matrix is $1\times1$. The study of web structure then amounts to investigating the properties of the mixing matrices $R_{DD'}$. Having introduced these concepts, we consider more examples in appendix~\ref{sec:more-3loop-ECF}. In the next section, we study general properties of the mixing matrices in more detail, starting from a four loop example.

\subsection{Webs in multi-parton scattering: four loops and general discussion\label{sec:fourloop}}

\begin{figure}[htb]
\begin{center}
\scalebox{.97}{\includegraphics{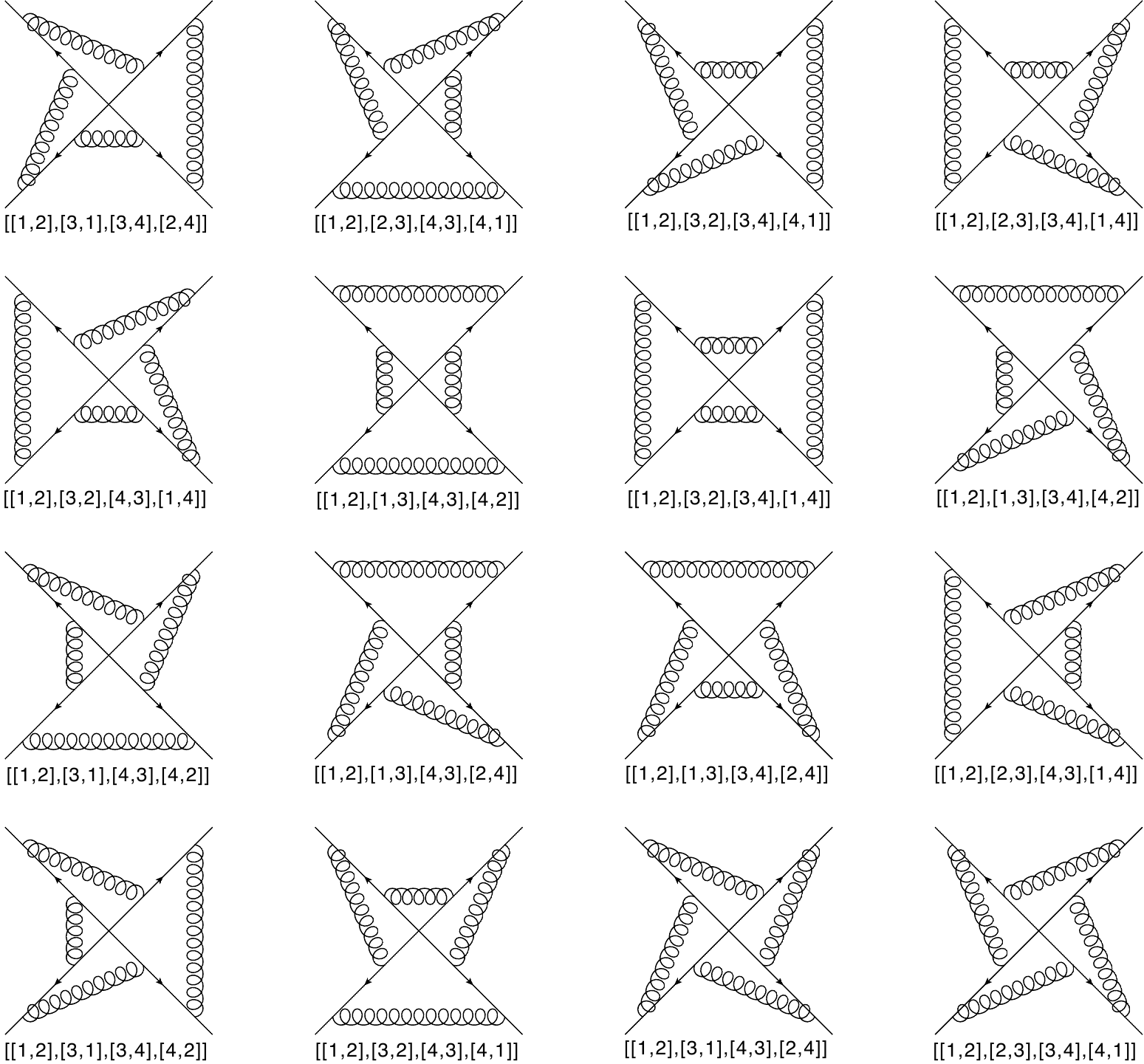}}
\caption{The four loop diagrams which give rise to eq.~(\ref{4loop_4lines}).}
\label{fourloopdiags}
\end{center}
\end{figure}

In this section, we consider the diagrams shown in figure~\ref{fourloopdiags}, consisting of four parton lines connected by four soft gluons. This set of 16 diagrams form a closed set {\it i.e.} mix only with each other in the exponent. We label these diagrams using the list notation introduced in section~\ref{sec:replica}. The mixing $R_{DD'}\, C(D')$ for these diagrams is given by

{\footnotesize
\begin{align}
%\begin{split}
\!\!\frac{1}{24}\left( \begin {array}{rrrrrrrrrrrrrrrr}
6&-6&2&2&-2&4&-4&2&-2&-2&-4&4
&-4&4&0&0\\ \noalign{\medskip}-6&6&-2&-2&2&-4&4&-2&2&2&4&-4&4&-4&0&0
\\ \noalign{\medskip}2&-2&6&-2&2&4&-4&-2&2&-6&4&4&-4&-4&0&0
\\ \noalign{\medskip}2&-2&-2&6&2&4&-4&-2&-6&2&-4&-4&4&4&0&0
\\ \noalign{\medskip}-2&2&2&2&6&4&-4&-6&-2&-2&4&-4&4&-4&0&0
\\ \noalign{\medskip}2&-2&2&2&2&4&-4&-2&-2&-2&0&0&0&0&0&0
\\ \noalign{\medskip}-2&2&-2&-2&-2&-4&4&2&2&2&0&0&0&0&0&0
\\ \noalign{\medskip}2&-2&-2&-2&-6&-4&4&6&2&2&-4&4&-4&4&0&0
\\ \noalign{\medskip}-2&2&2&-6&-2&-4&4&2&6&-2&4&4&-4&-4&0&0
\\ \noalign{\medskip}-2&2&-6&2&-2&-4&4&2&-2&6&-4&-4&4&4&0&0
\\ \noalign{\medskip}-2&2&2&-2&2&0&0&-2&2&-2&4&0&0&-4&0&0
\\ \noalign{\medskip}2&-2&2&-2&-2&0&0&2&2&-2&0&4&-4&0&0&0
\\ \noalign{\medskip}-2&2&-2&2&2&0&0&-2&-2&2&0&-4&4&0&0&0
\\ \noalign{\medskip}2&-2&-2&2&-2&0&0&2&-2&2&-4&0&0&4&0&0
\\ \noalign{\medskip}-18&-6&-6&-6&-18&12&12&-6&-18&-18&12&12&12&12&24&0
\\ \noalign{\medskip}-6&-18&-18&-18&-6&12&12&-18&-6&-6&12&12&12&12&0&
24\end {array} \right)
 \left( \begin {array}{c} C[[1,2],[3,1],[3,4],[2,4]]
\\ \noalign{\medskip}C[[1,2],[2,3],[4,3],[4,1]]\\ \noalign{\medskip}C[[1
,2],[3,2],[3,4],[4,1]]\\ \noalign{\medskip}C[[1,2],[2,3],[3,4],[1,4]]
\\ \noalign{\medskip}C[[1,2],[3,2],[4,3],[1,4]]\\ \noalign{\medskip}C[[1
,2],[1,3],[4,3],[4,2]]\\ \noalign{\medskip}C[[1,2],[3,2],[3,4],[1,4]]
\\ \noalign{\medskip}C[[1,2],[1,3],[3,4],[4,2]]\\ \noalign{\medskip}C[[1
,2],[3,1],[4,3],[4,2]]\\ \noalign{\medskip}C[[1,2],[1,3],[4,3],[2,4]]
\\ \noalign{\medskip}C[[1,2],[1,3],[3,4],[2,4]]\\ \noalign{\medskip}C[[1
,2],[2,3],[4,3],[1,4]]\\ \noalign{\medskip}C[[1,2],[3,1],[3,4],[4,2]]
\\ \noalign{\medskip}C[[1,2],[3,2],[4,3],[4,1]]\\ \noalign{\medskip}C[[1
,2],[3,1],[4,3],[2,4]]\\ \noalign{\medskip}C[[1,2],[2,3],[3,4],[4,1]]
\end {array} \right) \notag \\ 
\label{4loop_4lines}
%\end{split}
\end{align}}

\noindent
where we have included the vector of colour factors $C(D')$ in order to define the ordering of the rows of the matrix.

As in the three loop cases, there is a non-trivial structure to this matrix. We note here some interesting properties. There are two columns of the mixing matrix of eq.~(\ref{4loop_4lines}) which contain only a single non-zero entry: these are the last two columns. The meaning of this becomes clear on examining the last two diagrams in figure~\ref{fourloopdiags}. We refer to these as staircase diagrams, due to the visual similarity with the well-known Escher staircase. The last two columns of the matrix then tell us that the colour factors of the staircase diagrams do not enter the exponentiated colour factors of the non-staircase diagrams. It is easy to understand how this property arises from the replica trick. The replica ordering operator ${\cal R}$ can either leave gluon structures intact (in cases where all gluons have the same replica number, or are already replica ordered), or disentangle crossed gluons. It can never cross gluons that were previously uncrossed, as this results in an inconsistency between the replica orderings on different lines. In the present case, one may verify that the staircase diagrams are reordered to non-staircase diagrams if more than one replica is present. Furthermore, it is impossible to reorder a non-staircase diagram to make a staircase diagram.

The special feature of those staircase diagrams, which do not enter the exponentiated colour factors of other diagrams in the set, is that they have no ultraviolet subdivergences. To see this note that in these diagrams the four gluons are maximally linked: one cannot shrink any one gluon into the centre of the diagram (the hard vertex) without also pulling in all other gluons. This means that these diagrams have a single overall ultraviolet divergence, with no subdivergences\footnote{A systematic way to examine presence of subdivergences is presented in section \ref{sec:subdivergences}.}. 
It is easy to verify from figure~\ref{fourloopdiags} that this is not the case for any non-staircase diagram; these all have subdivergences. It follows that the staircase diagrams cannot be produced by the exponentiation of lower order diagrams. The ECF of any graph consists of the normal colour factor, plus terms which effectively subtract off those parts which arise from the exponentiation of lower order graphs. Thus, the colour factors for the staircase diagrams cannot enter the exponentiated colour factors of the non-staircase diagrams.

Note that there is also a column containing a single non-zero entry (unit entry) in the mixing matrix of eq.~(\ref{1-6mat}). In this case the diagram corresponding to this column is the graph (3B) of figure~\ref{3lsix}, and one again sees that this is maximally irreducible, in the sense that it has no ultraviolet subdivergences (it is not possible to shrink any gluon to the centre of the diagram independently of the others). Further examples of diagrams with no ultraviolet subdivergences are provided in Appendix \ref{sec:more-3loop-ECF}. These are the third diagram [[1],[2,1,2],[2],[]] in figure \ref{appendixfig2}, the first and the last diagrams in each of the figures \ref{appendixfig3} and \ref{appendixfig4} and all the diagrams in figure \ref{appendixfig7} but the second one. In each case there is a corresponding column containing a single non-zero entry in the mixing matrix.

Another remarkable property of the mixing matrices we computed is that they all have \emph{zero-sum rows}, that is the entries in any row of these matrices sum to zero\footnote{Note that this is not true for the trivial case in which $R$ is a $1\times1$ matrix. As discussed in section~\ref{sec:replica}, such diagrams have $\widetilde{C}(D)=C(D)$, and thus $R=1$.}.
In the notation of eq.~(\ref{setmix}) one has 
\begin{equation}
\sum_{D'}R_{DD'}=0\,, \qquad\quad \forall D.
\label{Rprop}
\end{equation}
This implies a symmetry of eq.~(\ref{setmix}), 
\begin{equation}
\widetilde{C}(D)\equiv\sum_{D'}R_{DD'}\,  C(D')\,=\,\sum_{D'}R_{DD'} \, \Big(C(D')+K\Big),
\end{equation}
namely that the exponent of the scattering amplitude is invariant under transformations of the colour factors
\begin{equation}
C(D')\rightarrow C(D')+K
\label{coltrans}
\end{equation}
for a given set, where $K$ does not depend on which member ($D'$) of the set one considers. Another way of saying this is that any contribution to each colour factor which is independent of the ordering of the gluon attachments on each parton line does not contribute to the exponent. This is the multiloop generalisation of a property already noted at two loops -- see eq.~(\ref{twoemcol7}) -- which we may illustrate as follows. Considering a line $i$ with two gluon attachments, these give a contribution to the colour factor
\begin{equation}
{\mathrm T}_i^A{\mathrm T}^B_i=\frac{1}{2}\left(\{{\mathrm T}_i^A,{\mathrm T}_i^B\}+[{\mathrm T}_i^A,{\mathrm T}_i^B]\right),
\label{Tpair}
\end{equation}
where ${\mathrm T}_i^A$ is a colour generator on line $i$ with adjoint index $A$, and on the right-hand-side we have written the pair of colour matrices as a sum of symmetric and antisymmetric parts\footnote{One may view this as a generalisation of the ``maximally non-abelian'' nature of webs discussed in~\cite{Gatheral:1983cz}, due to the fact that the abelian part of the colour factor does not care about the ordering of gluon emissions. }. The symmetric part does not contribute to the exponent, which we view here as a consequence of the symmetry of the exponent under transformations of the form of eq.~(\ref{coltrans}). That is, the symmetric part of the colour factor does not depend on the ordering of the gluon attachments. At higher loop orders, a string of three or more colour matrices may occur on any given line, and the above noted property of the mixing matrices again corresponds to the fact that the symmetric part of the colour factor does not contribute to the exponent of the scattering amplitude. 

In fact, there is more to say about the properties of the mixing matrix in terms of the cancellation of subdivergences in the exponent. This is the subject of the following section.

\section{Cancellation of subdivergences\label{sec:subdivergences}}

In the previous section we have seen that the exponent of the eikonal scattering amplitude is a sum of webs, where each web~(\ref{setmix}) represents a closed set of diagrams related by permuting gluon attachments on the eikonal lines. 
The mixing matrices of each web have zero-sum rows~(\ref{Rprop}), which one may understand as a consequence of the vanishing of the symmetric parts of the exponentiated colour factors. 
In this section, we explore these matrices from a different point of view, namely in terms of the cancellation of subdivergences in the exponent. 

As explained in section \ref{sec:exponentiation}, 
after performing running-coupling renormalization and absorbing any collinear divergences into jet factors, all remaining singularities in the exponent of an eikonal scattering amplitude must be associated  with the renormalization of the multi-eikonal vertex. The latter involves, at each order in the coupling, a single counter-term, which removes only simple poles, $\epsilon^{-1}$, in dimensional regularisation. 
This picture should be consistent with the fact that the kinematic factors ${\cal F}(D)$ corresponding to individual diagrams $D$ which appear in the exponent do contain, in general, higher-order poles, up to ${\cal O}(1/\epsilon^n)$ at $n$-loops. 
Thus, there must be a cancellation of all higher-order poles in the exponent. In the following section we perform an exploratory investigation of the mechanisms by which this cancellation is realised.

\subsection{Renormalization of the multi-eikonal vertex}
\label{renormme}

The general structure of singularities of a multi-leg eikonal amplitude is determined by the fact that it renormalizes multiplicatively. It would be very interesting to understand how this structure emerges in term of webs. 

As before, we consider a multi-leg eikonal amplitude, corresponding to the soft matrix 
 ${\cal S}(\epsilon)$ in (\ref{M_SH_fact}). 
To simplify the notation we now drop the colour flow index notation as well as the first two arguments of this function (the cusp angles and $\alpha_s$) but keep $\epsilon$, to stress the fact that this function is singular.
To be concrete we assume that the eikonal lines are all off the lightcone (in the case of lightline eikonal lines there are extra collinear singularities that need to be absorbed into colour-singlet jet functions before the following considerations can be applied).
We apply dimensional regularization and assume, in addition, that there is an infrared cutoff so we can systematically work with $D<4$ (or $\epsilon>0$).
We emphasize that in ${\cal S}(\epsilon)$ running-coupling renormalization was already applied, and thus all remaining singularities are related to the renormalization of the multi-leg eikonal vertex. 

Let us now examine the renormalization of the eikonal vertex along the lines of section~3 in \cite{MSS}. We introduce a renormalization factor $Z$ for the multi-leg eikonal vertex, which is a matrix in colour flow space, such that 
\begin{equation}
\label{S_ren}
{\cal S}_{\rm ren}(\mu)= {\cal S}(\epsilon) \,Z^{-1} (\epsilon,\mu)\,,
\end{equation}
where $Z^{-1}$ absorbs all the singularities of ${\cal S}(\epsilon)$, making the renormalized eikonal amplitude ${\cal S}_{\rm ren}(\mu)$ \emph{finite}, and at the same time $\mu$-dependent. The ordering of the factors on the right-hand side of eq.~(\ref{S_ren}) is important because of  their non-commuting nature. Since $Z^{-1}$ involves only a single counter-term at each order in the coupling, this equation constrains the singularity structure ${\cal S}(\epsilon)$ may have, thus effectively constraining the corresponding webs. 

Let us see how these constraints arise.
Both the eikonal amplitude ${\cal S}(\epsilon)$ and the $Z$ factor exponentiate:
\begin{equation}
{\cal S}(\epsilon)=\exp \left\{ \sum_n\alpha_s^nw^{(n)} (\epsilon) \right\} \qquad Z^{-1} (\epsilon,\mu) =\exp \left\{ 
\sum_n\alpha_s^n\zeta^{(n)}(\epsilon,\mu)  \right\}
\end{equation}
Here we have adopted the notation of~\cite{MSS}, in which $w^{(n)}(\epsilon)$ represents the sum of all
diagrams occurring in the exponent at ${\cal O}(\alpha_s^n)$\footnote{In~\cite{MSS}, $w^{(n)}$
is defined so as to implicitly include the factor $\alpha_s^n$. Here we make this explicit.}. 
Also, $\zeta^{(n)}$ is a counter-term, consisting of pure single pole terms (in the
minimal subtraction scheme) which removes, at each order, any remaining ${\cal O}(\epsilon^{-1})$ singularity
in $w^{(n)}$. As argued in \cite{MSS}, owing to the non-trivial matrix structure of the eikonal amplitude and 
 the $Z$ factor, the exponent of ${\cal S}_{\rm ren}(\mu)$ in (\ref{S_ren}) is not simply the sum of 
the exponents of ${\cal S}$ and $Z^{-1}$. Rather, defining
\begin{equation}
w=\sum_n\alpha_s^nw^{(n)},\quad \zeta=\sum_n\alpha_s^n\zeta^{(n)},
\label{wzdef}
\end{equation}
this sum is corrected by an infinite series of commutators
according to the Baker-Campbell-Hausdorff formula:
\begin{align}
\label{S_ren_BCH}
\begin{split}
{\cal S}_{\rm ren}(\mu)&=  \exp \left\{ w (\epsilon) \right\} \, \exp \left\{ 
\zeta(\epsilon,\mu)  \right\} \\&= 
\exp\left\{
w+\zeta 
+\frac12 [w,\zeta] 
+\frac{1}{12}\Big([w,[w,\zeta]] - [\zeta,[w,\zeta]]\Big)
-\frac{1}{24}[\zeta,[w,[w,\zeta]]] +\cdots
\right\}\,.
\end{split}
\end{align}
where for better clarity we did not write the arguments of $w$ and $\zeta$ in the second line.

Let us first note that if the $[w,\zeta]$ commutators vanish, as happens for example in the two-eikonal line case, the counter-terms $\zeta$ at any given order in the coupling must simply be the negative of the ${\cal O}(1/\epsilon)$ term in $w$. This clearly implies that $w$ is strictly free of subdivergences: its singularities are just simple poles. This, of course, goes hand in hand with the absence of subdivergences in webs for the case of eikonal amplitudes with two lines; we return to this point in the following section.  

In the case of many eikonal lines, things are more complicated. Diagrams which contribute to the exponent will indeed contain subdivergences. Yet, the counter-terms $\zeta^{(n)}$ at any order $n$,
which must render the exponent of ${\cal S}_{\rm ren}(\mu)$ in (\ref{S_ren_BCH}) finite, are strictly~${\cal O}(\epsilon^{-1})$. How is this possible?

We identify two distinct mechanisms which combine to facilitate the necessary cancellation of singularities:
\begin{enumerate}
\item The mixing matrices $R_{DD'}$ in (\ref{setmix}) generate particular linear combinations of kinematic functions in which certain subdivergences cancel.
This cancellation occurs within sets of graphs $W_i^{(n)}$ (webs of order $n$, where the subscript $i$ refers to a specific web) which build up\footnote{It looks natural to assume that there are no further cancellations between different webs $W_i^{(n)}$ and $W_j^{(n)}$, although we have not constructed a proof this never occurs. } the contribution to the exponent $w(\epsilon)$ of the unrenormalized eikonal amplitude ${\cal S}(\epsilon)$ at any given order $n$:
\begin{equation}
\label{small_w_big_W}
w(\epsilon)=\sum_n w^{(n)} (\epsilon)\alpha_s^n=\sum_n\,\sum_{i} W_i^{(n)} 
=\sum_n\,\sum_{i} \sum_{D,D'}{\cal F}(D) \,R_{DD'}^{(n,i)}\,C(D')
\,.
\end{equation}   
\item The commutator terms in eq.~(\ref{S_ren_BCH}) involving \emph{lower-order webs} and \emph{lower-order counter-terms} generate multiple epsilon poles.
\end{enumerate}

We emphasize that despite considerable cancellation through the mixing matrices, which we shall elucidate in the following subsections, webs are not entirely free of subdivergences. 
Indeed, as we shall see, starting at three loops, at least some multiple poles do indeed survive in the exponent of the unrenormalized eikonal amplitude (\ref{small_w_big_W}). This is where the second mechanism becomes essential.
To appreciate its role, let us examine eq.~(\ref{S_ren_BCH}) at the
first few orders in $\alpha_s$. 
First one may define 
\begin{equation}
S_{\rm ren}(\mu)=\exp\left[\sum_n\alpha_s^nw^{(n)}_{\rm ren}\right],
\label{wrendef}
\end{equation}
where $w_{\rm ren}^{(n)}$, as the notation suggests, is the renormalised version of $w^{(n)}$ collecting
all diagrams at ${\cal O}(\alpha_s^n)$. Each term $w_{\rm ren}^{(n)}$ is thus finite as 
$\epsilon\rightarrow 0$. Equating the exponent of eq.~(\ref{wrendef}) with that of 
eq.~(\ref{S_ren_BCH}) at successive perturbative orders, one finds
\begin{align}
w_{\rm ren}^{(1)}&=w^{(1)}+\zeta^{(1)};\label{webren1}\\
w_{\rm ren}^{(2)}&=w^{(2)}+\zeta^{(2)}+\frac{1}{2}\left[w^{(1)},\zeta^{(1)}\right];\label{webren2}\\
w_{\rm ren}^{(3)}&=w^{(3)}+\zeta^{(3)}+\frac{1}{2}\left(\left[w^{(1)},\zeta^{(2)}\right]
+\left[w^{(2)},\zeta^{(1)}\right]\right)+\frac{1}{12}\left[w^{(1)}-\zeta^{(1)},
\left[w^{(1)},\zeta^{(1)}\right]\right].\label{webren3}
\end{align}
One now sees explicitly that at first order, the counter-term is simply the negative of the single
pole in $w^{(1)}$. Evaluating eq.~(\ref{webren1}) at ${\cal O}(\epsilon^{-1})$ (using the fact that $w_{\rm ren}^{(1)}$
is manifestly finite as $\epsilon\rightarrow0$), one finds
\begin{equation}
\zeta^{(1)}=-w^{(1)}|_{\epsilon^{-1}}.
\label{zeta1}
\end{equation}

Things become less trivial already at two-loop order. Individual two-loop diagrams do indeed have subdivergences. A priori these could either cancel out in webs owing to the mixing matrices,  or  survive in webs, generating ${\cal O}(\epsilon^{-2})$ contributions to $w^{(2)}$ in eq.~(\ref{webren2}).
In fact, the latter possibility can be excluded since $\zeta^{(2)}={\cal O}(\epsilon^{-1})$ and the commutator term in (\ref{webren2}) cannot contribute at ${\cal O}(\epsilon^{-2})$: such a contribution could only come from the ${\cal O}(\epsilon^{-1})$ in both $\zeta^{(1)}$ and 
$w^{(1)}$, but these are proportional to each other owing to (\ref{zeta1}), and therefore commute.
There will, however, be a contribution from this commutator at ${\cal O}(\epsilon^{-1})$, coming from the ${\cal O}(\epsilon^0)$ part of $w^{(1)}$~\cite{MSS}. Consequently, the two-loop counter term $\zeta^{(2)}$ is \emph{not} the negative of the single pole of $w^{(2)}$, but rather,
\begin{equation}
\zeta^{(2)}=-w^{(2)}|_{\epsilon^{-1}}-
\frac{1}{2}\left[w^{(1)}|_{{\cal O}(\epsilon^0)},\zeta^{(1)}\right]\,.
\end{equation}

We saw that to conform with renormalization, $w^{(2)}$ must only have a single pole, despite the fact that some of the contributing diagrams have double poles owing to subdivergences. We therefore conclude that at two loops subdivergences must cancel entirely due to the mixing matrices $R_{DD'}$. 
An example is provided by the diagrams (2a) and (2b) of figure~\ref{2a-d}, which form a two-loop web $W_{(1,2,1)}$. 
The contribution of these diagrams to the exponent of the unrenormalised soft function is
\begin{equation}
W_{(1,2,1)}={\cal F}^T R C= 
\left(\begin{array}{c}{\cal F}(2a)\\{\cal F}(2b)\end{array}\right)^T
\frac{1}{2}\,\left(\begin{array}{rr}1&-1\\-1&1\end{array}\right)
\left(\begin{array}{c}C(2a)\\C(2b)\end{array}\right),
\label{colfacs2a2b}
\end{equation}
where we have expressed the results of section~\ref{W3l2l} in terms of the appropriate mixing
matrix. Multiplying the vector of kinematic factors by the mixing matrix, this in turn can be rewritten as
\begin{equation}
W_{(1,2,1)}=\frac{1}{2}\,\left(\begin{array}{c}{\cal F}(2a)-{\cal F}(2b)\\
{\cal F}(2b)-{\cal F}(2a)\end{array}\right)^T
\left(\begin{array}{c}C(2a)\\C(2b)\end{array}\right)
=\frac12 \,\left[{\cal F}(2a)-{\cal F}(2b)\right]\left[C(2a)-C(2b)\right]
\,.
\label{colfacs2a2b2}
\end{equation}
Both the diagrams
(2a) and (2b) contain ultraviolet subdivergences, as can be easily seen from the fact that one may shrink the innermost gluon to the origin independently of the outermost gluon. Thus, ${\cal F}(2a)\sim
{\cal F}(2b)\sim\epsilon^{-2}$. In fact, an explicit calculation shows that the $\epsilon^{-2}$ parts
of ${\cal F}(2a)$ and ${\cal F}(2b)$ are equal~\cite{Aybat:2006mzy,Mitov:2010xw}. One then finds immediately from the combination of kinematic factors appearing in eq.~(\ref{colfacs2a2b2}) that subdivergences cancel due to the mixing matrix.

It is important to remember that we derived the mixing matrices by considering how
exponentiated colour factors are related to the conventional colour factors of sets of graphs.
We now see that this same mixing matrix is responsible for conforming with renormalization through the cancellation of subdivergences, implying a relationship between the colour structure of graphs, and their kinematic information.
The two-loop example we analysed is a special case in that \emph{full} cancellation of subdivergences is achieved by the mixing matrix. It is clear that this cancellation occurs within each closed set of diagrams. 

At higher orders, the mixing matrix will only partially cancel subdivergences in general, 
with remaining higher order $\epsilon$ poles being mopped up by the  
commutator terms of eq.~(\ref{S_ren_BCH}). Because these are built out of lower order counter-term and webs, multiple pole terms in $w^{(n)}(\epsilon)$ remain highly constrained: according to eq.~(\ref{S_ren_BCH}) they can be \emph{fully determined} by lower order webs. 
 
We note in particular, that the cancellation of the leading subdivergence in webs, at any given order $n$, is \emph{complete}: $w^{(n)}(\epsilon)$ never contains an ${\cal O}(\epsilon^{-n})$ singularity.
We have already seen above that at two loop order subdivergences cancel within $w^{(2)}$
itself, so that $w^{(2)}$ is manifestly ${\cal O}(\epsilon^{-1})$. It follows that at ${\cal O}(\alpha_s^3)$ (i.e. in eq.~(\ref{webren3})), the only potential source of $\epsilon^{-3}$ poles is due to the nested commutator term
\begin{displaymath}
\left[w^{(1)}-\zeta^{(1)},\left[w^{(1)},\zeta^{(1)}\right]\right].
\end{displaymath}
This vanishes at ${\cal O}(\epsilon^{-3})$ however, from eq.~(\ref{zeta1}). It follows that
$w^{(3)}$ must have no $\epsilon^{-3}$ singularities, and thus that the cancellation of the leading
subdivergence relies solely on the mixing matrices $R_{DD'}$ at this order. It is straightforward to iterate this argument to higher orders. At ${\cal O}(\alpha_s^n)$ the leading subdivergence is ${\cal O}(\epsilon^{-n})$. Due to the cancellation of the leading subdivergence in $w^{(j)}$ for any $j\leq n-1$, the only potential source of $\epsilon^{-n}$ poles in the commutator terms is via maximally nested commutators of the form
\begin{displaymath}
\left[w^{(1)},\left[w^{(1)},\ldots\left[w^{(1)},\zeta^{(1)}\right]\ldots\right]\right]
\end{displaymath}
(plus similar terms in which any of the factors $w^{(1)}$ is replaced by $\zeta^{(1)}$).
However, these all vanish from eq.~(\ref{zeta1}).
We have thus proven that at any order $n$,
\begin{equation}
\label{order_epsilon_cancellation}
w^{(n)}|_{{\cal O}(\epsilon^{-n})} =0\,,
\end{equation}
so that the cancellation of the leading subdivergence at any order is purely due to the mixing matrix, rather than the lower-order counter-terms.

Three loop order is the first order where multiple poles survive in webs. To see this consider the cancellation of the ${\cal O}(\epsilon^{-2})$ contribution on the r.h.s. of eq.~(\ref{webren3}).
We obtain, after some algebra,
\begin{equation}
\label{order_epsilon_minus_2_3loops}
w^{(3)}|_{{\cal O}(\epsilon^{-2})} =-\frac{1}{12}\left[\zeta^{(1)},\left[w^{(1)}|_{{\cal O}(\epsilon^{0})},\zeta^{(1)}\right]\right]
\end{equation}
which need not vanish, in general. It is interesting to note, however, that this three-loop contribution can be deduced directly from the one-loop calculation.
This is of course not so for the three-loop counter-term, 
\begin{align}
\begin{split}
\zeta^{(3)}=-w^{(3)}|_{{\cal O}(\epsilon^{-1})} &-
\frac{1}{2}\left(\left[w^{(1)}|_{{\cal O}(\epsilon^{0})},\zeta^{(2)}\right]
+\left[w^{(2)}|_{{\cal O}(\epsilon^{0})},\zeta^{(1)}\right]\right)\\
&-\frac{1}{12}
\left[w^{(1)}|_{{\cal O}(\epsilon^{0})},
\left[w^{(1)}|_{{\cal O}(\epsilon^{0})},\zeta^{(1)}\right]\right]\,,
\end{split}
\end{align}
which requires both a two-loop and a three-loop calculation to be performed down to the ${{\cal O}(\epsilon^{0})}$ and ${{\cal O}(\epsilon^{-1})}$ contributions, respectively.

To summarise, one may identify two mechanisms for the cancellation of subdivergences.
There are cancellations encoded by the mixing matrices $R_{DD'}$, as well as ones due to nested
commutators of counter-terms and lower order webs. 
For the leading subdivergence, only the mixing matrix is relevant, rendering the highest singularity of ${\cal O}(\alpha_s^n)$ webs of order $\epsilon^{n-1}$.
The full interplay of these two cancellation mechanisms clearly deserves further study. For now, however, we focus on elucidating the role of the mixing matrices and their properties in facilitating the necessary cancellations within webs, and postpone full investigation of the counter-term structure to future work.

\subsection{Identifying subdivergences}

To make our discussion of subdivergences more precise, we begin by describing a systematic approach to characterising ultraviolet divergences in the relevant type of graphs.
Let us first reconsider the case of two eikonal lines. Then the absence of subdivergences in webs is easily proven~\cite{Laenen:2000ij}, using the general rule that the superficial degree of divergence $D$ of any (sub-)diagram is given by
\begin{equation}
D=4-\sum_{f}E_f(s_f+1)-\sum_iN_i\Delta_i,
\label{Ddef}
\end{equation}
where $E_f$ is the number of external lines of type $f$ (bosons or fermions), and $s_f=0,\frac12$ for bosons or fermions respectively. Also, $N_i$ is the number of vertices of type $i$, with
\begin{equation}
\Delta_i=4-d_i-\sum_f(s_f+1)n_{i,f},
\label{Deltadef}
\end{equation}
where $d_i$ is the number of derivatives in vertex $i$, and $n_{i,f}$ is the number of fields of type $f$ at that vertex. Note that eikonal lines count as fermions, thus have $s_f=\frac12$. 

The values of $\Delta_i$ for various different vertex types relevant to our analysis are collected in table~\ref{tab:Delta}.
\begin{table}[htb]
\begin{center}
\begin{tabular}{c|c}
Vertex& $\Delta_i$\\
\hline
$EEg$ & 0\\
$3g$ & 0\\
$4g$ & 0\\
$nE$ & $4-\frac{3}{2}\,n$
\end{tabular}
\end{center}
\caption{Vertex factors of eq.~(\ref{Deltadef}) for various cases. Here $E$ denotes a generic eikonal line, $3g$ and $4g$ are the three and four gluon vertices respectively, and $nE$ denotes the coupling of  $L$ eikonal lines by a hard interaction.}
\label{tab:Delta}
\end{table}
One sees that $\Delta_i=0$ for all vertices we will encounter except the coupling of the eikonal lines by a hard interaction. Denoting this simply by $\Delta$, one may simplify eq.~(\ref{Ddef}) to give
\begin{equation}
D=4-\sum_f E_f(s_f+1)-\Delta.
\label{Ddef2}
\end{equation}

\begin{figure}[htb]
\begin{center}
\scalebox{1.0}{\includegraphics{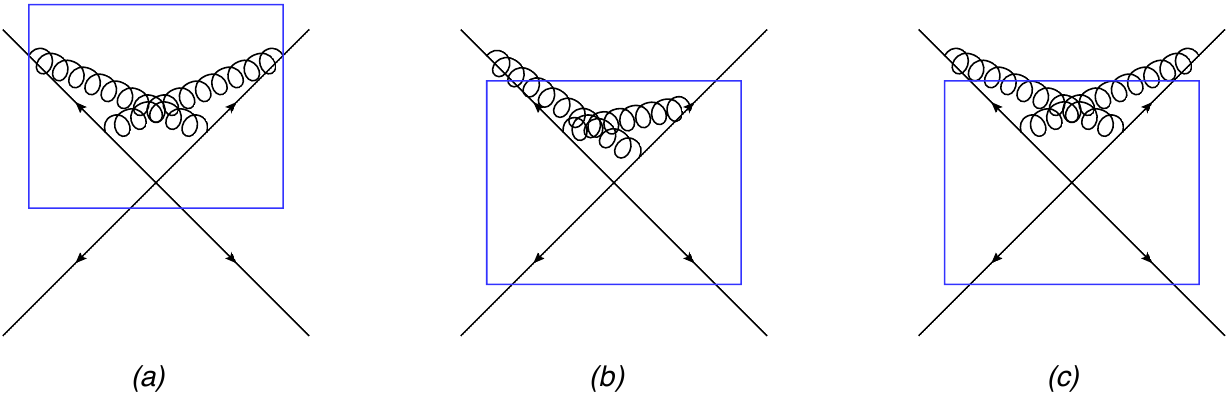}}
\caption{The formation of possible subdiagrams from a diagram in which two eikonal lines
are connected by soft gluon exchanges.}
\label{boxweb}
\end{center}
\end{figure}

Now consider the diagram of figure~\ref{boxweb}, which one may split into subdiagrams by drawing a box around parts of the diagram as shown\footnote{We consider here only boxes which include the multi-eikonal vertex.
Subdivergences in boxes which do not include this vertex are dealt with by renormalisation of the strong
coupling constant.}.
As can be seen from the figure, a subdiagram of the web may consist of the whole web, or subdiagrams in which one or more gluons become external. From eq.~(\ref{Ddef}), one finds that figure~\ref{boxweb}(a) has a superficial degree of divergence $D=0$, indicating the fact that an overall ultraviolet divergence is present. The two subgraphs of figures~\ref{boxweb}(b) and (c) have $D=-2$ and $D=-1$ respectively {\it i.e.} no subdivergences are present. The generalisation of this analysis to higher-order webs is straightforward. The trivial subgraph consisting of the whole web always has $D=0$, as one expects from the overall ultraviolet divergence. The fact that no subdivergences are present follows directly from two-eikonal line irreducibility, the defining characteristic of webs: in any given subgraph, one or more gluons will become external such that $D<0$. This is not the case for reducible graphs, as one may consider either of the distinct pieces as a subgraph by itself, which has $D=0$. 

Now let us return the $L$ parton case. For a given subdiagram with $L$ eikonal lines and $n_g$ external gluon lines, we get
\[
\sum_f E_f(s_f+1)=\frac32 L+n_g
\]
while for the hard interaction vertex we have $\Delta=4-\frac32 L$.  Thus (\ref{Ddef2}) yields:
\begin{equation}
D=-n_g\leq 0. 
\label{Dleq0}
\end{equation}
It is then clear that all diagrams in the exponent have $D=0$ when the whole diagram is considered, reflecting the fact that all diagrams have an overall ultraviolet divergence. Furthermore, one may again classify diagrams as reducible or irreducible, where the former have colour factors which may be decomposed as $C(G)=C(G_1)C(G_2)$ for some $G_1$, $G_2$. Irreducible diagrams have no subdivergences by a similar argument to the two-eikonal line case. That is, any box which selects a subgraph of an irreducible diagram will make at least one gluon external, such that the degree of divergence becomes negative by eq.~(\ref{Dleq0}).

\subsection{The mixing matrix as a projection operator}

Reducible diagrams will indeed have subdivergences. A full classification of their structure requires detailed investigation of their kinematic parts, and is beyond the scope of this paper. However, as we have
seen already in section~\ref{renormme}, cancellation of subdivergences involves an interplay
between commutators of counter-terms and lower order webs, and the mixing matrices $R_{DD'}$.
It is thus useful to examine some of the properties of the latter, which appear to be completely general\footnote{These properties have been observed in all examples considered throughout the paper (several sets of two, three and four-loop of diagrams). We will not prove that they apply for all mixing matrices at any order, although we expect they do. }. We then interpret them in the context of cancellation of subdivergences, and demonstrate this by examining a couple of examples. Finally we complete the picture by discussing the related combinatorics of colour factors subject to (\ref{Rprop}).

First, we observe that the mixing matrices $R$ are \emph{idempotent}, namely, they satisfy
\begin{equation}
\label{idempotence}
R^2=R \,,
\end{equation}
strongly suggesting that they are projection operators. This is indeed the case, as we explain below. 
The idempotence property of $R$ immediately implies that:
\begin{itemize}
\item{} $R$ is diagonalizable.
\item{} $R$ has only two distinct eigenvalues: $1$ and $0$, with degeneracies $r$ and $(d-r)$ respectively, where $d$ and $r$ denote the dimension and rank of $R$. 
\end{itemize}
One may then construct the diagonalising matrix $Y$ whose rows are the left-eigenvectors $v^{\mathrm T}$ of $R$ (or, equivalently, $v$ are the right-eigenvectors of $R^{\mathrm{T}}$). That is:
\begin{equation}
\label{diagonalization}
v^{\mathrm T} R  = v^{\mathrm T} \lambda_n\,\,; \qquad \quad Y\,R\,Y^{-1}={\rm diag}(\lambda_1,\lambda_2,\ldots,\lambda_d)   \,,
\end{equation}
where $\lambda_n=1$ for $n=1,2,\ldots, r$, and $\lambda_n=0$ for $n=(r+1),\ldots, d$.

Let us consider now the vector of kinematic functions ${\cal F}(D)$ corresponding to the set of diagrams $\left\{D\right\}$ in a given web. We may denote this vector by ${\cal F}$, and express it via the linear combinations formed by the diagonalising matrix,
${\cal F}^{\mathrm T}= \Big({\cal F}^{\mathrm T}\,Y^{-1}\Big)\, Y$, in such a way that the web in (\ref{setmix}) can be written as:
\begin{align}
\label{setmix_matrix}
\begin{split}
W_{(n_1,n_2,\ldots,n_L)}\,=\, {\cal F}^{\mathrm T}\,\widetilde{C} 
=&\,\,{\cal F}^{\mathrm T} \,R\, C\\
=&\,\,\Big({\cal F}^{\mathrm T} Y^{-1}\Big) \, Y \,R\, Y^{-1}\, \Big( Y \,C\Big)\\
=\,\,&\Big({\cal F}^{\mathrm T} Y^{-1}\Big) \, {\rm diag}(\lambda_1,\lambda_2,\ldots,\lambda_d)  \,\Big( Y \,C\Big)\\
=\,\,&\sum_{H=1}^{r}\Big({\cal F}^{\mathrm T} Y^{-1}\Big)_H \, \,\Big( Y \,C\Big)_H\,.
\end{split}
\end{align}
From the third line it is clear that owing to the presence of zero eigenvalues, certain entries in the vector ${\cal F}^{\mathrm T}\,Y^{-1}$, containing particular linear combinations of the kinematic functions ${\cal F}(D)$, do not enter the expression for $W$, {\it i.e.} do not contribute to the exponent.
As explicitly stated in the last line, only the first $r$ components of ${\cal F}^{\mathrm T}\,Y^{-1}$, corresponding to unit eigenvalues, enter the exponent.

It should be noted that owing to the degeneracy of the eigenvalues there is some freedom in choosing the particular basis of eigenvectors: any linear combination of eigenvectors with eigenvalue $1$ ($0$) is also an eigenvector with eigenvalue $1$\,($0$). 

Before proceeding it is worthwhile clarifying the precise sense in which $R$ is a projection operator. We wish to explain the physical interpretation of the eigenvalues $1$ and $0$, and the fact that these are the only ones that appear. To this end it is useful to rewrite the expression in the last line in (\ref{setmix_matrix}) as follows
\begin{align}
\label{setmix_matrix_alt}
\begin{split}
W_{(n_1,n_2,\ldots,n_L)}\,%=\,\,&\sum_{D=1}^{d}\Big({\cal F}^{\mathrm T} Y^{-1}\Big)_D \, \,\Big( Y \,C\Big)_D\,\theta(D\leq r)
%\\
=\,\,&\sum_{D=1}^{d}\Big({\cal F}^{\mathrm T} Y^{-1}\Big)_D \, \,\Big( Y \,C\Big)_D\,\Big[1-\theta(D> r)\Big]
\\
=\,\,&{\cal F}^{\mathrm T}  \,C\,\,-\,
\,\sum_{D=r+1}^{d}\Big({\cal F}^{\mathrm T} Y^{-1}\Big)_D \, \,\Big( Y \,C\Big)_D\,\,,
\end{split}
\end{align}
where we used the fact that upon summing over all $D=1,2,\ldots,d$ one recovers the unit matrix $\sum_{D=1}^{d} Y^{-1}_{GD}\,Y_{DH} =\delta_{GH}$.
The first term in the last line in (\ref{setmix_matrix_alt}) is easily recognised as the full non-exponentiated result for this class of diagrams. It becomes clear then that what is being subtracted from it, {\it i.e.} the sum over the subspace corresponding to the zero eigenvalue of $R$, are all the contributions that are generated by the exponentiation of lower-order graphs. This is analogous to how eq.~(\ref{ABrel}) was used in the previous section to extract the exponentiated contributions $B_n$ at any given order $n$, by subtracting the result of exponentiating the lower orders from $A_n$.
The conclusion is then that the zero eigenvalue is associated with contributions that are generated by the exponentiation of lower orders, while the unit eigenvalue is associated with new contributions 
to the exponent. This is how $R$ projects out the subspace of contributions which appear in the exponent.

\subsection{Cancellation of subdivergences: three-loop examples}

In the previous section we have seen that the mixing matrices $R$ that define webs in multi-leg eikonal amplitudes are very special in that they are idempotent and act as projection operators on the vector of kinematic functions (or, equivalently, on the vector of colour factors) corresponding to the different diagrams in the web. In this section, we make the connection between this property and the cancellation of subdivergences. To this end we consider again several sets of diagrams, which are sufficiently complex to feature these cancellations, and yet simple enough to facilitate explicit calculations.

First of all, we wish to illustrate how the leading subdivergence 
i.e. ${\cal O}(\epsilon^{-n})$ at ${\cal O}(\alpha_s^n)$ cancels out within webs, 
a property we deduced based on the renormalization structure of the eikonal amplitudes in
section~\ref{renormme} (see eq.~(\ref{order_epsilon_cancellation}) there). 
More generally, we expect that the particular linear combinations of kinematic functions, the ${\cal F}^{\mathrm T}\,Y^{-1}$ entries\footnote{In contrast, no cancellations are required in the $(d-r)$ entries in ${\cal F}^{\mathrm T}\,Y^{-1}$ which correspond to the eigenvalue $0$, as these do not contribute to the exponent.} corresponding to the eigenvalue~$1$, would have a singularity structure that is consistent with (\ref{S_ren_BCH}), where all multiple poles are given by commutators between counter-terms and lower order webs. 

The first indication that the linear combinations of kinematic functions corresponding to the unit eigenvalue work to remove subdivergences is that diagrams which are by themselves subdivergence-free reside in this subspace.
In other words considering a set of diagrams containing a particular diagram $D_0$ that has no subdivergences, we note that a basis of eigenvectors with eigenvalue $1$ may be chosen such that 
one of the first~$r$ columns of $Y^{-1}$ would have a single non-zero entry in the row corresponding to the subdivergence-free diagram, so as to pick the corresponding kinematic function ${\cal F}(D_0)$ alone:
\begin{equation}
({\cal F}^{\mathrm T}\,Y^{-1})_D ={\cal F}(D_0) \qquad\qquad \exists D\,|\,D\leq r\,.
\end{equation}
For example considering figure~\ref{fourloopdiags} with the mixing matrix $R_{DD'}$ of eq.~(\ref{4loop_4lines}), where the eigenvalue 1 appears with degeneracy five, we find that two of the five linear combinations $({\cal F}^{\mathrm T}\,Y^{-1})_D$ become trivial, and correspond to selecting one of the subdivergence-free staircase diagrams, namely ${\cal F}([[1,2],[3,1],[4,3],[2,4]])$ or ${\cal F}([[1,2],[2,3],[3,4],[4,1]])$, respectively.
We conclude that in general some of the entries in $({\cal F}^{\mathrm T}\,Y^{-1})_D$ with $D\leq r$ can be immediately recognised to be subdivergence-free, since they correspond to picking individual subdivergence-free diagrams, whereas in other entries genuine cancellations must take place. 

To observe these cancellations let us reconsider first the example of the web $W_{(1,2,2,1)}$ 
corresponding to the diagrams of figure~\ref{17-20}. 
These are all reducible, and contain subdivergences. By considering sequential shrinking of the gluons towards the hard vertex, or alternatively by applying eq.~(\ref{Dleq0}) to subdiagrams, one finds that
each and every one of these four diagrams has two subdivergences in addition to the overall divergence, making them ${\cal O}(\epsilon^{-3})$ in total. According to
section~\ref{renormme} this maximally-divergent contribution must cancel out entirely within the web.

The mixing matrix in this case is given by eq.~(\ref{17-20mat}), and its eigenvalues are 
\begin{equation}
\{\lambda\}_{(1,2,2,1)}=\{1,0,0,0\},
\label{ev1221}
\end{equation}
so here the dimension and the rank are, respectively, $d=4$ and $r=1$. Determining the eigenvectors we get the diagonalising matrix, which forms the following representation of the vector of kinematic factors ${\cal F}(D)$ for this set of diagrams:
\begin{align}
{\cal F}(D)&=\sum_{G,E}{\cal F}(G)\,Y^{-1}_{GE}Y_{ED}\notag\\
&=\frac{1}{6}\left(\begin{array}{c}{\cal F}(3a)-2{\cal F}(3b)-2{\cal F}(3c)+{\cal F}(3d)\\-{\cal F}(3a)+2{\cal F}(3b)+2{\cal F}(3c)+5{\cal F}(3d)\\{\cal F}(3a)-2{\cal F}(3b)+4{\cal F}(3c)+{\cal F}(3d)\\{\cal F}(3a)+4{\cal F}(3b)-2{\cal F}(3c)+{\cal F}(3d)\end{array}\right)^{\text{T}}\left(\begin{array}{rrrr}1&-1&-1&1\\-1&0&0&1\\2&0&1&0\\2&1&0&0\end{array}\right).
\label{17-20mat2}
\end{align}
Here we have explicitly written the matrix of left-eigenvectors $Y$ in the second line, and absorbed the inverse matrix $Y^{-1}$ into the vector of kinematic parts. Note that the first row of the eigenvector matrix corresponds to the eigenvalue $\lambda=1$; all other rows have $\lambda=0$. As we have seen in eq.~(\ref{setmix_matrix}), this tells us that, after acting with $R_{DD'}$ on the right-hand side, only the first kinematic combination in eq.~(\ref{17-20mat2}) contributes to the exponent of the scattering amplitude. The other combinations do not contribute to $W_{(1,2,2,1)}$, because they are associated with zero eigenvalues of the mixing matrix. 
Note that we have already seen this in eq.~(\ref{17-20rep}), where the very same combination of kinematic functions corresponding to the uppermost entry of eq.~(\ref{17-20mat2}) appears. We have now identified it as the unique combination corresponding to a non-vanishing eigenvalue of the mixing matrix. 

As discussed above, all the diagrams in figure~\ref{17-20} are expected to have a leading ${\cal O}(\epsilon^{-3})$ singularity. 
However, the coefficient of this leading pole should be different in the four cases. In diagram (3a), one may shrink the innermost gluon to a point, merging it with the cusp at the hard vertex. There are then two ways in which one can shrink the remaining gluons, given that they connect different lines. Likewise, in diagram (3d) one may start by shrinking either of the two innermost gluons before shrinking the right-hand gluon. Thus, there are again two ways of forming the leading $\epsilon$ pole. Things are different for diagrams (3b) and (3c), where the order of shrinking the gluons is prescribed by the structure of the diagram. In (3b), for instance, one must shrink the upper, right-hand and lower gluons in that order. Thus, (3b) and (3c) have only one way of forming subdivergences. These combinatoric factors are precisely such that the leading epsilon pole cancels in the uppermost entry of eq.~(\ref{17-20mat2}).

The above argument indicates that the leading subdivergence indeed cancels due to the properties of the mixing matrix, provided the subdivergences are related by the simple combinatoric factors derived above by considering the sequential shrinking of gluons. In fact, this reasoning is expected to work for the leading pole, but not beyond this. To see this in more detail, let us consider the web $W_{(2,3,1)}$ corresponding to the six diagrams in figure~\ref{3lsix}. This is a more complicated set of diagrams than those
of figure~\ref{17-20}, in that not all diagrams have a leading subdivergence. 
Diagram (3B) has no subdivergences, just an overall ${\cal O}(\epsilon^{-1})$ singularity; diagrams (3A) and (3C) have one subdivergence, {\it i.e.} a leading ${\cal O}(\epsilon^{-2})$ singularity; and the last three diagrams, (3D) through (3F), have two subdivergences {\it i.e.} a leading ${\cal O}(\epsilon^{-3})$ singularity.

The eigenvalues of the mixing matrix appearing in eq.~(\ref{1-6mat}) are
\begin{equation}
\{\lambda\}_{(2,3,1)}=\{1,1,1,0,0,0\}.
\label{ev231}
\end{equation}
We again see that the only eigenvalues are 1 and 0, although here both occur with a nontrivial degeneracy. In this case the dimension and the rank of $R$ are, respectively, $d=6$ and $r=3$. Constructing the matrix of left eigenvectors, the analogue of eq.~(\ref{17-20mat2}) in this case is:

{\footnotesize
\begin{align}
{\cal F}^{\mathrm T}=&\frac{1}{6}\left(\begin{array}{c}4 {\cal F}(3A) + {\cal F}(3B) - 2 {\cal F}(3C) + {\cal F}(3D) - 2 {\cal F}(3E) + {\cal F}(3F)\\ -2 {\cal F}(3A) - 2 {\cal F}(3B) - 2 {\cal F}(3C) - 2 {\cal F}(3D) + 
   4 {\cal F}(3E) - 2 {\cal F}(3F)\\ -3 {\cal F}(3A) - 3 {\cal F}(3B) + 3 {\cal F}(3C)\\ 
   -4 {\cal F}(3A) - {\cal F}(3B) + 2 {\cal F}(3C) - {\cal F}(3D) + 2 {\cal F}(3E) + 5 {\cal F}(3F) \\2 {\cal F}(3A) + 2 {\cal F}(3B) + 2 {\cal F}(3C) + 
   2 {\cal F}(3D) + 2 {\cal F}(3E) + 2 {\cal F}(3F)\\ 2 {\cal F}(3A) - {\cal F}(3B) - 4 {\cal F}(3C) + 5 {\cal F}(3D) + 2 {\cal F}(3E) - {\cal F}(3F)\end{array}\right)^{\mathrm T}\left(\begin{array}{rrrrrr}2& -2& 0& -1& 0& 1\\1& -1& 0& -1& 1& 0\\1& -2& 1& 0& 0& 0\\ 
   -\frac{1}{2}& 0& -\frac{1}{2}& 0& 0& 1\\ 1& 0& 1& 0& 1& 0\\ -\frac{1}{2}& 
  0& -\frac{1}{2}& 1& 0& 0\end{array}\right)\,.
\label{1-6mat2}
\end{align}
}
As before, the matrix on the right-hand side is the diagonalizing matrix $Y$ introduced in (\ref{diagonalization}). Its rows are the left eigenvectors of $R$, and we have absorbed the inverse matrix $Y^{-1}$ into the vector of kinematic functions, forming the linear combinations ${\cal F}^{\mathrm T}\,Y^{-1}$ which are explicitly displayed here. 
The ordering is such that the upper three entries of this vector correspond to kinematic combinations with eigenvalue $\lambda=1$, and the lower three entries to $\lambda=0$. 

The structure of subdivergences for these diagrams is not as straightforward as in the case of figure~\ref{17-20}, due principally to the fact that not every diagram has a leading divergence. However, we may easily see that the leading $\epsilon$ poles cancel. These come only from diagrams (3D) through (3F), where there are no crossed gluons. Furthermore, one expects the coefficient of the highest $\epsilon$ singularity, ${\cal O}(\epsilon^{-3})$,  to be the same in each of these three diagrams:
in each of them the order of shrinking the gluons towards the hard vertex is uniquely prescribed by the structure of the diagram. This is also verified by an explicit calculation~\cite{Sterman}.

One may then check that the upper three entries of the kinematic vector in eq.~(\ref{1-6mat2}), the ones corresponding to unit eigenvalue, have no leading ${\cal O}(\epsilon^{-3})$ singularities, as these cancel in the combination ${\cal F}(3D) - 2 {\cal F}(3E) + {\cal F}(3F)$, whereas this is not true for the lower three rows. The latter evidently contain ${\cal O}(\epsilon^{-3})$ singularities, but because they correspond to a vanishing eigenvalue of the mixing matrix, they do not contribute to the $W_{(2,3,1)}$ web. As expected, the exponent remains free of such singular terms.

It is instructive to examine what happens with subleading divergences in this three-loop web. Writing the $\epsilon$ poles 
of each kinematic factor as
\begin{equation}
{\cal F}(D)=\sum_{m=1}^{m=3}\frac{1}{\epsilon^m}{\cal F}_D^{(-m)},
\end{equation}
we may write the higher pole terms ($m=2,3$, where the $m=3$ terms must cancel within the web) 
in the upper three rows~$F_i$ of eq.~(\ref{1-6mat2}) as:
\begin{align}
F_1&=\frac{1}{\epsilon^3}\left({\cal F}_{3D}^{(-3)}-2{\cal F}_{3E}^{(-3)}+{\cal F}_{3F}^{(-3)}\right)\notag\\
&\quad+\frac{1}{\epsilon^2}\left({\cal F}_{3D}^{(-2)}-2{\cal F}_{3E}^{(-2)}+{\cal F}_{3F}^{(-2)}
+4{\cal F}_{3A}^{(-2)}-2{\cal F}_{3C}^{(-2)}\right)\label{row1};\\
F_2&=\frac{1}{\epsilon^3}\left(-2{\cal F}_{3D}^{(-3)}+4{\cal F}_{3E}^{(-3)}-2{\cal F}_{3F}^{(-3)}\right)\notag\\
&\quad+\frac{1}{\epsilon^2}\left(-2{\cal F}_{3D}^{(-2)}+4{\cal F}_{3E}^{(-2)}-2{\cal F}_{3F}^{(-2)}
-2{\cal F}_{3A}^{(-2)}-2{\cal F}_{3C}^{(-2)}\right)\label{row2};\\
F_3&=\frac{1}{\epsilon^2}\left(-3{\cal F}_{3A}^{(-2)}+3{\cal F}_{3C}^{(-2)}\right)\label{row3}.
\end{align}
We have already argued that one expects ${\cal F}_{3D}^{(-3)}={\cal F}_{3E}^{(-3)}={\cal F}_{3F}^{(-3)}$,
so the ${\cal O}(\epsilon^{-3})$ poles do indeed vanish in eqs.~(\ref{row1}) and (\ref{row2}). Considering the subleading
($\epsilon^{-2}$) poles, an explicit calculation~\cite{Sterman}\footnote{We have explicitly verified
this calculation~\cite{Sterman} using our own methods, and found agreement.} reveals that the ${\cal O}(\epsilon^{-2})$
poles of diagrams (3A) and (3C) are related by
\begin{equation}
{\cal F}_{3A}^{(-2)}\,=\,2\,{\cal F}_{3C}^{(-2)}\,.
\label{eps2poles}
\end{equation} 
Thus, we see explicitly, for example, from eq.~(\ref{row3}) that the subleading subdivergence does not cancel due to the mixing matrix alone: an ${\cal O}(\epsilon^{-2})$ singularity survives in
the web $W_{(2,3,1)}$, as indeed required by (\ref{order_epsilon_minus_2_3loops}). 
This illustrates the interplay between the web structure and the commutator of lower order counter-terms and webs in eq.~(\ref{S_ren_BCH}) for subleading subdivergences. 
A full exposition of this interplay is beyond the scope of this paper.

\subsection{Discussion}

Let us then recapitulate what we have learnt, and present our general conjecture. 
Considering the web in (\ref{setmix}), involving a linear combination of kinematic functions ${\cal F}(D)$ and colour factors $C(D')$ of any of the $d$ diagrams in the set (which are mutually related by permutations of the gluon attachments along the eikonal lines) we have identified a very interesting structure, which we believe to be completely general. 
This structure is dictated by the fact that the mixing matrix $R_{DD'}$ is idempotent, having only two eigenvalues, $1$ and $0$, with degeneracy $r$ and $(d-r)$, respectively.
The unit eigenvalue corresponds to new contributions to the exponent, while the zero eigenvalue corresponds to contributions that are discarded because they have been already accounted for by the exponentiation of lower-order graphs.

Writing the web in this basis in eq.~(\ref{setmix_matrix}) we deduce that the mixing matrix can be expressed as follows\footnote{Note that the sum over $H$ in (\ref{R_in_terms_of_Y}) extends only over the range $1$ to $r$. Clearly if it had extended to $d$ instead one would have obtained the identity matrix $\sum_{H=1}^{d} Y^{-1}_{GH}\,Y_{HD} =\delta_{GD}$. },  
\begin{equation}
\label{R_in_terms_of_Y}
R_{DD'}=\sum_{H=1}^{r} Y^{-1}_{DH}\,Y_{HD'}\,.
\end{equation}
Using this representation of $R_{DD'}$ in (\ref{setmix}), and summing over $D$, we get $r$ independent linear combinations ($H=1,2,\ldots,r$) of kinematic functions, $\sum_D {\cal F}(D) Y^{-1}_{DH}$. Similarly, summing over $D'$ we get corresponding linear combinations of colour factors,~$\sum_{D'}Y_{HD'} C(D')$.

Based on our results, we conjecture that each individual order-$n$ web (closed set of order-$n$ diagrams) has a singularity structure which is consistent with the renormalization properties of the eikonal vertex:
\begin{itemize}
\item{} The leading singularities, ${\cal O}(\epsilon^{-n})$, must conspire to cancel exactly between diagrams, such that all linear combinations of kinematic functions which enter the exponent of the unrenormalized eikonal amplitude ($\sum_D {\cal F}(D) Y^{-1}_{DH}$ corresponding to 
eigenvalue~1), have no ${\cal O}(\epsilon^{-n})$ singularity.
\item{} When renormalizing the eikonal amplitude, all surviving multiple-pole terms ${\cal O}(\epsilon^{-j})$ with $j\leq n-1$ in these linear combinations exactly cancel the contributions of commutators of lower-order terms as dictated by eq.~(\ref{S_ren_BCH}). 
\end{itemize}
The renormalization structure of the multi-leg eikonal vertex is highly constrained, implying that the mixing matrices must encode significant cancellations at any order. It remains for future work to determine whether the renormalization procedure can be applied on a web-by-web basis, and establish that individual webs do indeed conform with eq.~(\ref{S_ren_BCH}).

Our discussion here focused on the kinematic dependence, where appropriate cancellations must occur. The structure of eq.~(\ref{setmix_matrix}) implies that $Y$ acts also on the 
colour factors. Here another general property of this matrix becomes important: 
 each of the left-eigenvectors $v^{\mathrm T}$ corresponding to a unit eigenvalue, {\it i.e.} each of the first $r$ rows of the diagonalizing matrix $Y$ have entries that sum up to zero, namely
\begin{equation}
\sum_{D}Y_{HD}=0\,, \qquad\quad \forall H\,\vert\, 1\leq H\leq r\,.
\label{Yprop}
\end{equation}
Note that the zero-sum property does not necessarily hold for those eigenvectors that correspond to a zero eigenvalue $r+1 \leq H\leq d$, as can be checked in the previous examples.

It is straightforward to see that the zero-sum property in (\ref{Yprop}) is related to the zero-sum property of the rows in the matrix $R$ itself, eq.~(\ref{Rprop}) above. 
Indeed, expressing $R_{DD'}$ using $Y$ as in (\ref{R_in_terms_of_Y}) and summing over $D'$, for any fixed $D$, we have
\begin{equation}
\sum_{D'=1}^{d}R_{DD'}=\sum_{D'=1}^{d}\sum_{H=1}^{r} Y^{-1}_{DH}\,Y_{HD'}=
\sum_{H=1}^{r} Y^{-1}_{DH}\,\underbrace{\Big(\sum_{D'=1}^{d} Y_{HD'}\Big)}_{0} =0\,,
\end{equation}
where we changed the order of summation and used (\ref{Yprop}), recovering eq.~(\ref{Rprop}) above.

Eq.~(\ref{Yprop}) has a clear physical interpretation: according to (\ref{setmix_matrix}) only specific linear combinations of colour factors $(Y\,C)_H$ enter the exponent ($r$ such combinations in total, $H=1,2,\ldots, r$). 
What characterized these linear combinations is that they are free of any 
component which is independent of the ordering of the gluons along the line, as imposed by the symmetry in (\ref{coltrans}), namely
\begin{equation}
\sum_{D}Y_{HD}\,  C(D)\,=\,\sum_{D}Y_{HD} \, \Big(C(D)+K\Big)\,,
\end{equation}
where $K$ is $D$-independent.
We thus see that the linear combinations $(Y\,C)_H$ (for $H=1,2,\ldots,r$) encode the  generalization of the antisymmetrisation property familiar from the two-loop case (see {\it e.g.} eq.~(\ref{twoemcol7})). 

In conclusion we have seen that webs have a very special structure, summarized by eq.~(\ref{setmix_matrix}). This implies that the mixing matrices, or equivalently, the corresponding diagonalising matrix $Y$, is simultaneously responsible for
\begin{itemize} 
\item{} forming linear combinations of kinematic functions of different diagrams, $({\cal F}^{\mathrm T} Y^{-1})_H$, that conform with the singularity structure implied by the renormalization of the multi-eikonal vertex. In particular, these combinations are 
 free of the leading subdivergence (eq.~(\ref{order_epsilon_cancellation})), and all their lower-order multiple poles correspond to commutators of lower-order webs and counter-terms. 
\item{} forming corresponding linear combinations of the colour factors $(YC)_H$ which are antisymmetric with respect to permutations of gluons along any of the lines, as required by (\ref{Yprop}).
\end{itemize}
Thus, these two operations on kinematics and on colour space are intimately related.

So far we have considered the properties of strictly eikonal gluon emissions or, equivalently, Wilson lines. In the next section, we discuss how the notion of webs can be generalised further to include contributions in scattering amplitudes arising from subleading powers in gluon momentum.

\section{Next-to-eikonal webs\label{sec:next_to_eikonal}}

In this section, we consider the application of the results of this paper to the extension of resummation methods in multiparton scattering to include corrections beyond the eikonal approximation in the momentum of the emitted gluons. 

In more detail, the perturbative expansions of inclusive differential cross-sections (at the parton level) have the generic form
\begin{equation}
\frac{d\hat{\sigma}}{dz}=\sum_{m,n}^\infty \alpha_s^n\left[a_{nm}
\left(\frac{\ln^m(1-z)}{1-z}\right)_{+}\,+\,
v_{nm}\delta(1-z)+ b_{nm}\ln^m(1-z)+\ldots\right],
\label{xsec}
\end{equation} 
where $1-z$ is a dimensionless combination related to the energy carried by soft gluons, and the ellipsis denotes terms which are suppressed by powers of $(1-z)$ and thus non-singular as $z\rightarrow 1$. With the exception of collinear singularities, the first set of terms on the right-hand-side of eq.~(\ref{xsec}) can be obtained from the eikonal approximation, in which the four-momentum of each emitted gluon goes to zero. The ``$+$'' prescription and the $\delta(1-z)$ terms represent the contribution of purely virtual diagrams. 
Finally, the last set of terms, suppressed by one power of $(1-z)$ compared to the first set, arises from the {\it next-to-eikonal} approximation in which gluon momenta may occur to first order in the amplitude. When the logarithms in eq.~(\ref{xsec}) become large, fixed-order perturbation theory breaks down and one must resum the enhanced terms to all orders in $\alpha_s$.

In addition to the large body of work on eikonal resummation, there have also been a number of studies focusing on next-to-eikonal (NE) effects. Early approaches included subleading effects in the collinear evolution kernel entering the renormalisation group equations for the resummation, thus effecting a partial resummation of subeikonal corrections~\cite{Kramer:1996iq,Catani:2001ic,Harlander:2001is,Catani:2003zt,Kidonakis:2007ww,Basu:2007nu}. More recently, efforts have been made to systematically construct NE resummation formulae, such as the approach of~\cite{Grunberg:2009yi,Grunberg:2009vs}, the physical evolution kernel approach of~\cite{Moch:2009mu,Moch:2009my,Moch:2009hr,Soar:2009yh,Vogt:2010cv}, and the application of a modified evolution equation for parton distributions~\cite{Dokshitzer:2005bf} within the context of threshold resummation~\cite{Laenen:2008ux}. So far none of these approaches attempts to classify the diagrams which underpin NE resummation to all orders. 

This was undertaken in~\cite{Laenen:2008gt}, which used path integral methods to rewrite the problem of soft gluon resummation in terms of a field theory for the soft gauge field. A set of effective Feynman rules was obtained for emission of soft photons or gluons up to next-to-eikonal order, and it was shown for the case of scattering involving two partons that up to NE order, a given hard scattering amplitude ${\cal M}_0$ dressed by soft gluons has the following schematic form:
\begin{equation}
{\cal M}={\cal M}_0\exp\left[\sum_{D_{\text{E}}}\widetilde{C}(D_\text{E}){\cal F}(D_{\text{E}})+\sum_{D_{\text{NE}}}\widetilde{C}(D_{\text{NE}}){\cal F}(D_{\text{NE}})\right](1+{\cal M}_r).
\label{ampstruc}
\end{equation}
Here the $D_{\text{E}}$ are eikonal webs, and $D_{\text{NE}}$ are next-to-eikonal webs {\it i.e.} two-particle irreducible subdiagrams which contain one NE Feynman rule, with all other emissions eikonal. These formally exponentiate, and have exponentiated colour factors as do the eikonal webs. The term ${\cal M}_r$ is a remainder term which does not exponentiate, but which has an iterative structure to all orders in perturbation theory. It collects contributions from diagrams in which an eikonal gluon is emitted from an external line, and lands inside the hard subamplitude ${\cal M}_0$. Such contributions have already arisen in the literature as part of the Low-Burnett-Kroll theorem~\cite{Low:1958sn,Burnett:1967km}, which was generalised by Del Duca~\cite{DelDuca:1990gz} to correctly include collinear singularities. The presence of next-to-eikonal webs in the case of amplitudes involving two coloured particles suggests that once webs have been generalised to the multiparton case, a subset of next-to-eikonal contributions exponentiates as well. 

The proof of web exponentiation in the two parton case of~\cite{Laenen:2008gt} used the path integral method for eikonal and next-to-eikonal resummation. This involves non-trivial combinatorics in the case of non-abelian theories, due to the noncommuting nature of the source terms in the field theory obtained for the soft gauge field. Nevertheless, a classification of NE webs was possible using a replica trick argument. This suggests that the generalisation of the replica trick presented in the present paper can be used to study next-to-eikonal contributions to multiparton scattering. 
That is, the multiparton version of eq.~(\ref{ampstruc}) takes the form:
\begin{equation}
{\cal M}={\cal M}_0\exp\left[\sum_{D_{\text{E}},D'_{\text{E}}}
{\cal F}(D_{\text{E}}) \,R^{\text{E}}_{DD'}\, C(D'_{\text{E}})
\,\,+
\sum_{D_{\text{NE}},D'_{\text{NE}}}{\cal F}^{\text{NE}}(D_{\text{NE}})\,R^{\text{NE}}_{DD'}\,C(D'_{\text{NE}})\,
\right](1+{\cal M}_r).
\label{multi-leg_NE}
\end{equation}
where the first sum in the exponent goes over the eikonal webs of eq.~(\ref{setmixintro}), and the second represents the next-to-eikonal analogue thereof, where each of the diagrams $D_{\text{NE}}$ contains one NE Feynman rule. Here ${\cal F}(D)$ denotes the kinematic part of a diagram, $R_{DD'}$ the block-diagonal mixing matrix (where each block corresponds to a closed set of diagrams forming a given web), and~$C(D')$ the conventional colour factor of diagram~$D'$. Given the different set of diagrams, the next to eikonal mixing matrix $R^{\text{NE}}_{DD'}$ is of course different from its eikonal counterpart $R^{\text{E}}_{DD'}$.

In the remainder of this section, we argue that one indeed expects next-to-eikonal diagrams
in which all gluons are external to the hard interaction to exponentiate. Our aim is not to undertake a fully comprehensive calculation of NE logarithms, nor to classify the resulting mixing matrices. Rather, we wish to clarify the structure of NE corrections in multiparton scattering, in line with the comments made for the two parton case in~\cite{Laenen:2008gt}. 

Let us begin by briefly reviewing the path integral formalism we use. The starting point is to consider a scattering amplitude for the production of $L$ final state hard particles, each of which may emit
further soft radiation.  One may then separate the gauge field into hard and soft modes. For brevity, and to present the argument in its simplest form, we consider explicitly scalar emitting particles. 
The case of fermionic emitting particles is similar, as explained in~\cite{Laenen:2008gt}. In abelian gauge theory, the scattering amplitude has the factorised form
\begin{equation} {\cal M}(p_1,\ldots,p_n)=\int{\cal D}A^\mu_s
  H(x_1,\ldots, x_L)\prod_{k=1}^L\langle
  p_k|(S-{\mathrm i}\varepsilon)^{-1}|x_k\rangle(p_k^2+m^2)e^{{\mathrm i}S[A_s]}.
  \label{amp1}
\end{equation}
Here $A^\mu_s$ is the soft gauge field with action $S[A_s]$, and
$H(x_1,\ldots x_L)$ the hard interaction producing the emitting
scalar particles at positions $x_i$. The factors $\langle
p_k|(S-{\mathrm i}\varepsilon)^{-1}|x_k\rangle$ represent propagators for a
scalar particle in soft background gauge field -- with
$S$ denoting the quadratic operator for the scalar field in the
Lagrangian -- sandwiched between states of given initial position
(the points $x_k$ at which the particles are created by the hard
interaction) and given final momentum $p_k$. 
Both $x_k$ and $p_k$ are 4-vectors. 
The explicit factors of $(p_k^2+m^2)$ in eq.~\eqref{amp1} truncate the free propagators
associated with the external legs, and there is an implicit
integration over the positions $x_k$.
The propagator factors in
eq.~(\ref{amp1}) can be represented as first-quantised path
integrals~\cite{Strassler:1992zr,vanHolten:1995ds}. One finds
\begin{equation}
  (p_k^2+m^2)\langle p_k|(S-{\mathrm i}\varepsilon)^{-1}|x_k\rangle=e^{-{\mathrm i}p_k\cdot x_k}f_k(\infty),
  \label{path1}
\end{equation}
where
\begin{align}
  f_k(\infty)&=\int_{y_k(0)=0}{\cal D}y_k\exp\left[{\mathrm i}\int_0^\infty dt\left(\frac{1}{2}\dot{y}_k^2+(p_k+\dot{y}_k)\cdot A_s\left(x_k+p_kt+y_k(t)\right)\right.\right.\notag\\
  &\hspace*{100pt}\left.\left.+\frac{{\mathrm i}}{2}\partial\cdot
      A_s\left(x_i+p_ft+y_k(t)\right)\right)\right].
  \label{path2}
\end{align}
Here $y_k$ is the fluctuation about the classical path associated with
the scalar particle $k$, and the path integral is over all such
fluctuations subject to the boundary conditions of given initial
position $x_k$ and final momentum $p_k$. Substituting this result into
eq.~(\ref{amp1}) yields
\begin{align}
  {\cal M}(p_1,\ldots,p_L)&=\int{\cal D}A^\mu_s \int{\cal D}y_1\ldots \int{\cal D}y_LH(x_1,\ldots ,x_L)e^{{\mathrm i}S[A_s]}e^{-{\mathrm i}(x_1\cdot p_1+\ldots + x_L\cdot p_L)}\notag\\
  &\quad \times \prod_k\exp\left[{\mathrm i}\int_0^\infty
    dt\left(\frac{1}{2}\dot{y}_k^2+(p_k+\dot{y}_k)\cdot
      A_s+\frac{{\mathrm i}}{2}\partial\cdot A_s\right)\right].
  \label{amp2}
\end{align}
We see that the scattering amplitude has taken the form of a generating functional for a quantum field theory
for the soft gauge field $A^\mu_s$. The exponent in the second line
contains terms linear in $A_s^\mu$, which act as sources for $A_s^\mu$.
These sources are located on the external lines, so that
the path integral over $A_s^\mu$ generates all subdiagrams
that connect the sources on the external lines, which can be connected
or disconnected (they are subdiagrams because they do not contain the
emitting particles).  Thus formulated, the exponentiation of soft photon
corrections (in terms of connected subdiagrams) is precisely
equivalent to the well-known exponentiation of connected diagrams in
quantum field theory.

The advantage of the above approach stems from its clear physical
interpretation in terms of the worldline trajectories of the emitting
particles. To see which diagrams exponentiate, one must calculate the
soft gauge field Feynman rules that result after carrying out the path
integrations over $y_k$ in eq.~(\ref{amp2}). This can be done by systematically expanding about the
classical straight-line trajectory of eq.~(\ref{class}). Note that the classical
trajectory corresponds to the eikonal approximation in which the
emitting particles do not recoil. We introduce a scaling variable
$\lambda$ such that ${\cal  O}(\lambda^0)$ and ${\cal O}(\lambda^{-1})$ constitute the eikonal
and the next-to-eikonal approximations, respectively.

Up to NE accuracy we can then write eq.~\eqref{amp2} as
\begin{align}
{\cal M}_{b_1\ldots b_n}(p_1,\ldots,p_L)=\int d^dx_1\ldots\int d^dx_LH_{a_1\ldots a_n}(x_1,\ldots,x_L)\notag\\
\times\int\left[ {\cal D}{A}^\mu_s\right]\,e^{{\mathrm i}S[A^\mu_s]}\prod_ke^{-{\mathrm i}x_k\cdot p_k}f^{(k)}_{a_kb_k}(A^\mu,\beta_k),
\label{amppathNE}
\end{align}
where\footnote{For the explicit derivation of this result, we refer the reader to appendix
B of~\cite{Laenen:2008gt}.}
\begin{align}
\begin{split}
f_k(A^\mu,\beta_k)&={\cal P}\exp\Bigg\{
{\mathrm i}g_s\int_0^\infty dt\bigg[\beta_k\cdot {A}_s(\beta_kt)+\frac{{\mathrm i}}{2\lambda}\partial\cdot {A}_s(\beta_kt)+\frac{{\mathrm i}t}{2\lambda}\beta_{k\mu}\,\Box {A}^\mu_s(\beta_kt)\bigg]
\\
&-g_s^2\frac{{\mathrm i}}{\lambda}\int_0^\infty dt\int_0^\infty dt'\bigg[\frac{1}{2}\delta(t-t')\,g_{\mu\nu}\,{A}_s^\mu(\beta_kt){A}_s^\nu(\beta_kt')
\\
&\hspace*{100pt}
+\theta(t-t')\,\beta_{k\mu}\,g_{\sigma\nu}\big[\partial^\sigma {A}^\mu_s(\beta_kt)\big]\big[{A}^\nu_s(\beta_kt')\big]
\\&\hspace*{100pt}
+\frac{1}{2}\text{min}(t,t')\,\beta_{k\mu}\,\beta_{k\nu}\, g^{\sigma\tau}
\big[\partial_\sigma {A}^\mu_s(\beta_kt)\big]\big[\partial_\tau {A}^\nu_s(\beta_kt')\big]\bigg]\Bigg\}
\end{split}
\label{fkdef}
\end{align}
\\
Here we have explicitly written the integrals over the positions $x_k$, $k=1\ldots L$. 
As explained in~\cite{Laenen:2008gt}, these can be carried out by expanding the hard interaction
and $e^{-{\mathrm i}x_k\cdot p_k}$ factors about $x_k=0$, and lead to next-to-eikonal
contributions which depend only on the quantum numbers of a single parton leg. That is,
such contributions do not involve non-trivial colour flows between the external legs,
thus may be ignored in the following discussion. 

One recognises the first term in the exponent as the Wilson line exponent of eq.~(\ref{amppath}). The other terms, by analogy with the eikonal term, act as source terms for the soft gauge field. They generate next-to-eikonal Feynman rules that couple the soft gauge field to the outgoing parton lines. Carrying out the path integral in eq.~(\ref{amppathNE}) generates all soft gluon subdiagrams which connect the external lines, where each diagram contains at most one NE Feynman rule. We do not write down the NE Feyman rules here, but refer the reader to~\cite{Laenen:2008gt} for more details. Here we merely note that the NE Feynman rules contain both one and two gluon vertices. Furthermore, the precise form of the vertices depends upon the spin of the external lines (as is expected for corrections to the eikonal approximation, which is itself insensitive to spin effects). The case of scalar particles is shown above.
 
\begin{figure}[htb]
\begin{center}
\scalebox{0.8}{\includegraphics{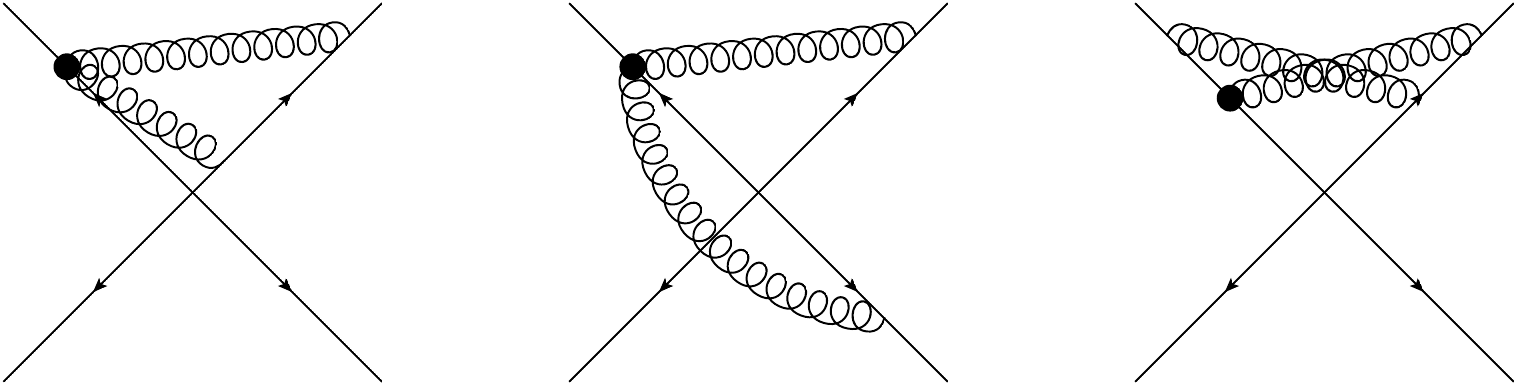}}
\caption{Examples of diagrams entering the exponent at next-to-eikonal order, in which $\bullet$ denotes a NE Feynman rule.}
\label{2loopNE}
\end{center}
\end{figure}
The power of the path integral approach used to derive eq.~(\ref{amppathNE}) is that the argument for exponentiation of the scattering amplitude in terms of webs is the same at both eikonal and NE order. By analogy with equation~(\ref{Zdef2}), one may define a generating functional
\begin{equation}
{\cal Z}^{\rm NE}=\int\left[ {\cal D}{A}^\mu_s\right]\,e^{{\mathrm i}S[A^\mu_s]}\,\,{\bf f}^{(1)}\otimes {\bf f}^{(2)}\otimes {\bf f}^{(3)}\ldots \otimes {\bf f}^{(L)},
\label{ZdefNE}
\end{equation} 
where ${\bf f}^{(k)}$ is the generalised Wilson line operator defined in eq.~(\ref{fkdef}), and there are $L$ parton lines. Constructing again a replicated theory with $N$ identical copies of the soft gauge field, one has
\begin{align}
\left({\cal Z}^{\rm NE}\right)^N&=\int{\cal D}{A}_\mu^1\ldots{A}_\mu^N \,e^{{\mathrm i}\sum_i S[A^i_\mu]}\,\,
\left({\bf f}^{(1)}_1{\bf f}^{(1)}_2\ldots{\bf f}^{(1)}_N\right)\otimes \left({\bf f}^{(2)}_1\ldots{\bf f}^{(2)}_N \right)
\otimes\notag\\
&\hspace*{130pt}\ldots\otimes\left( {\bf f}^{(L)}_1\ldots{\bf f}^{(L)}_N\right)\,,
\label{ZdefrepNE}
\end{align}
where ${\bf f}^{(k)}_i$ is the generalised Wilson line operator associated with parton $k$ and replica number $i$. Each parton line carries a product of generalised Wilson lines. Using the replica ordering operator ${\cal R}$, such a product may be written as
\begin{align}
\begin{split}
{\bf f}_1^{(k)}{\bf f}_2^{(k)}\!\ldots{\bf f}_N^{(k)}
&={\cal R}{\cal P}\exp\Bigg\{{\mathrm i}g_s\sum_{i=1}^N\int_0^\infty \!dt\bigg[\beta_k\cdot {A}^i(\beta_kt)+\frac{{\mathrm i}}{2\lambda}\partial\cdot {A}^i(\beta_kt)
+\frac{{\mathrm i}t}{2\lambda}\beta^{k\mu}\Box {A}^i_\mu(\beta_kt)\bigg]
\\&-g_s^2\frac{{\mathrm i}}{\lambda}
\sum_{i=1}^N\int_0^\infty dt\int_0^\infty dt'
\bigg[\frac{1}{2}
\delta(t-t')\,g^{\mu\nu}\,{A}^i_\mu(\beta_kt){A}^i_\nu(\beta_kt')\\
&\hspace*{100pt}
+\theta(t-t')\,\beta^\mu_{k}\,g^{\sigma\nu}
\left[\partial_\sigma {A}^i_\mu(\beta_kt)\right]
\left[{A}^i_\nu(\beta_kt')\right]
\\&\hspace*{100pt}+\frac{1}{2}\text{min}(t,t')\,\beta^\mu_{k}\,\beta^\nu_{k} g^{\sigma\tau}
\left[\partial_\sigma {A}^i_\mu(\beta_kt)\right]
\left[\partial_\tau {A}^i_\nu(\beta_kt')\right]\bigg]\Bigg\}.
\end{split}
\label{fprod}
\end{align}

Note in particular that in the two gluon vertex terms, the gluons emerging from the vertex must have the same replica number. Diagrams in the replicated theory are similar to that obtained in the eikonal case. That is, their kinematic parts are the same as in the unreplicated theory, but the colour factors correspond to the replica ordered diagrams, rather than the original diagrams. One may again write
\begin{equation}
({\cal Z}^{\rm NE})^N=1+N\ln{{\cal Z}^{\rm NE}}+{\cal O}(N^2),
\label{taylorNNE}
\end{equation}
so that it remains true that diagrams which are linear in the replica number contribute to the exponent of the scattering amplitude. Finally, one extracts the coefficient of $N^1$ as before and derives the exponentiation of the scattering amplitude up to next-to-eikonal order. The exponentiated colour factors are again given by the colour factors of graphs in the replicated theory, where the modification is understood as arising from the action of the ${\cal R}$ operator. Examples of next-to-eikonal diagrams which enter at two loop order are shown in figure~\ref{2loopNE}. The first two examples enter singly in the exponent i.e. do not mix with other diagrams. The third example mixes with similar diagrams in which a different eikonal coupling is replaced by the next-to-eikonal vertex.

As already discussed for the case of two eikonal lines, NE webs are not the only source of next-to-eikonal corrections. There are also diagrams in which an eikonal gluon is emitted from an external line, and lands inside the hard interaction. These are related to derivatives of the hard interaction and jet functions with no emission, using the appropriate formulation of the Low-Burnett-Kroll theorem~\cite{Low:1958sn,Burnett:1967km,DelDuca:1990gz}. Nevertheless, the results of this section show that a subset of NE corrections indeed formally exponentiates.

\section{Conclusions\label{discuss}}

In this paper we have investigated the generalisation of soft-gluon exponentiation in terms of webs from two-parton scattering to the $L$-parton case. We found that the idea of webs survives in multiparton scattering, but that there are important differences with respect to the two-eikonal-line case. Our main result is that one finds closed sets of diagrams, related by permuting gluon attachments along each of the eikonal lines, which mix with each other in the exponent. The relevant mixing matrices have zero-sum rows, reflecting the fact that symmetric combinations of colour matrices do not contribute to the exponent. This is the appropriate generalisation of the similar property in the two-eikonal line case, originally known as the maximally non-abelian nature of webs.
We note that in contrast to the two-line case, the topology of those diagrams which contribute to the exponent is not necessarily irreducible, as has been demonstrated here in numerous examples. 

If many reducible diagrams are present in the exponent, what is the point of classifying webs? 
At the very least, being able to calculate the exponentiated colour factor of any diagram allows one to work directly with the exponent. This produces a considerable simplification in {\it e.g.} calculations of soft anomalous dimension matrices for soft-gluon resummation. 
Furthermore, classifying the structure of webs has allowed us to generalise the arguments given in~\cite{Laenen:2008gt} for the formal exponentiation of a subclass of next-to-eikonal corrections, namely those which arise from diagrams in which all soft gluon emissions are external to the hard interaction. 

Using the replica trick we have established an algorithm by which the exponentiated colour factor of any multi-loop (and multi-leg) diagram can be directly computed. We have also derived  a corresponding closed-form expression (eq.~(\ref{modcol2})) in terms of colour factors of subdiagrams.
We have demonstrated the application of the method, considering several two, three and four loop examples, where in each case we have identified the closed set of diagrams that mix with each other and determined their mixing matrix.
Furthermore, considering two non-trivial three-loop classes of diagrams we reproduced the results obtained by the replica trick by explicitly exponentiating lower-order diagrams, and subtracting these from the full unexponentiated sum of diagrams. This alternative calculation required manipulating products of colour factors as well as products of integrals of one- and two-loop diagrams, bringing them into the form of recognisable three-loop expressions.
These examples made clear the drastic simplification afforded by the replica-trick based method. 

Beyond introducing a new calculational method, our generalization of the notion of webs to the multi-leg case is a conceptual step forward. The conventional understanding of exponentiation through evolution equations and anomalous dimensions has been very effective in clarifying the general singularity structure of amplitudes in the dimensional regularization parameter, in explaining the way running coupling effects can be incorporated to all orders, and in constraining the kinematic and colour structure of the exponent in the massless case~\cite{Gardi:2009qi,Becher:2009qa}.
However, in general, in the massive case and beyond the planar limit, not much is known about the kinematic structure or the colour structure of the exponent in multiparton scattering beyond the two-loop level, the state-of-the-art of explicit calculations~\cite{Mitov:2009sv,Kidonakis:2009ev,Becher:2009kw,Beneke:2009rj,Czakon:2009zw,Ferroglia:2009ep,Ferroglia:2009ii,Chiu:2009mg,Mitov:2010xw,Ferroglia:2010mi}. It is clear though that the exponent has a much simpler structure than the amplitude as a whole, opening a window to the all-order structure of perturbation theory. We have shown that webs provide a complementary way to understand exponentiation in multiparton scattering, revealing intriguing connections between kinematics and colour space.

Comparing the conventional picture, of exponentiation in terms of the soft anomalous dimension, to the picture of webs is very insightful. The former tells us that in dimensional regularization, after removing collinear and running-coupling singularities, the eikonal amplitude can be renormalized by introducing a single ${\cal O}(1/\epsilon)$ counter-term at any order $n$ in perturbation theory.
In contrast, a straightforward diagrammatic analysis reveals the presence of multiple subdivergences, leading to much stronger singularities up to ${\cal O}(1/\epsilon^n)$. 
In the two-eikonal-line case, no difficulty arises owing to the fact that only very special diagrams contribute to the exponent, those which have no subdivergences. These are precisely the diagrams that have irreducible colour structure, the ones we call webs.

In the multi-leg case a major difficulty arises: individual diagrams that contribute to the exponent present subdivergences. In this paper we showed that the problem is solved upon generalizing the notion of webs in the multi-leg case to be closed sets of diagrams rather than individual ones. These sets naturally arise upon computing the exponentiated colour factor of any given diagram as a linear combination of the colour factors of other diagrams. This gives rise to mixing as described by eq.~(\ref{setmix}), where any given set includes all the diagrams that are mutually related by permuting the gluon attachments to the eikonal lines. 
We argue that the compound object that is formed upon adding up the contributions of all 
the diagrams in the set, is the proper generalization of a web: 
it admits a generalised antisymmetry property with respect to permutations 
(realised through the zero-sum row property of eq.~(\ref{Rprop})) and it is expected to have a singularity structure that conforms with the renormalization of the multi-leg eikonal amplitude.  
  
The mechanism by which multiple-pole terms associated with subdivergences cancel out was investigated here in section \ref{sec:subdivergences}, revealing a very interesting picture. A first level of cancellation occurs within each individual web contributing to the unrenormalized amplitude. This cancellation is facilitated by the fact that the mixing matrices are idempotent, and therefore have just two eigenvalues $1$ and $0$. Thus only particular linear combinations of kinematic functions, those corresponding to eigenvalue $1$, enter the exponent; these combinations turn out to be very special: for example, for any web of order $n$, these combinations are free of the leading ${\cal O}(1/\epsilon^n)$ singularity. 
A second level of cancellation takes place upon renormalizing the eikonal amplitude. In the exponent of the renormalised amplitude, all remaining multiple-pole terms, ${\cal O}(1/\epsilon^j)$ with $j\leq n-1$
at ${\cal O}(\alpha_s^n)$ get cancelled by nested commutators of lower order webs and lower order counter-terms, such that a single ${\cal O}(1/\epsilon)$ counter-term will suffice. For webs to match this structure is highly non-trivial. In particular, all remaining multiple-pole terms in webs must be given by nested commutators of lower-order webs.

It is important to emphasize that this picture is still largely conjectural: first, the properties of multi-leg webs have not been proven, but rather been deduced by considering certain sets of diagrams. 
Second, it is not obvious that  renormalization of the multi-leg eikonal amplitude can indeed be performed on a web-by-web basis. In any case, further study is needed to fully expose the interplay between the singularity structure of individual multi-leg webs and the renormalization of the corresponding amplitude.

We observed that the structure that emerges in webs (see eq.~(\ref{setmix_matrix})) presents an intriguing relation between kinematics and colour space: the same diagonalising matrix $Y$ which generates antisymmetric combinations of colour factors (owing to the zero-sum property in the rows corresponding to unit eigenvalue) also acts on the kinematic functions to remove subdivergences. 
This beautiful structure clearly calls for an in-depth mathematical analysis, which will hopefully establish that the above results are indeed completely general, and shed light on the mechanism of cancellation of subdivergences.

We have obtained our results using the replica trick, albeit a simplified form of this idea, which was inspired by statistical physics methods~\cite{Replica}. Hence, as a concluding thought, it is interesting to ponder whether there are any other unexplored\footnote{For some existing applications, see~\cite{Arefeva:1983sv,Fujita:2008rs,Akemann:2000df,Damgaard:2000gh,Damgaard:2000di}.} uses of this idea in high energy physics, given the many other similarities that exist between the fields of high energy and statistical physics.

\newpage

\noindent
{\bf Note added:} As the first version of the present paper became public, A. Mitov, G. Sterman and I. Sung have completed a study of related topics~\cite{MSS}. While section 2 in~\cite{MSS} provides 
an alternative way to derive exponentiation in terms of webs, entirely consistent with our results in section \ref{sec:inverted_formula}, section 3 in~\cite{MSS} addresses the renormalization of the multi-leg eikonal vertex, elucidating the
role of commutators of lower order webs and counter-terms, which we overlooked in our original preprint. In this (published) version of the paper we confirm some of the results of this section, and furthermore use it to clarify the emerging picture of the singularity structure of webs.

\vspace*{30pt}

\acknowledgments

We are grateful to George Sterman and Lorenzo Magnea for inspiring discussions. 
We would like to thank the Particle Physics group at the University of Manchester for organising a  workshop on eikonal physics (November 2009), at which some of the ideas of this paper were first discussed. 
EL is supported by the Netherlands Foundation for Fundamental Research of Matter (FOM), and the National Organization for Scientific Research (NWO). CDW is funded by the STFC postdoctoral fellowship ``Collider Physics at the LHC''. He is grateful to Claude Duhr, Pietro Falgari and Paul Heslop for useful discussions. We have used JaxoDraw~\cite{Binosi:2008ig,Binosi:2003yf} throughout the paper.

\appendix

\section{Alternative derivation of ECF using eq.~(\ref{modcol2})\label{app-col2}}

In eqs.~(\ref{17-20rep}) and~(\ref{1-6rep}) we have given the exponentiated colour factors for the three loop diagrams of figures~\ref{17-20} and~\ref{3lsix} respectively, derived using the replica trick argument as explained in section~\ref{sec:replica}. The aim of this appendix is to rederive these results using the explicit combinatoric formula for exponentiated colour factors given in eq.~(\ref{modcol2}). This serves to demonstrate the application of this formula, as well as to illustrate how it encapsulates the result of the replica trick. 

Consider first the diagrams of figure~\ref{17-20}. The decompositions of figure~\ref{17-20}(a) are shown in figure~\ref{partex2}. There are five of them, and eq.~(\ref{modcol2}) explicitly gives
\begin{align}
&1\times C(3a)-\frac{1}{2}\left[\{C(2a),C(1c)\}+\{C(2e),C(1b)\}+\{C(2d),C(1a)\}\right]+\frac{1}{3}\left[C(1a)C(1b)C(1c)\right.\notag\\
&\left.+\text{perms}\right]\notag\\
&=C(3a)-\frac{1}{2}\left[C(3c)+C(3a)+C(3a)+C(3d)+C(3b)+C(3a)\right]\notag\\
&\quad+\frac{1}{3}\left[C(3c)+C(3a)+C(3d)+C(3d)+C(3a)+C(3b)\right]\notag\\
&=\frac{1}{6}\left[C(3a)-C(3b)-C(3c)+C(3d)\right],
\label{partex2res}
\end{align}
where we have used the labels of figures~\ref{1a-c}, \ref{2a-d} and \ref{2e}. The result indeed agrees with the result of eq.~(\ref{17rep}). 

Figure~\ref{17-20}(b) has decompositions $(3b)$, $(2b,1c)$, $(2e,1b)$, $(2d,1a)$ and $(1a,1b,1c)$, where we use a notation $(g_1,\ldots,g_n)$ labelling the diagrams $g_i$ in each decomposition. Applying eq.~(\ref{modcol2}) gives 
\begin{align}
&C(3b)-\frac{1}{2}\left[\{C(2b),C(1c)\}+\{C(2e),C(1b)\}+\{C(2d),C(1a)\}\right]+\frac{1}{3}\left[C(1a)C(1b)C(1c)\right.\notag\\
&\left.+\text{perms}\right]\notag\\
&=\frac{1}{3}\left[-C(3a)+C(3b)+C(3c)-C(3d)\right],
\label{18rep2}
\end{align}
agreeing with eq.~(\ref{18rep}). The corresponding results for figures \ref{17-20}(c) and \ref{17-20}(d) are
\begin{align}
&C(3c)-\frac{1}{2}\left[\{C(2a),C(1c)\}+\{C(2e),C(1b)\}+\{C(2c),C(1a)\}\right]+\frac{1}{3}\left[C(1a)C(1b)C(1c)\right.\notag\\
&\left.+\text{perms}\right]\notag\\
&=\frac{1}{3}\left[-C(3a)+C(3b)+C(3c)-C(3d)\right]
\label{19rep2}
\end{align}
and
\begin{align}
&C(3d)-\frac{1}{2}\left[\{C(2b),C(1c)\}+\{C(2e),C(1b)\}+\{C(2c),C(1a)\}\right]+\frac{1}{3}\left[C(1a)C(1b)C(1c)\right.\notag\\
&\left.+\text{perms}\right]\notag\\
&=\frac{1}{6}\left[C(3a)-C(3b)-C(3c)+C(3d)\right],
\label{20rep2}
\end{align}
respectively, again agreeing with the previous results of eqs.~(\ref{19rep}) and (\ref{20rep}).

One may also consider the diagrams of figure~\ref{3lsix}. Figure~\ref{3lsix}(a) has decompositions $(3A)$, $(2g,1b)$, $(2a,1a)$ (repeated twice) and $(1a,1a,1b)$. Applying eq.~(\ref{modcol2}), taking care with the repeated decomposition gives
\begin{align}
&C(3A)-\frac{1}{2}\left[\{C(2g),C(1b)\}+2\{C(2a),C(1a)\}\right]+\frac{1}{3}\left[C(1a)C(1a)C(1b)+\text{perms}\right]\notag\\
&=\frac{1}{6}\left[3C(3A)-3C(3C)-2C(3D)-2C(3E)+4C(3F)\right],
\label{1rep2}
\end{align}
which agrees with the result of eq.~(\ref{resA}). Equivalent formulae for the other diagrams in figure~\ref{3lsix} are
\begin{align}
&C(3B)-\frac{1}{2}\left[\{C(2g),C(1b)\}+\{C(2a),C(1a)\}+\{C(2b),C(1a)\}\right]\notag\\
&\quad+\frac{1}{3}\left[C(1a)C(1a)C(1b)+\text{perms}\right]\notag\\
&=\frac{1}{6}\left[-3C(3A)+6C(3B)-3C(3C)+C(3D)-2C(3E)+C(3F)\right];
\label{2rep2}\\
&C(3C)-\frac{1}{2}\left[\{C(2g),C(1b)\}+2\{C(2b),C(1a)\}\right]+\frac{1}{3}\left[C(1a)C(1a)C(1b)+\text{perms}\right]\notag\\
&=\frac{1}{6}\left[-3C(3A)+3C(3C)+4C(3D)-2C(3E)-2C(3F)\right];
\label{3rep2}\\
&C(3D)-\frac{1}{2}\left[\{C(2f),C(1b)\}+2\{C(2a),C(1a)\}\right]+\frac{1}{3}\left[C(1a)C(1a)C(1b)+\text{perms}\right]\notag\\
&=\frac{1}{6}\left[C(3D)-2C(3E)+C(3F)\right];
\label{4rep2}\\
&C(3E)-\frac{1}{2}\left[\{C(2f),C(1b)\}+\{C(2a),C(1a)\}+\{C(2b),C(1a)\}\right]\notag\\
&\quad+\frac{1}{3}\left[C(1a)C(1a)C(1b)+\text{perms}\right]\notag\\
&=\frac{1}{3}\left[-C(3D)+2C(3E)-C(3F)\right];
\label{5rep2}\\
&C(3F)-\frac{1}{2}\left[\{C(2f),C(1b)\}+2\{C(2b),C(1a)\}\right]+\frac{1}{3}\left[C(1a)C(1a)C(1b)+\text{perms}\right]\notag\\
&=\frac{1}{6}\left[C(3D)-2C(3E)+C(3F)\right];
\label{6rep2}
\end{align}
all of which agree with the results of eqs.~(\ref{resB}-\ref{resF}).

\section{Example applications of the generalised Gatheral formula\label{app-gatheral}}

Here, for completeness, we provide some examples of the application of the generalised Gatheral formula, eq.~(\ref{colfacfin}), which expresses the conventional colour factor of a graph $G$ in terms of lower order exponentiated colour factors. First, let us consider diagram (3a) of figure~\ref{17-20}. The decompositions of this diagram are given in figure~\ref{partex2}, and applying eq.~(\ref{colfacfin}) gives
\begin{align}
C(3a)&=\widetilde{C}(3d)+\frac{1}{2}\left[\widetilde{C}(2a)\widetilde{C}(1c)+\widetilde{C}(1c)\widetilde{C}(2a)+\widetilde{C}(2e)\widetilde{C}(1b)+\widetilde{C}(1b)\widetilde{C}(2e)+\widetilde{C}(2d)\widetilde{C}(1a)\right.\notag\\
&\left.\quad+\widetilde{C}(1a)\widetilde{C}(2d)\right]+\frac{1}{6}\left[\widetilde{C}(1a)\widetilde{C}(1b)\widetilde{C}(1c)+\text{perms}\right],
\label{3agengath}
\end{align}
where we have used the labels introduced in figures~\ref{1a-c}, \ref{2a-d} and \ref{2e}. In obtaining this result, we have summed over all decompositions, and used the fact that the multiplicity factor $N_{G|\{m_H\}}=1$ in all cases. Upon substituting $\widetilde{C}(1a)=C(1a)$ etc. and using the results of eqs.~(\ref{twoemcol7}-\ref{20rep}), one may combine the colour factors on the right-hand side of eq.~(\ref{3agengath}) to obtain $C(3D)$, thus showing that eqs.~(\ref{modcol2}) and~(\ref{colfacfin}) are consistent. 

As a second example, let us consider diagram (3A) of figure~\ref{3lsix}. This is an interesting case, as it involves nontrivial examples of the multiplicity factors $N_{G|\{m_H\}}$. Applying eq.~(\ref{colfacfin}) for this graph gives
\begin{align}
C(3A)&=\widetilde{C}(3A)+\frac{1}{2}\left[\widetilde{C}(2g)\widetilde{C}(1b)+\widetilde{C}(1b)\widetilde{C}(2g)+2\widetilde{C}(2a)\widetilde{C}(1a)+2\widetilde{C}(1a)\widetilde{C}(2a)\right]\notag\\
&\quad+\frac{1}{6}\left[\widetilde{C}(1a)\widetilde{C}(1a)\widetilde{C}(1b)+\text{perms}\right],
\label{3Agengath}
\end{align}
where we have used the diagram labels in figures~\ref{1a-c} and~\ref{2f2g} on the right-hand side. Here the three terms correspond to decompositions containing 1, 2 and 3 diagrams respectively. In the first and second terms, we have used the fact that the multiplicity factor $N_{G|\{m_H\}}=1$ in all cases. In the third term, we have used the fact that $N_{G|\{m_H\}}=2$, and also instated the combinatoric factor $(3!2!)^{-1}$ from the prefactor $(n!\prod m_H!)^{-1}$ in eq.~(\ref{colfacfin}). One may show using previously obtained results for the modified colour factors that the right-hand side is indeed equal to $C(3A)$, again demonstrating the consistency of eqs.~(\ref{modcol2}) and (\ref{colfacfin}).

\section{Additional examples of three and four loop webs\label{sec:more-3loop-ECF}}

In this appendix, we give additional examples of three and four loop webs. Together with the other examples in this paper, this completes the survey of all three loop webs which occur in the case when the external particles are massless (we do not discuss here connected diagrams with a single attachment to each eikonal line: in this case there is no mixing, each such diagram is a web by itself and its exponentiated colour factor is equal to the original one). We also give a couple of examples of sets of graphs containing an eikonal-line self-energy, which is present only in the case of massive external particles. Finally we give another interesting four-loop example.
We adopt the notation for diagrams introduced in section~\ref{sec:replica}, and present the results for the mixing matrices $R_{DD'}$ of eq.~(\ref{setmix}) together with the column vector $C(D')$ on which it acts.

\begin{figure}[htb]
\begin{center}
\scalebox{1.0}{\includegraphics{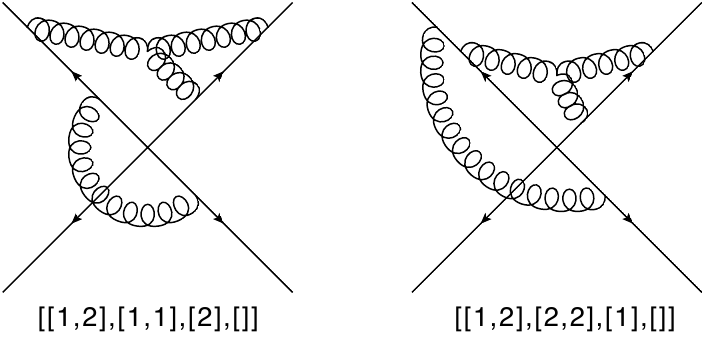}}
\caption{Diagrams contributing to eq.~(\ref{appex1}).}
\label{appendixfig1}
\end{center}
\end{figure}

Let us begin with an example with a three gluon vertex, shown in figure~\ref{appendixfig1}.
There is a doublet of diagrams with (2,2,1) attachments, where by this notation we denote the number of gluon attachments on lines 1, 2 and 3 respectively. The mixing matrix is given by
\begin{align}
\frac{1}{6}\left( \begin {array}{rr} 3&-3\\ \noalign{\medskip}-3&3\end {array}
 \right) 
 \left( \begin {array}{c} C[[1,2],[1,1],[2],[]]\\ \noalign{\medskip}C[[1
,2],[2,2],[1],[]]\end {array} \right),
\label{appex1}
\end{align}
where we also show the vector of colour factors on the right-hand side, so as to define the ordering of rows in the matrix.

\begin{figure}[htb]
\begin{center}
\scalebox{1.0}{\includegraphics{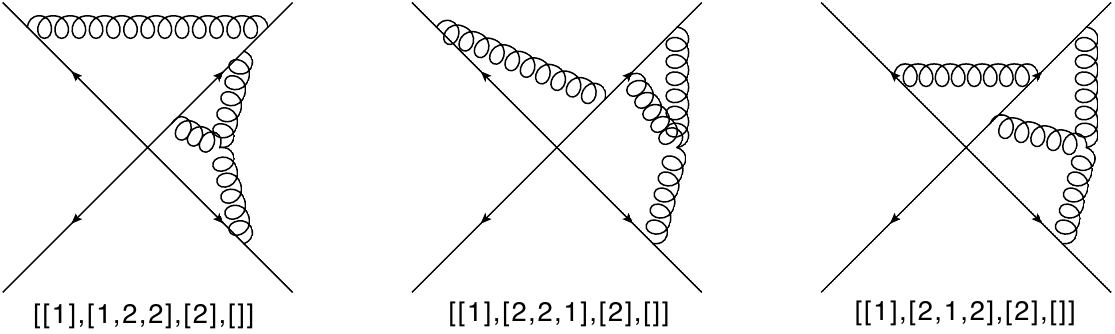}}
\caption{Diagrams contributing to eq.~(\ref{appex2}).}
\label{appendixfig2}
\end{center}
\end{figure}
Next, we consider the diagrams of figure~\ref{appendixfig2} {\it i.e.} a triplet with (1,3,1) attachments:
\begin{align}
\frac{1}{6}\left( \begin {array}{rrr} 3&-3&0\\ \noalign{\medskip}-3&3&0
\\ \noalign{\medskip}-3&-3&6\end {array} \right) 
 \left( \begin {array}{c} C[[1],[1,2,2],[2],[]]\\ \noalign{\medskip}C[[1
],[2,2,1],[2],[]]\\ \noalign{\medskip}C[[1],[2,1,2],[2],[]]\end {array}
 \right)
\label{appex2}
\end{align}

\begin{figure}[htb]
\begin{center}
\scalebox{0.9}{\includegraphics{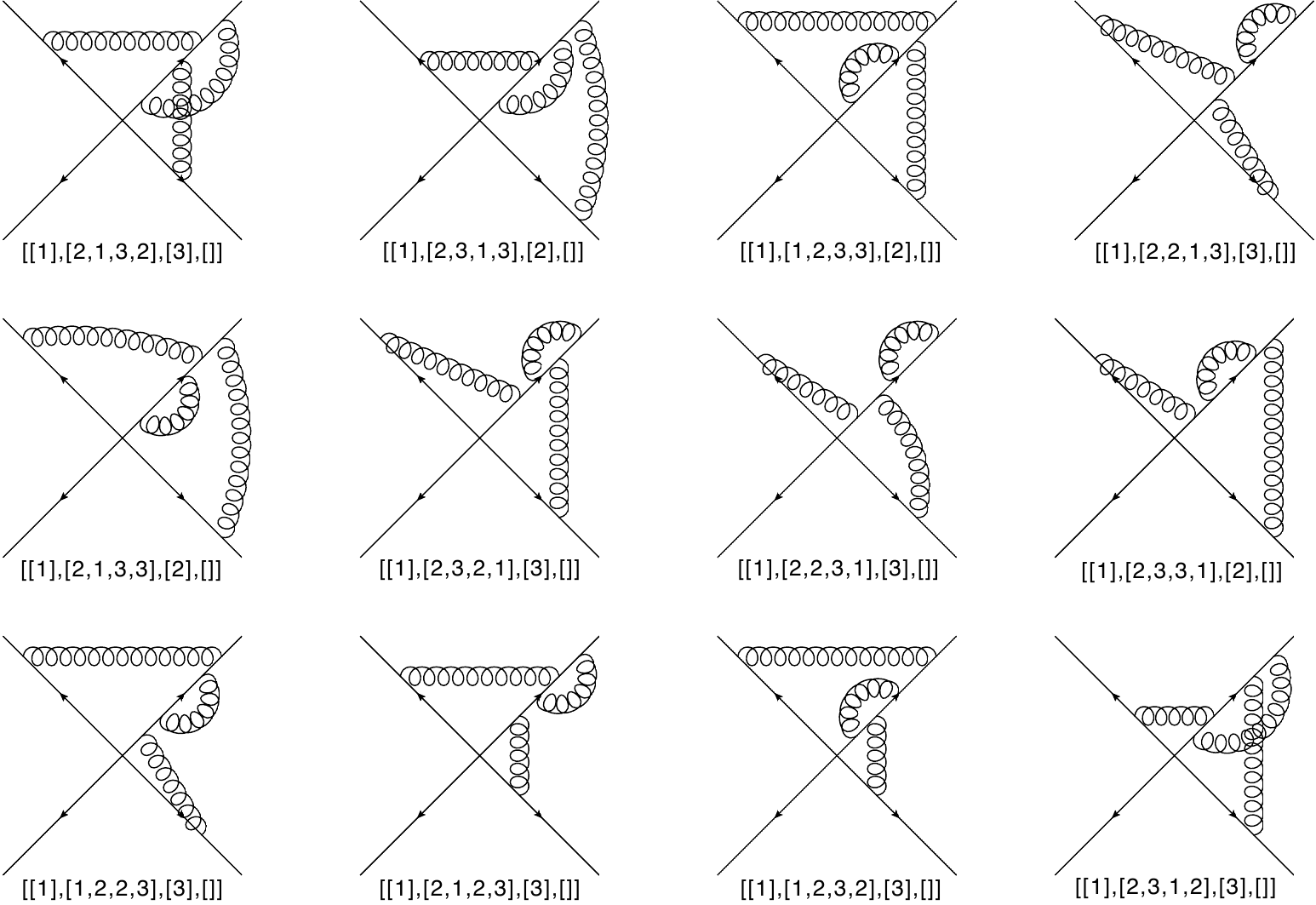}}
\caption{Diagrams contributing to eq.~(\ref{appex3}).}
\label{appendixfig3}
\end{center}
\end{figure}

\begin{figure}[htb]
\begin{center}
\scalebox{1.0}{\includegraphics{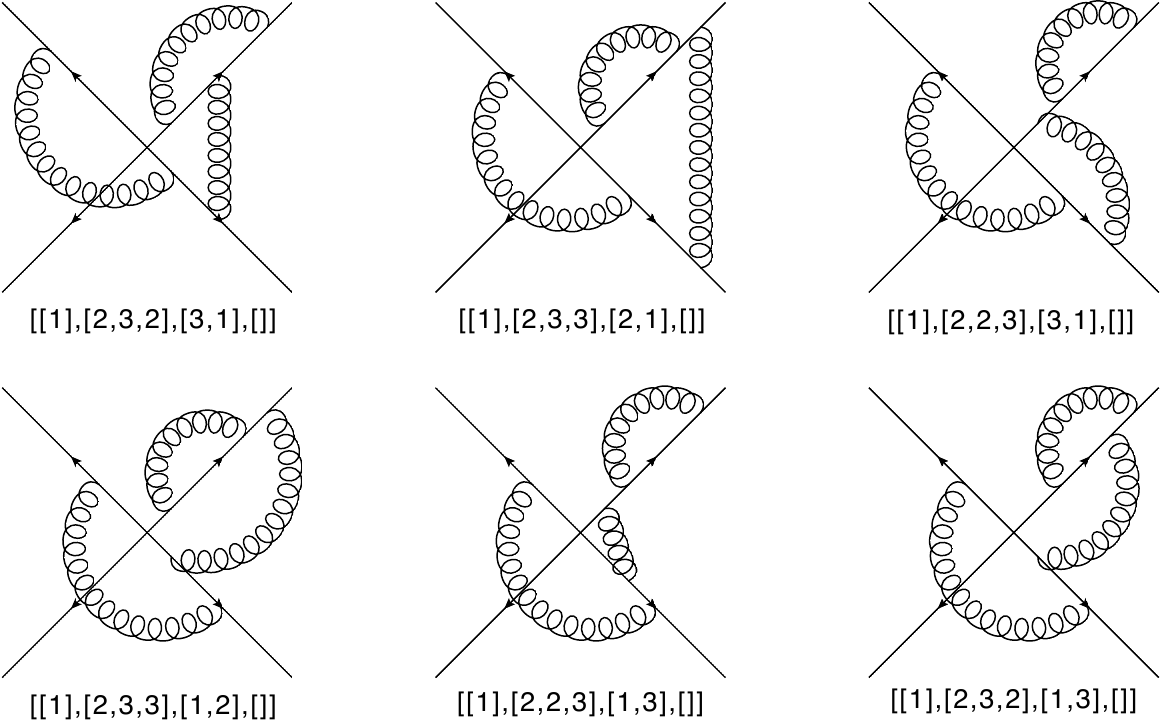}}
\caption{Diagrams contributing to eq.~(\ref{appex6}).}
\label{appendixfig6}
\end{center}
\end{figure}

In figures \ref{appendixfig3} and \ref{appendixfig6} we show a couple of three-loop sets of diagrams where one of the gluons is attached to the same line at both ends, an eikonal-line self-energy type graph. As remarked above, these diagrams vanish in the massless (lightlike eikonal line) case but not so in the massive case. The mixing matrix is
\begin{align}
\frac{1}{6}
\left( \begin {array}{rrrrrrrrrrrr} 6&-3&-1&-1&2&-3&2&2&2&-3&-3&0
\\ \noalign{\medskip}0&3&-1&2&-1&0&-1&-1&2&-3&0&0\\ \noalign{\medskip}0
&0&2&-1&-1&0&2&-1&-1&0&0&0\\ \noalign{\medskip}0&0&-1&2&2&0&-1&-1&-1&0
&0&0\\ \noalign{\medskip}0&0&-1&2&2&0&-1&-1&-1&0&0&0
\\ \noalign{\medskip}0&0&2&-1&-1&3&-1&-1&2&0&-3&0\\ \noalign{\medskip}0
&0&2&-1&-1&0&2&-1&-1&0&0&0\\ \noalign{\medskip}0&0&-1&-1&-1&0&-1&2&2&0
&0&0\\ \noalign{\medskip}0&0&-1&-1&-1&0&-1&2&2&0&0&0
\\ \noalign{\medskip}0&-3&-1&-1&2&0&-1&2&-1&3&0&0\\ \noalign{\medskip}0
&0&-1&-1&-1&-3&2&2&-1&0&3&0\\ \noalign{\medskip}0&-3&2&2&-1&-3&-1&2&2&
-3&-3&6\end {array} \right) 
 \left( \begin {array}{c} C[[1],[2,1,3,2],[3],[]]\\ 
\noalign{\medskip}C[[1],[2,3,1,3],[2],[]]\\ 
\noalign{\medskip}C[[1],[1,2,3,3],[2],[]]\\ 
\noalign{\medskip}C[[1],[2,2,1,3],[3],[]]\\ 
\noalign{\medskip}C[[1],[2,1,3,3],[2],[]]\\ 
\noalign{\medskip}C[[1],[2,3,2,1],[3],[]]\\ 
\noalign{\medskip}C[[1],[2,2,3,1],[3],[]]\\ 
\noalign{\medskip}C[[1],[2,3,3,1],[2],[]]\\ 
\noalign{\medskip}C[[1],[1,2,2,3],[3],[]]\\ 
\noalign{\medskip}C[[1],[2,1,2,3],[3],[]]\\ 
\noalign{\medskip}C[[1],[1,2,3,2],[3],[]]\\ 
\noalign{\medskip}C[[1],[2,3,1,2],[3],[]]
\end {array} \right)
\label{appex3}
\end{align}

We have used the notation $(n_1,n_2,\ldots,n_L)$ throughout the paper for indicating the number of gluon attachments on each external line. We emphasize that this notation does not uniquely specify a given set of diagrams, as can be seen from the following example of a (1,3,2) web involving a self-energy. The relevant diagrams are shown in figure~\ref{appendixfig6}, and their topologies are different to those obtained by rotating the diagrams of figure~\ref{3lsix} to form (1,3,2) diagrams. The mixing matrix in this case is given by
\begin{equation}
\frac{1}{6}
 \left( \begin {array}{rrrrrr} 3&-2&-1&2&1&-3\\ \noalign{\medskip}0&1&
-1&-1&1&0\\ \noalign{\medskip}0&-2&2&2&-2&0\\ \noalign{\medskip}0&-2&2
&2&-2&0\\ \noalign{\medskip}0&1&-1&-1&1&0\\ \noalign{\medskip}-3&1&2&-
1&-2&3\end {array} \right) 
 \left( \begin {array}{c} C[[1],[2,3,2],[3,1],[]]\\ \noalign{\medskip}C[
[1],[2,3,3],[2,1],[]]\\ \noalign{\medskip}C[[1],[2,2,3],[3,1],[]]
\\ \noalign{\medskip}C[[1],[2,3,3],[1,2],[]]\\ \noalign{\medskip}C[[1],[
2,2,3],[1,3],[]]\\ \noalign{\medskip}C[[1],[2,3,2],[1,3],[]]
\end {array} \right).
\label{appex6}
\end{equation}

Now we have a couple of examples where all three gluons connect different lines, first with (2,2,2) attachments, shown in  figure~\ref{appendixfig4}:
\begin{figure}[htb]
\begin{center}
\scalebox{0.93}{\includegraphics{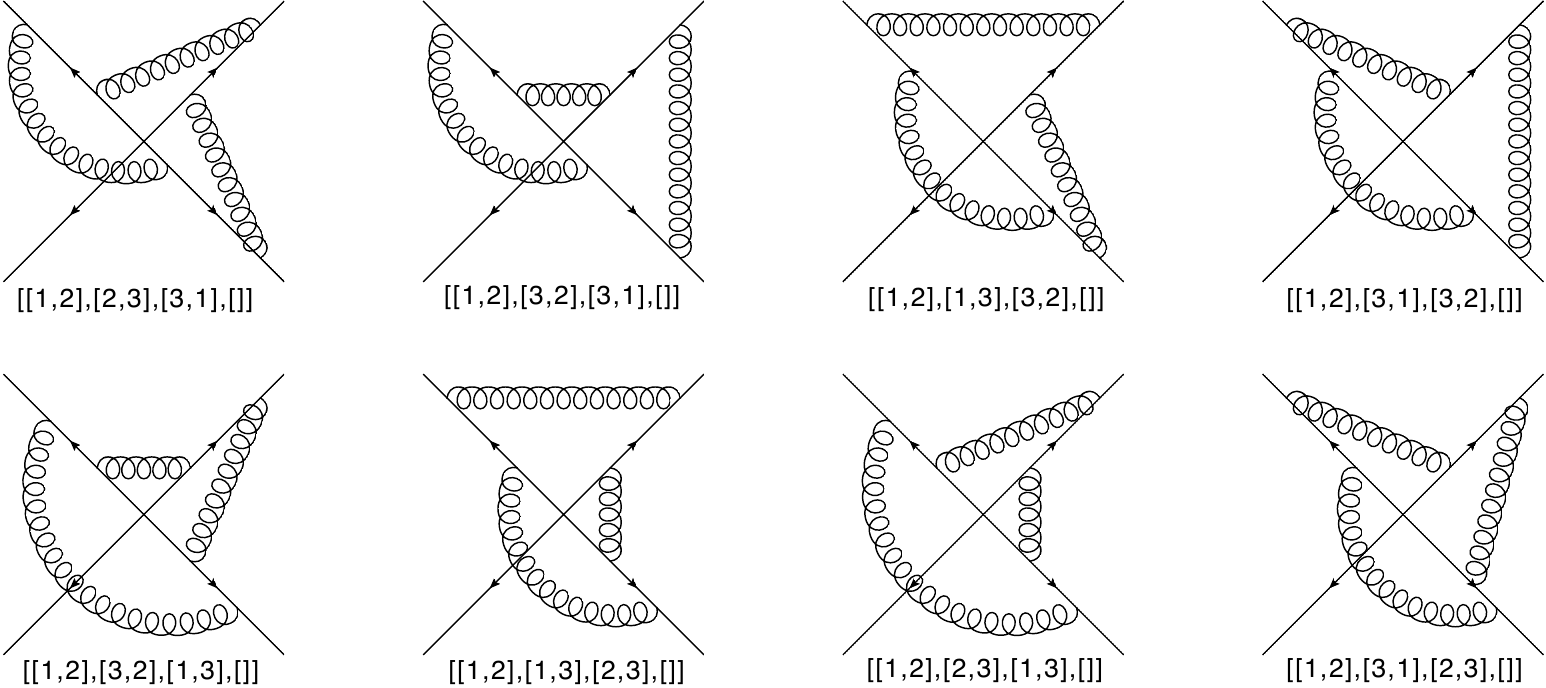}}
\caption{Diagrams contributing to eq.~(\ref{appex4}).}
\label{appendixfig4}
\end{center}
\end{figure}
\begin{align}
\frac{1}{6}\left( \begin {array}{rrrrrrrr} 6&-4&-4&2&2&2&-4&0
\\ \noalign{\medskip}0&2&-1&-1&-1&2&-1&0\\ \noalign{\medskip}0&-1&2&-1
&2&-1&-1&0\\ \noalign{\medskip}0&-1&-1&2&-1&-1&2&0
\\ \noalign{\medskip}0&-1&2&-1&2&-1&-1&0\\ \noalign{\medskip}0&2&-1&-1
&-1&2&-1&0\\ \noalign{\medskip}0&-1&-1&2&-1&-1&2&0
\\ \noalign{\medskip}0&2&2&-4&-4&-4&2&6\end {array} \right)
 \left( \begin {array}{c} C[[1,2],[2,3],[3,1],[]]\\ 
\noalign{\medskip}C[[1,2],[3,2],[3,1],[]]\\ 
\noalign{\medskip}C[[1,2],[1,3],[3,2],[]]\\ 
\noalign{\medskip}C[[1,2],[3,1],[3,2],[]]\\
\noalign{\medskip}C[[1,2],[3,2],[1,3],[]]\\ 
\noalign{\medskip}C[[1,2],[1,3],[2,3],[]]\\
\noalign{\medskip}C[[1,2],[2,3],[1,3],[]]\\ 
\noalign{\medskip}C[[1,2],[3,1],[2,3],[]]\end {array} \right) 
\label{appex4}
\end{align}
and then with (1,3,1,1) attachments, shown in figure \ref{appendixfig5}:
\begin{figure}[htb]
\begin{center}
\scalebox{1.0}{\includegraphics{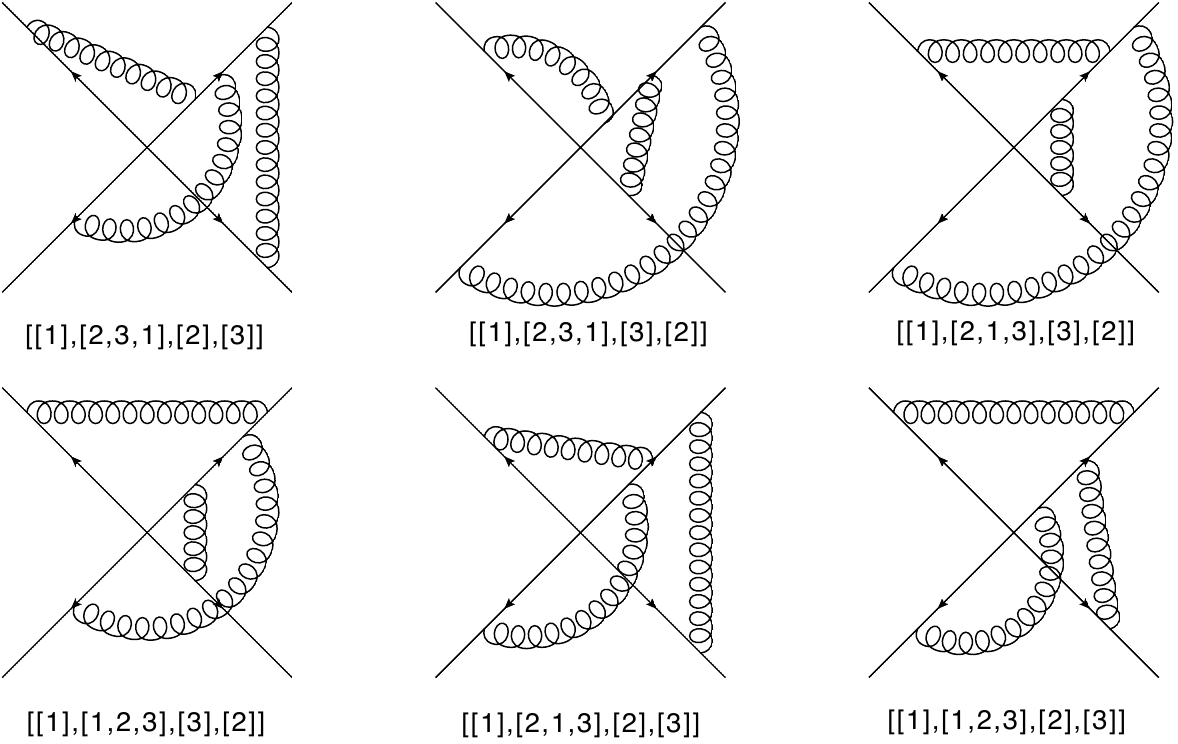}}
\caption{Diagrams contributing to eq.~(\ref{appex5}).}
\label{appendixfig5}
\end{center}
\end{figure}
\begin{align}
\frac{1}{6}\left(  \begin {array}{rrrrrr} 2&-1&-1&2&-1&-1\\ \noalign{\medskip}-1&
2&-1&-1&-1&2\\ \noalign{\medskip}-1&-1&2&-1&2&-1\\ \noalign{\medskip}2
&-1&-1&2&-1&-1\\ \noalign{\medskip}-1&-1&2&-1&2&-1
\\ \noalign{\medskip}-1&2&-1&-1&-1&2\end {array} \right) 
 \left( \begin {array}{c} C[[1],[2,3,1],[2],[3]]\\ \noalign{\medskip}C[[
1],[2,3,1],[3],[2]]\\ \noalign{\medskip}C[[1],[2,1,3],[3],[2]]
\\ \noalign{\medskip}C[[1],[1,2,3],[3],[2]]\\ \noalign{\medskip}C[[1],[2
,1,3],[2],[3]]\\ \noalign{\medskip}C[[1],[1,2,3],[2],[3]]\end {array}
 \right) 
\label{appex5}
\end{align}

\begin{figure}[htb]
\begin{center}
\scalebox{1.12}{\includegraphics{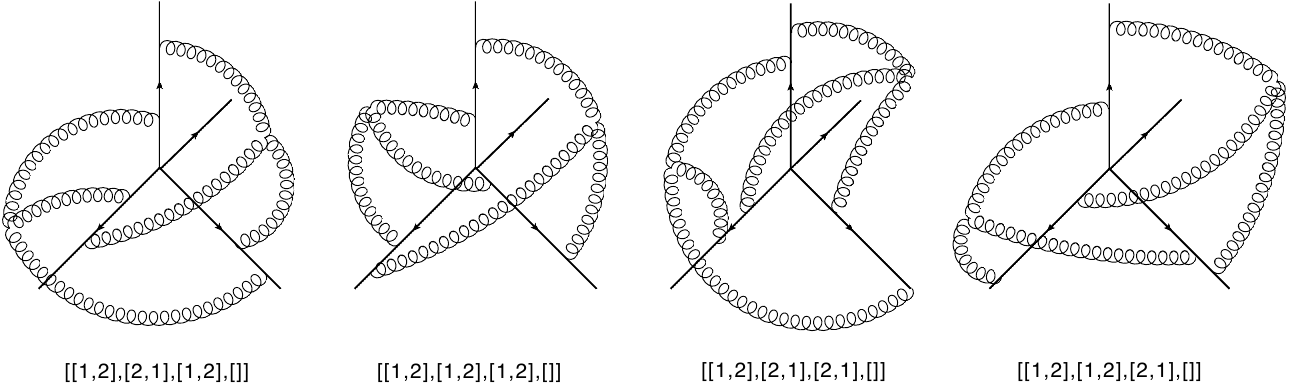}}
\caption{Diagrams contributing to eq.~(\ref{appex7}).}
\label{appendixfig7}
\end{center}
\end{figure}

Finally, we consider the four-loop example of figure~\ref{appendixfig7}. 
The mixing matrix in this case is given by:
\begin{equation}
 \frac{1}{24}\,\left( \begin {array}{cccc} 24&-24&0&0\\ \noalign{\medskip}0&0&0&0
\\ \noalign{\medskip}0&-24&24&0\\ \noalign{\medskip}0&-24&0&24
\end {array} \right) 
 \left( \begin {array}{c} C[[1,2],[2,1],[1,2],[]]\\ \noalign{\medskip}C[[1,
2],[1,2],[1,2],[]]\\ \noalign{\medskip}C[[1,2],[2,1],[2,1],[]]
\\ \noalign{\medskip}C[[1,2],[1,2],[2,1],[]]\end {array}
 \right) 
\label{appex7}
\end{equation}
Note that here the three-eikonal-line reducible diagram 
[[1,2],[1,2],[1,2],[]] does not appear in the exponent, while all three others, in which the gluons are entangled, do.

\bibliographystyle{JHEP}
\bibliography{refs}

\end{document}